\documentclass[aps,prd,twocolumn,floatfix,showpacs,eqsecnum]{revtex4-1}

\pdfoutput=1
\usepackage[centertags]{amsmath}
\usepackage{amssymb}
\usepackage{latexsym}
\usepackage{enumerate}
\usepackage{graphicx}
\usepackage{mathrsfs}
\usepackage{hyperref}
\usepackage{stmaryrd}

\usepackage{soul}

\newcommand{\bi}{\begin{itemize}}
\newcommand{\ei}{\end{itemize}}

\newcommand{\be}{\begin{equation}}
\newcommand{\ee}{\end{equation}}

\newcommand{\ba}{\begin{eqnarray}}
\newcommand{\ea}{\end{eqnarray}}

\newcommand{\nn}{\nonumber}

\newcommand{\g}{\gamma}

\newcommand{\mnras}{MNRAS}

\newcommand{\aap}{A\&A}

\newcommand{\nature}{Nature}

\begin{document}

\title{Relativistic effects in the tidal interaction between a white dwarf 
and a massive\\ black hole in Fermi normal coordinates} 

\author{Roseanne M. Cheng}
\email{rcheng@physics.gatech.edu}
\affiliation{Center for Relativistic Astrophysics, School of Physics, 
Georgia Institute of Technology, Atlanta, Georgia 30332, USA}
\affiliation{Department of Physics and Astronomy, University of North 
Carolina, Chapel Hill, North Carolina 27599, USA}

\author{Charles R. Evans}
\email{evans@physics.unc.edu}
\affiliation{Department of Physics and Astronomy, University of North 
Carolina, Chapel Hill, North Carolina 27599\\
{\rm(Received 17 March 2013; published 6 May 2013)}}

\begin{abstract}

We consider tidal encounters between a white dwarf and an intermediate mass 
black hole.  Both weak encounters and those at the threshold of disruption 
are modeled.  The numerical code combines mesh-based hydrodynamics, 
a spectral method solution of the self-gravity, and a general 
relativistic Fermi normal coordinate system that follows the star and 
debris.  Fermi normal coordinates provide an expansion of the black hole tidal 
field that includes quadrupole and higher multipole moments and relativistic 
corrections.  We compute the mass loss from the white dwarf that occurs 
in weak tidal encounters.  Secondly, we compute carefully the energy 
deposition onto the star, examining the effects of nonradial and radial 
mode excitation, surface layer heating, mass loss, and relativistic orbital 
motion.  We find evidence of a slight relativistic suppression in tidal 
energy transfer.  Tidal energy deposition is compared to orbital energy 
loss due to gravitational bremsstrahlung and the combined losses are used 
to estimate tidal capture orbits.  Heating and partial mass stripping will 
lead to an expansion of the white dwarf, making it easier for the star to 
be tidally disrupted on the next passage.  Finally, we examine angular 
momentum deposition.  By including the octupole tide, we are able for 
the first time to calculate deflection of the center of mass of the star and 
debris.  With this observed deflection, and taking into account orbital 
relativistic effects, we compute directly the change in orbital angular 
momentum and show its balance with computed spin angular momentum deposition. 

\end{abstract}

\pacs{04.25.Nx, 04.40.Nr, 04.70.Bw, 98.62.Js \hfill 
DOI: 10.1103/PhysRevD.87.104010}

\maketitle

\section{Introduction}
\label{sec:intro}

The tidal disruption of a star can serve as a diagnostic for the presence of 
a dormant black hole in a distant galaxy~\cite{rees1988,komo2004}.  The 
theoretically predicted rate is $10^{-5}$ to $10^{-4}$ yr$^{-1}$ for 
galaxies like the Milky Way~\cite{mago1999,wang2004}.  While such tidal 
disruption events (TDEs) are rare, they give rise to powerful flares 
of emission at and above Eddington 
luminosity~\cite{cart1982,rees1988,phin1989,evan1989,stru2009}, with spectral 
features and time scales that might reveal both the type of star and the mass 
(and perhaps spin) of the black hole.  For main sequence stars disruption may 
occur in close encounters with supermassive black holes (SMBH) of mass 
$M \lesssim 10^8 (R_*/R_{\odot})^{3/2}(M_*/M_{\odot})^{-1/2} M_{\odot}$, 
where $M_*$ and $R_*$ are the mass and radius of the star.  With a black hole 
of higher mass the star will cross the event horizon before disrupting.  The 
upper mass limit is increased if the black hole's spin is near 
maximal~\cite{belo1992,kesd2012}.  Higher black hole masses are relevant for 
stripping red giant envelopes.  Disrupting a white dwarf in contrast 
requires an intermediate mass black hole (IMBH) with 
$M \lesssim 1.8 \times 10^5 (R_*/0.012 R_{\odot})^{3/2}
(M_*/0.6 M_{\odot})^{-1/2} M_{\odot}$.  

In excess of a dozen TDE candidates have been discovered so far.  A number 
of these have been observed in x-ray~\cite{grup1999,komo1999,grei2000,
komo2004,halp2004,esqu2007,capp2009,komo2009,maks2010,cumm2011,cenk2011}, 
with others picked up in the optical/UV~\cite{renz1995,geza2008,geza2009,
vanv2011,geza2012} and radio~\cite{zaud2011}.  Two of the most recent TDEs 
were detected by the {\it Swift} satellite: {\it Swift} J164449.3+573451 
(hereafter {\it Sw} J1644+57)~\cite{cumm2011,leva2011,bloo2011,burr2011,
zaud2011} and {\it Swift} J2058.4+0516~\cite{cenk2011}.  These two events 
have high-energy features 
and coincident radio emission that imply the presence of collimated 
relativistic jets (i.e., blazarlike activity).  Another object, PS1-10jh, 
discovered~\cite{geza2012} in the Pan-STARRS1 Medium Deep Survey, appears 
to have been the disruption of a helium-rich stellar core, whose red giant 
envelope was presumably stripped in some preceding tidal event.

A star is disrupted if its orbit reaches within the tidal radius of the 
black hole, given by $R_t \simeq R_* (M/M_*)^{1/3}$.  For a parabolic 
orbit that just reaches the tidal radius at pericenter, $R_t = R_p$, 
theoretical considerations indicate the star should disrupt with 
approximately half the debris bound to the hole and half ejected from the 
system.  The free streaming of bound debris implies a late-time mass return 
rate that decays as $t^{-5/3}$~\cite{rees1988}.  Early numerical models 
confirmed the expectations, showing a rapid rise in the mass return rate 
well above Eddington level before the power-law decay set 
in~\cite{evan1989}.  The precise form of the early rise and plateau is 
sensitive to stellar structure and effects of hydrodynamic shocks as 
the star undergoes disruption~\cite{loda2009}.  The picture is altered if 
the star approaches on an already bound orbit~\cite{haya2012}, which affects 
the late time-dependence of returning mass.  Moreover, stars may have orbits 
that result in a weak tidal encounter or that penetrate well inside the tidal 
radius~\cite{cart1982,koba2004}.  The strength of the encounter is often 
parametrized by $\eta = (R_p^3/M)^{1/2}(M_*/R_*^3)^{1/2}$~\cite{pres1977}, 
or by the penetration factor $\beta = \eta^{-2/3}$ where $R_p = R_t/\beta$.  
In the deep plunging case $\beta \gg 1$, strong tidal compression leads to 
break out of a shock and a prompt x-ray flare, which comes in advance of 
the flare produced as captured gas streams back to the black hole after 
one or more orbits.  Other potential observational signatures include 
supernovalike remnant structures associated with the ejection of 
debris~\cite{khok1996,kase2010}, coincident electromagnetic and gravitational 
wave signals~\cite{koba2004}, and possible thermonuclear runaway from severe 
compression of white dwarfs~\cite{cart1982,rath2005,rosA2009}.

Tidal disruption has been investigated with various numerical 
means, including smoothed particle hydrodynamics 
~\cite{nolt1982,bick1983,evan1989,lagu1993,ayal2000,koba2004,loda2009}, 
mesh-based finite difference or spectral 
methods~\cite{khok1993,khok1993strong,frol1994,marc1996,dien1997,guil2009}, and
affine models~\cite{cart1982,cart1983,lumi1986,koch1992,ivan2001,ivan2003}.  
Numerical modeling of radiative processes is also important for
understanding how TDEs appear in multiple wave bands~\cite{bogd2004} and in
accounting for nonadiabatic effects on the dynamics.  In TDEs that involve 
white dwarfs, the close approach to the black hole requires handling
general relativistic effects~\cite{frol1994,dien1997,haas2012}.  Moreover, 
new observations continue to bring surprises, such as the unanticipated 
rapid formation of relativistic jets and synchrotron radio 
emission~\cite{bloo2011} associated with {\it Sw} 1644+57.  These 
observations point to the need to include magnetohydrodynamics in models 
as well (see arguments in Ref.~\cite{shch2012}).

The {\it Sw} 1644+57 event is generally thought to represent tidal
disruption of a main sequence star by a $10^6$ to $10^7 M_{\odot}$
SMBH~\cite{bloo2011,burr2011,leva2011,zaud2011}.  An alternative
model, however, posits that the observed short time scales and
multiple bursts can be best explained via the disruption of a white
dwarf by an IMBH~\cite{krol2011,krol2012} in multiple passes.
Whatever the case may be with {\it Sw} 1644+57, because of the shorter
time scales and higher mass return rate, it has been argued that white
dwarf TDEs may end up being frequently observed in future flux-limited
surveys~\cite{shch2012}.

In this paper our focus is on encounters between white dwarfs and IMBHs, 
with primary attention devoted to events at threshold for disruption 
($\eta \simeq 1$) or weaker ($\eta >1$).  We are particularly interested in 
(1) computing partial mass loss in weaker encounters, (2) calculating 
accurately energy and angular momentum deposition, (3) observing relativistic 
effects, and (4) determining the capture orbits of white dwarfs after their 
initial passage.  To this end, we have constructed a new numerical code 
whose central feature is use of Fermi normal coordinates 
(FNCs)~\cite{ishi2005}.  In FNCs the spacetime geometry of the black hole 
is expanded in the vicinity of a timelike geodesic that approximately tracks 
the center of mass (CM) motion of the star (and debris).  Our approach is 
similar to Ref.~\cite{frol1994} but differs in the use of a higher-order 
expansion of the tidal field, which includes higher moments and relativistic 
corrections.  

Despite relativistic orbital motion, these coordinates allow use of Newtonian 
hydrodynamics and self-gravity within the FNC domain.  By not including 
post-Newtonian (PN) corrections to the stellar self-gravity and hydro, a 
floor is set on the accuracy ($\simeq 10^{-4}$) of the model.  Nevertheless, 
we show that octupole and $l = 4$ moments can be significant, as well as 
several orbital PN corrections.  Hydrodynamics is computed using a 
piece-wise parabolic method (PPM) Lagrangian remap code (PPMLR).  The 
self-gravitational field is obtained using three-dimensional fast Fourier 
transforms.  The code runs on cluster computers and each of our 
three-dimensional simulations is run at several resolutions to confirm 
numerical convergence.  Since the size of the FNC domain is necessarily 
limited, there is a limit on how long a disrupted star or stripped gas can 
be followed.  In principle, however, our simulations can serve as initial 
conditions for a second code that would calculate the return of gas to 
the black hole and formation of an accretion disk.

This paper is organized as follows.  In Sec. \ref{sec:formalism} the 
formalism is presented for including relativistic effects and higher 
moments in tidal interactions.  We give an orbital PN expansion of the tidal 
field and consider the order of magnitude significance of various terms.  
Tidal field moments through $l =4$ might be significant.  Similarly, 
(orbital) PN corrections to $l = 2$ and $l = 3$, as well as the 
gravitomagnetic potential, might be significant.  The fluid equations of 
motion are given with this level of approximation.  We discuss our numerical 
results in Secs.~\ref{sec:models-orbits} through~\ref{sec:tidal-angular}, 
relegating to Appendix~\ref{sec:method} description of the numerical 
hydrodynamics and self-gravity methods.  Section~\ref{sec:models-orbits} 
details our initial stellar model and the range of inbound orbits we consider.  
Section~\ref{sec:features-massloss} discusses the overall hydrodynamic 
features and shows the amount of mass loss from a white dwarf during 
various weak tidal encounters.  For TDEs considered here, we found 
the $l = 4$ moment to have negligible impact.  The situation with the 
gravitomagnetic potential is subtle, and we intend to address it in a 
subsequent paper.  In Sec.~\ref{sec:tidal-energy} we consider energy 
deposition onto the star and energy loss from the orbit.  We discuss the 
combined effects of nonradial and radial mode excitation, surface layer 
heating, mass loss, relativistic orbital motion, and gravitational 
bremsstrahlung.  Section~\ref{sec:tidal-angular} presents results on 
deposition of angular momentum (spin) onto the star.  By including the 
octupole tide, we show in Sec.~\ref{sec:tidal-cm} the computed tidal 
deflection of the CM and, with relativistic effects accounted for, relate 
it to the reduction of orbital angular momentum.  In 
Sec.~\ref{sec:conclusions} we summarize our conclusions.  
Appendix~\ref{sec:cijk} gives the form of the octupole tidal field.

Throughout this paper, where relativity is concerned, we use the sign 
conventions and notation of Misner {\it et al}. \cite{misn1973}.  
We use geometrical units in which $c = G = 1$ and scale all dimensional 
quantities relative to the black hole mass $M$, except where otherwise 
indicated in discussing astrophysical consequences.  

\section{Formalism}
\label{sec:formalism}

Fermi normal coordinates provide a convenient local moving frame for
calculating relativistic tidal encounters.  The FNC formalism was
first developed by Manasse and Misner \cite{mana1963} (see additional
work by Mashhoon \cite{mash1975} and Marck \cite{marc1983}).  In more
recent work, the metric was expanded through fourth order in the
spatial distance from the geodesic by Ishii {\it et al}. \cite{ishi2005}.
The result is a tidal field rich in multipoles and relativistic terms.
Here we specialize to a Schwarzschild background and summarize the
coordinate system and the tidal field expansion.  We also consider
physical scales associated with application to white dwarf/IMBH
encounters and address use of Newtonian self-gravity and hydrodynamics
in the FNC frame.  This approximation, adequate for white dwarfs,
affects which terms in the relativistic tidal field are worth
consistently retaining.

\subsection{Fermi normal coordinates}
\label{sec:FNC}

Fermi normal coordinates rely upon using local flatness in the vicinity of a 
freely falling observer over an extended period of time.  Consider an 
arbitrary spacetime with coordinates $X^{\mu}$ and a timelike geodesic 
$\mathcal{G}$ described by $X^{\mu}(\tau)$, parametrized by proper time 
$\tau$.  The tangent vector is $\mathbf{u}=\partial/\partial\tau$.  Greek 
indices label these arbitrary four-dimensional coordinates and coordinate 
components of tensors in this system.  The second (FNC) coordinate system 
$x^a = (\tau, x^i)$ has its spatial origin fixed to move along the trajectory 
$\mathcal{G}$ \cite{mana1963,pois2004}, with the conditions 
\be
\label{Fermi_conditions} 
g_{ab} |_\mathcal{G} = \eta_{ab}, \qquad 
\Gamma^a_{\ bc} |_\mathcal{G}=0 , 
\ee 
enforced at all times.  Here latin indices beginning with $a,b,c,\ldots$ 
label the four new coordinates and components in this frame.  Latin indices 
beginning with $i,j,k,\ldots$ are reserved for FNC spatial coordinates.  Let 
$\mathcal{P}_0$ be the single event on the geodesic $\mathcal{G}$ at 
$\tau=0$.  Construct an orthonormal tetrad 
$\boldsymbol\lambda_a = (\boldsymbol\lambda_0,
\boldsymbol\lambda_1,\boldsymbol\lambda_2,\boldsymbol\lambda_3)$ at this 
point.  Choose $\boldsymbol\lambda_0=\mathbf{u}$, making the tangent vector 
the timelike member of the tetrad at $\mathcal{P}_0$.  The remaining tetrad 
elements at $\mathcal{P}_0$ are spacelike.  Now extend the basis by parallel 
transporting the tetrad $\boldsymbol\lambda_a$ along $\mathcal{G}$.  One
condition is identically satisfied, since 
$\nabla_\mathbf{u} \boldsymbol\lambda_0 = \nabla_\mathbf{u} \mathbf{u} = 0$.  
Imposing $\nabla_\mathbf{u} \boldsymbol\lambda_i = 0$ on the spacelike 
elements $\boldsymbol\lambda_i$ defines the tetrad in the future and past 
of $\mathcal{P}_0$.  With the moving tetrad defined on $\mathcal{G}$, each 
element $\boldsymbol\lambda_i$ of the triad is used at any constant time 
$\tau$ to launch spacelike geodesics from $\mathcal{P}(\tau)$.  The proper 
distance along each of these three curves defines the spatial coordinates 
$x^i$.  Thus $\boldsymbol\lambda_i = \partial/\partial x^i$.  The proper 
time $\tau$ along the geodesic completes the coordinate system: $x^0=\tau$.  

The conditions \eqref{Fermi_conditions} further imply that all time 
derivatives of the connection and of the first derivative of the metric 
vanish along $\mathcal{G}$: $\Gamma^a_{\ bc,0} = \Gamma^a_{\ bc,00} = 0$, 
etc, and $g_{ab,c0} = g_{ab,c00}=0 $, etc.  This implies 
\cite{mana1963,pois2004,ishi2005} that the metric may be expanded in a
power series in spatial distance of the form
\ba
\label{FNC_metric_expansion} 
g_{ab} 
\nn & = & \eta_{ab} + \tfrac{1}{2}
g_{ab,ij}(\tau)x^i x^j + \tfrac{1}{6}g_{ab,ijk}(\tau)x^i x^j x^k \\
& & +
\tfrac{1}{24} g_{ab,ijkl}(\tau) x^i x^j x^k x^l + \mathcal{O}(x^5) .
\ea
The coefficients involve spatial derivatives of the metric only and are 
functions of just the FNC time coordinate $\tau$.

The specific form of the Taylor coefficients depends on how the tetrad is 
extended away from $\mathcal{G}$ but will in any case involve the Riemann 
tensor and its derivatives evaluated on the geodesic as functions of $\tau$.  
The expansion was derived to quadratic order by Manasse and Misner 
\cite{mana1963} and was extended to fourth order by Ishii {\it et al.} \cite{ishi2005}.  Gathering results in the latter paper, we find
\begin{widetext}
\ba
\label{eqn:metric00}
g_{00} &=& - 1 - C_{ij}\, x^i x^j 
-\frac{1}{3} C_{ijk}\, x^i x^j x^k  -\frac{1}{12}
\left( C_{ijkl} + 4C_{(ij}C_{kl)} - 4 B_{(kl|n|} {B_{ij)}}^n \right) 
x^i x^j x^k x^l + \mathcal{O}\left(|\vec{x}|^5/\mathcal{R}^5\right) ,
\\
\label{eqn:metric0m}
\nn
g_{0m} &=& \frac{2}{3} B_{ijm}\, x^i x^j + \frac{1}{4} R_{m(ij|0|;k)} 
\, x^i x^j x^k 
\\
& & \qquad + \frac{1}{135} \left( 9R_{m(ij|0|;kl)} - 6R_{m(ij}^{\ \ \ \ \ 0} 
R_{|0|kl)0} - 2R_{m(ij}^{\ \ \ \ \ n} R_{|n|kl)0} \right) 
x^i x^j x^k x^l + \mathcal{O}\left(|\vec{x}|^5/\mathcal{R}^5\right) ,
\\
\label{eqn:metricmn}
\nn
g_{mn} &=& \delta_{mn} + \frac{1}{6} \left(R_{imnj} + R_{inmj}\right)x^i x^j
-\frac{1}{36} \left( R_{injm;k} + R_{inkm;j} + R_{jnim;k} 
+ R_{knim;j} + R_{knjm;i} + R_{jnkm;i} \right) x^i x^j x^k
\\
& & \qquad + \frac{1}{180} \left( 9R_{m(ij|n|;kl)} 
- 6R_{m(ij}^{\ \ \ \ \ 0}R_{|n|kl)0} 
- 2R_{m(ij}^{\ \ \ \ \ p}R_{|n|kl)p} \right) 
x^i x^j x^k x^l + \mathcal{O}\left(|\vec{x}|^5/\mathcal{R}^5\right) ,
\ea
\end{widetext}
where the following tidal tensor definitions,
\begin{alignat}{2}
\label{eqn:tidaltensors}
\nn C_{ij} &\equiv R_{0i0j} , &\qquad 
C_{ijk} & \equiv R_{0(i|0|j;k)} , 
\\
C_{ijkl} &\equiv R_{0(i|0|j;kl)} , &\qquad 
B_{ijk} &\equiv R_{k(ij)0} ,
\end{alignat}
have been used.  We refer to $C_{ij}$, $C_{ijk}$, and $C_{ijkl}$ as the 
quadrupole, octupole, and $l =4$ tides, respectively.  The quadrupole 
tide is the electric part $\mathscr{E}_{ij}$ of the Riemann tensor, while 
$B_{ijk}$ is related to the magnetic part $\mathscr{B}_{ij}$ of the Riemann 
tensor.  The latter further gives rise to the gravitomagnetic potential 
\cite{thor1985}
\be
A_k = \tfrac{2}{3} B_{ijk} x^i x^j ,
\ee
which will appear in the fluid equations of motion.  In the metric expansion, 
$\mathcal{R}$ is a (smallest) length scale associated with the inhomogeneity 
and curvature length scales of the surrounding spacetime.  Also, the usual 
notation has been used \cite{misn1973} that parentheses bracketing 
indices indicate symmetrization with respect to all enclosed indices, e.g., 
\ba
A_{(ij)}
& = & \tfrac{1}{2} (A_{ij} + A_{ji} ) , \\
A_{(ijk)}
\nn & = & \tfrac{1}{6} 
\left( A_{ijk} + A_{jik} + A_{jki} + A_{kji} + A_{kij} + A_{ikj} \right) ,
\ea
excluding, however, any that are enclosed by vertical strokes.

We see that this coordinate system provides an expansion of the metric
provided the Riemann tensor $R_{abcd}$ and its derivatives in the FNC
are known.  Fortunately, the Riemann tensor and derivatives are only
required along $\mathcal{G}$ and need only be known in some coordinate
system (e.g., $R_{\mu\alpha\nu\beta}$).  The two are linked by a
coordinate transformation and the original coordinate components of
the tetrad vectors yield the Jacobian matrix along $\mathcal{G}$:
$\lambda_a^{\ \mu} = \partial X^{\mu}/\partial x^a |_{\mathcal{G}}$.
Thus we easily find \ba
\label{eqn:projection}
R_{abcd} 
& = & R_{\mu\alpha\nu\beta}\ \lambda^{\ \mu}_{a}\ 
\lambda^{\ \alpha}_{b}\ \lambda^{\ \nu}_{c}\ \lambda^{\ \beta}_{d} , \\
R_{abcd;e} 
\nn & = & R_{\mu\alpha\nu\beta;\rho}\ 
\lambda^{\ \mu}_{a}\ \lambda^{\ \alpha}_{b}\ \lambda^{\ \nu}_{c}\ 
\lambda^{\ \beta}_{d}\ \lambda^{\ \rho}_{e} , \\
R_{abcd;ef} 
\nn & = & R_{\mu\alpha\nu\beta;\rho\sigma}\ 
\lambda^{\ \mu}_{a}\ \lambda^{\ \alpha}_{b}\ \lambda^{\ \nu}_{c}\ 
\lambda^{\ \beta}_{d}\  \lambda^{\ \rho}_{e}\  \lambda^{\ \sigma}_{f} .
\ea

\subsection{Geodesic motion on Schwarzschild spacetime and construction 
of the FNC frame}
\label{sec:geodesic_motion_on_Schw}

The coordinates $X^{\mu}$ can be taken to be standard Schwarzschild 
coordinates $(t,r,\theta,\phi)$.  The line element is
\be\label{eqn:schwarzschild_metric}
ds^2 = - f dt^2 + f^{-1} dr^2 + r^2 
\left(d\theta^2 + \sin^2\theta d\phi^2 \right) ,
\ee
with $f(r)=1-2M/r$.  Test body orbits have two constants of motion
\be
-\tilde{E} = U_{t} , \qquad \qquad \tilde{L} = U_{\phi} ,
\ee
where $\tilde{E}$ is the specific orbital energy and $\tilde{L}$ is the 
specific angular momentum.  We confine the motion to the equatorial plane 
and have first-order equations of motion 
\be
\label{eqn:firstordereqns}
\frac{dt}{d\tau} = \frac{\tilde{E}}{f(r)}, \quad 
\frac{d\phi}{d\tau} = \frac{\tilde{L}}{r^2}, \quad 
\left(\frac{dr}{d\tau} \right)^2 = \tilde{E}^2 - V(r) ,
\ee
where $V\equiv f (1+\tilde{L}^2/r^2)$ is the effective potential for radial
motion.  Marginally bound orbits have $\tilde{E}=1$.

To integrate an arbitrary geodesic we use the parametrization of Darwin 
\cite{darw1959} (see also \cite{cutl1994}).  A semilatus rectum $p$ and 
eccentricity $e$ are defined, along with a radial phase angle $\chi$, 
defined by
\be
\label{eqn:darwin1}
r(\chi) = \frac{pM}{1+e\cos\chi} .
\ee
Let $r_1$ represent periastron and $r_2$ be apastron.  We find
\be
r_1 = \frac{pM}{1+e}, \qquad \qquad r_2 = \frac{pM}{1-e} .
\ee
Either pair of parameters can be used to specify a bound orbit.  Similarly, 
we can make the connection 
\ba\label{eq:specific_orbital_parameters}
\tilde{E}^2 &=& \frac{(p-2-2e)(p-2+2e)}{p(p-3-e^2)} , \\
\tilde{L}^2 &=& \frac{p^2M^2}{p-3-e^2} .
\ea
In terms of $\chi$, Eqs. (\ref{eqn:firstordereqns}) take the form
\ba
\label{eq:dtdChi}
\frac{dt}{d \chi} &=& \frac{p^2 M}{(p-2-2 e \cos\chi)(1 + e \cos\chi)^2} \nn \\
& & \quad \times 
\left[ \frac{(p-2)^2 - 4 e^2}{p - 6 - 2 e \cos \chi} \right]^{1/2} , \\
\label{eqn:darwin3}
\frac{d\phi}{d\chi} &=& p^{1/2} \left(p-6-2 e\cos\chi\right)^{-1/2} , \\
\label{eqn:darwin2}
\frac{d\tau}{d\chi} &=& \frac{p^{3/2}M}{\left (1+e\cos\chi \right )^2} 
\left ( \frac{p-3-e^2}{p-6-2e\cos\chi} \right )^{1/2} . \quad
\ea
The benefit of the curve parameter $\chi$ is in removing singularities from 
the integration at radial turning points.  We are thus able to consider 
tidal encounters of a system already in a bound orbit or, with suitable 
redefinition of parameters, hyperbolic systems with $\tilde{E}>1$.  We are
principally interested in $\tilde{E}=1$ orbits.  For these we have 
$e=1$ and define $r_1 \equiv R_p$.  Then $r_2\rightarrow \infty$ and 
\be
pM = 2R_p, \qquad \qquad \tilde{L}^2 = \frac{p^2 M^2}{p-4} .
\ee

Once an orbit is adopted we can construct the Fermi normal frame vectors. 
The vectors must satisfy the orthonormality condition 
$\lambda_{a}^{\ \mu} \lambda_{b}^{\ \nu} g_{\mu\nu} = \eta_{ab}$.  After 
selecting $\lambda_0^{\ \mu}=U^{\mu}$, a second natural choice is to take 
one spatial vector pointing out of the equatorial plane:
$\lambda_2^{\ \mu}=(0,0,1/r,0)$.  Following Marck \cite{marc1983}, we 
can construct two more vectors, $\tilde\lambda_1^{\ \mu}$ and
$\tilde\lambda_3^{\ \mu}$, that make an orthonormal set with 
$\lambda_0^{\ \mu}$ and $\lambda_2^{\ \mu}$,
\ba
\lambda_0^{\ \mu} &=& 
\left(\frac{\tilde{E}}{f}, U^r,0,\frac{\tilde{L}}{r^2} \right) , 
\\
\nn
\tilde\lambda_1^{\ \mu} &=& 
\left(\frac{U^r r}{f\sqrt{r^2+\tilde{L}^2}},
\frac{r \tilde{E}}{\sqrt{r^2+\tilde{L}^2}},0,0\right) ,
\\
\nn 
\lambda_2^{\ \mu} &=& \left( 0,0,\frac{1}{r},0 \right) ,
\\
\nn
\tilde\lambda_3^{\ \mu} &=& 
\left(\frac{\tilde{E}\tilde{L}}{f\sqrt{r^2+\tilde{L}^2}},
\frac{U^r \tilde{L}}{\sqrt{r^2+\tilde{L}^2}},0,
\frac{\sqrt{r^2+\tilde{L}^2}}{r^2} \right) .
\ea

While orthonormal, it may be shown that $\tilde\lambda_1^{\ \mu}$ and 
$\tilde\lambda_3^{\ \mu}$ do not satisfy the parallel transport condition 
\be
\label{parallel_transport_condition} 
U^{\nu} \partial_{\nu} \lambda_a^{\ \mu} 
+ \Gamma^{\mu}_{\ \alpha\beta} U^{\alpha} \lambda_a^{\ \beta} = 0 .  
\ee
We can, however, form two new vectors, $\lambda_1^{\ \mu}$ and 
$\lambda_3^{\ \mu}$, via a purely spatial rotation, 
\ba
\nn
\lambda_1^{\ \mu} &=& 
\tilde\lambda_1^{\ \mu} \cos\Psi - \tilde\lambda_3^{\ \mu}\sin\Psi ,
\\
\lambda_3^{\ \mu} &=& 
\tilde\lambda_1^{\ \mu} \sin\Psi + \tilde\lambda_3^{\ \mu}\cos\Psi ,
\ea
and then attempt to enforce the parallel transport condition on these.  
This proves possible as long as the frame precesses at a rate given by 
\be
\label{eqn:FNC_frame_rotation}
\frac{d\Psi}{d\tau} = \frac{\tilde{E}\tilde{L}}{r^2 + \tilde{L}^2} .
\ee
For a chosen value of $\tilde{L}$ and $\tilde{E} = 1$, we have an 
orthonormal tetrad parallel transported along a parabolic geodesic.

\subsection{Tidal tensor components and orbital PN expansion}
\label{subsec:tidal_tensor_comp_PN_expansion}

In Schwarzschild coordinates the components of the Riemann tensor are
\begin{alignat}{2}
\label{Riemann_Schwarzschild}
\nn
R_{trtr} &= -\frac{2M}{r^3} , & R_{t\theta t \theta} &= \frac{Mf}{r} , 
\\
\nn
R_{t\phi t\phi} &= \frac{Mf}{r} \sin^2\theta , 
& R_{r\theta r\theta} &= -\frac{M}{rf} ,
\\
R_{r\phi r\phi} &= -\frac{M}{rf} \sin^2\theta , 
&\quad R_{\theta\phi\theta\phi} &= 2Mr \sin^2\theta ,
\end{alignat}
with other nonzero components following from the symmetries
$R_{\mu\alpha\nu\beta} = -R_{\mu\alpha\beta\nu} = -
R_{\alpha\mu\nu\beta} = +R_{\nu\beta\mu\alpha}$ and with all other elements 
vanishing.  The first and second covariant derivatives are readily computed.  
We may then use Eq. (\ref{eqn:projection}) to project the Riemann tensor 
and its covariant derivatives (and thus the various tidal tensors).  In fact, 
we have a choice.  We can use $\{\boldsymbol\lambda_0,\boldsymbol{\lambda_1},
\boldsymbol\lambda_2,\boldsymbol{\lambda_3}\}$ to express components in the
FNC frame or use $\{\boldsymbol\lambda_0,\boldsymbol{\tilde\lambda_1},
\boldsymbol\lambda_2,\boldsymbol{\tilde\lambda_3}\}$ to cast tensor 
components into the noninertial frame.  It is convenient to consider both.

To distinguish tidal tensors in the FNC frame from those in the noninertial 
frame, the latter carry a tilde: $\tilde{C}_{ij}$, $\tilde{C}_{ijk}$, 
$\tilde{C}_{ijkl}$, and $\tilde{B}_{ijk}$).  To obtain $\tilde{A}_k$, we 
also have to rotate the coordinates,
\ba
\nn
\tilde{x}^1 &=& x^1 \cos\Psi + x^3 \sin\Psi , 
\\
\nn
\tilde{x}^2 &=& x^2 , 
\\
\tilde{x}^3 &=& -x^1\sin\Psi + x^3\cos\Psi .
\ea

The tilde frame, while not parallel-propagated, affords a simpler form for 
the tidal tensors.  The quadrupole tidal tensor in the noninertial frame is 
diagonal, with
\ba
\label{tilde_tidal_tensor_Cij}
\nn
\tilde{C}_{11} &=& -\frac{2 M}{r^3}\left(1+\frac{3\tilde{L}^2}{2 r^2}\right) ,
\\
\nn
\tilde{C}_{22} &=& \frac{M}{r^3} \left(1+\frac{3 \tilde{L}^2}{r^2}\right) ,
\\
\tilde{C}_{33} &=& \frac{M}{r^3} .
\ea
In contrast, the nonzero components in the FNC frame are
\ba
\label{tidal_tensor_Cij}
\nn
C_{11} &=& \frac{M}{r^3}\left[\left(1-3\cos^2\Psi \right) - 
\frac{3 \tilde{L}^2}{r^2}\cos^2\Psi\right] ,
\\
\nn
C_{13} &=& C_{31} 
= -\frac{3 M}{r^3}\left(1+\frac{\tilde{L}^2}{r^2}\right)\sin\Psi \cos\Psi ,
\\
\nn
C_{22} &=& \frac{M}{r^3}\left(1 + \frac{3 \tilde{L}^2}{r^2}\right) ,
\\ 
C_{33} &=& \frac{M}{r^3}\left[\left(1-3\sin^2\Psi \right) - 
\frac{3 \tilde{L}^2}{r^2} \sin^2\Psi \right] .
\ea
Likewise, the nonzero components of the octupole tidal tensor are simpler 
in the noninertial frame,
\ba
\label{tilde_tidal_tensor_Cijk}
\nn
\tilde{C}_{111} &=& \frac{6 M}{r^4} \left(1+\frac{3\tilde{L}^2}{2 r^2}\right) 
V_2^{-1} ,
\\
\nn
\tilde{C}_{131} &=& \tilde{C}_{311} = \tilde{C}_{113}
= \frac{4 M}{r^4} \frac{\tilde{L}}{r} U^r 
\left(1+\frac{5\tilde{L}^2}{4 r^2}\right) 
V_2^{-1} ,
\\
\nn
\tilde{C}_{122} &=& \tilde{C}_{212} = \tilde{C}_{221} 
= -\frac{3 M}{r^4} \left(1+\frac{7\tilde{L}^2}{3 r^2}\right)
V_2^{-1} ,
\\
\nn 
\tilde{C}_{133} &=& \tilde{C}_{313} = \tilde{C}_{331}  
= -\frac{3 M}{r^4} \left(1+\frac{2 \tilde{L}^2}{3 r^2}\right)
V_2^{-1} ,
\\
\nn 
\tilde{C}_{322} &=& \tilde{C}_{232} = \tilde{C}_{223} 
= -\frac{M}{r^4} \frac{\tilde{L}}{r} U^r 
\left(1+\frac{5\tilde{L}^2}{r^2}\right) 
V_2^{-1} ,
\\
\tilde{C}_{333} &=& -\frac{3 M}{r^4} \frac{\tilde{L}}{r} U^r \,
V_2^{-1} ,
\ea
where $V_2 \equiv \sqrt{1 + \tilde{L}^2/r^2}$ (notation of Ishii {\it et al}. 
\cite{ishi2005}).  The octupole tidal tensor in the FNC frame is somewhat 
more complicated and we list its components in Appendix B.

We have also obtained the lengthy expressions for $\tilde{C}_{ijkl}$
and $C_{ijkl}$ using MATHEMATICA but in the interests of brevity omit
reproducing them here.  We confirm the expressions found in
\cite{ishi2005} with the exception of $\tilde{C}_{2233}$ in Eq. (B22), which should not have an overall minus sign.

The nonzero components of $\tilde{B}_{ijk}$ in the noninertial frame 
are given by
\ba
\nn
\tilde{B}_{131} &=& \tilde{B}_{311} = -\tilde{B}_{232} 
= -\tilde{B}_{322} = -\tfrac{1}{2} \tilde{B}_{113} = 
\tfrac{1}{2} \tilde{B}_{223} \\
&=& -\frac{3 M}{2 r^3}\, \frac{\tilde{L}}{r}\, V_2 .
\ea
In the FNC frame we have instead
\ba
\nn
B_{131} &=& B_{311} = -B_{232} = -B_{322} = -\tfrac{1}{2} B_{113} = 
\tfrac{1}{2} B_{223} 
\\
&=& -\frac{3 M}{2 r^3}\, \frac{\tilde{L}}{r}\, V_2 \cos\Psi , 
\\
\nn
B_{122} &=& -B_{133} = B_{212} = -\tfrac{1}{2} B_{221} 
= -B_{313} = \tfrac{1}{2} B_{331} 
\\
&=& -\frac{3 M}{2 r^3}\, \frac{\tilde{L}}{r}\, V_2 \sin\Psi .
\ea
From these we derive the components of the gravitomagnetic potential.  
In the FNC frame we find
\ba
\label{eqn:gravitomagneticpotential}
\nn
A_1 &=& -\frac{2 M}{r^3}\, \frac{\tilde{L}}{r}\, V_2\, 
\Big \{ x^1 x^3 \cos\Psi + \left[ (x^3)^2-(x^2)^2\right] \sin\Psi \Big \} ,
\\
A_2 &=& \frac{2 M}{r^3}\, \frac{\tilde{L}}{r}\, V_2\, x^2 
\left(x^3 \cos\Psi - x^1 \sin\Psi \right) ,
\\
\nn
A_3 &=& \frac{2 M}{r^3}\, \frac{\tilde{L}}{r}\, V_2\, 
\Big \{\left[ (x^1)^2 - (x^2)^2 \right]\cos\Psi + 
x^1 x^3 \sin\Psi\Big \} .
\ea

For all but the most relativistic orbits, the spatial components of 
the four-velocity will be small compared to unity, 
$|U^r| \sim |r\, U^{\phi}| \lesssim 1$.  We can denote the maximum velocity 
scale at pericenter by $\delta$ and set
\be
\label{eqn:deltadef}
\delta = \left(\frac{M}{R_p}\right)^{1/2} .
\ee
Assuming $\delta$ is sufficiently small, we can use it as the basis 
for making an orbital PN expansion.  The various tidal tensors have 
been written in a suggestive way, since 
$\tilde{L}/r \sim |U^{\hat{i}}|\lesssim \mathcal{O}(\delta)$.  
We can expand the tidal tensors in the following way
\ba
C_{ij} &=& C_{ij}^{(0)} + C_{ij}^{(1)} ,
\\
C_{ijk} &=& C_{ijk}^{(0)} + C_{ijk}^{(1)} + C_{ijk}^{(2)} + \cdots ,
\\
C_{ijkl} &=& C_{ijkl}^{(0)} + C_{ijkl}^{(1)} + C_{ijkl}^{(2)} + \cdots ,
\ea
where the leading terms represent the Newtonian limit (e.g., 
$C_{ij}^{(0)} = \partial_i \partial_j \Phi_{bh}$, 
$C_{ijk}^{(0)} = \partial_i \partial_j \partial_k \Phi_{bh}$, etc).  
Higher-order terms ($C_{ij\ldots}^{(n)}$) are orbital PN corrections and
are $\mathcal{O}(\delta^{2 n})$ relative to the Newtonian limit.  A similar 
story holds for $B_{ijk}$ and $A_k$ except their expansions start at 
$\mathcal{O}(\delta)$,
\ba
B_{ijk} &=& B_{ijk}^{(0.5)} + B_{ijk}^{(1.5)} + \cdots ,
\\
A_{k} &=& A_{k}^{(0.5)} + A_{k}^{(1.5)} + \cdots .
\ea
We see that we could reexpress the metric given in Eqs. 
(\ref{eqn:metric00})--(\ref{eqn:metricmn}) as simultaneous power series in 
$\nu = |\vec{x}|/\mathcal{R}$ and $\delta$.

\subsection{Self-gravity of the star combined with the external tidal field}
\label{subsec:self_grav_tidal_field}

Ishii {\it et al}. \cite{ishi2005} derived the third- and fourth-order 
terms in the tidal field, which are summarized above, and used the expansion 
along with a Newtonian stellar model to study tidal effects on a star in 
circular orbit about a Kerr black hole.  In this section and the next we 
provide a justification for use of Newtonian self-gravity and hydrodynamics 
(see also \cite{fish1973}), estimate the resulting errors, and determine 
what parts of the tidal field expansion should be consistently retained.  
This approximation is adequate for main sequence stars and white dwarfs, but 
much less so for neutron stars.

Consider a star of mass $M_*$ that encounters a more massive black hole of 
mass $M$.  We are concerned with mass ratios in the range
\be
\label{eqn:massratio}
\mu \equiv \frac{M_*}{M} \sim 10^{-5} \text{--} 10^{-3} ,
\ee
corresponding to black holes with masses $M \sim 10^3-10^5 M_\odot$.  Let the 
stellar radius be $R_*$.  The strength of the tidal encounter is determined by
\be
\label{eqn:etadefinition}
\eta = \left(\frac{R_p^3}{M} \frac{M_*}{R_*^3}\right)^{1/2} .
\ee
In this paper we restrict attention to stars that just reach the tidal 
radius at pericenter and disrupt ($\eta \simeq 1$) and to weaker, partially 
disruptive encounters ($\eta \sim 2-6$). 

The metric given in Eqs. (\ref{eqn:metric00})--(\ref{eqn:metricmn}) would
serve to compute near $\mathcal{G}$ motion of test bodies or of a fluid of 
negligible mass.  A star with finite mass will necessarily alter the 
geometry.  Even for a star with strong gravity, the tidal field expansion 
is still useful provided the star is sufficiently isolated.  This requires 
a buffer region whose radius is small compared to the characteristic length 
scale of the tidal field but large compared to the compact object, so that
self-gravity is also weak within the buffer region \cite{thor1985}.  If the 
star is approximately Newtonian, the latter condition is satisfied throughout 
the star and the self-gravity ($\Phi_{*}$) and tidal fields will linearly 
superpose to lowest order and can be computed separately.  We only require 
then that the domain of interest have a size $\mathcal{L} \gtrsim R_*$ small 
compared to the characteristic length scale of the tidal field.  The assumption 
(\ref{eqn:massratio}) on the mass ratio makes this possible, since
\be
\frac{R_*}{r(t)} \lesssim \frac{R_*}{R_p} = \mu^{1/3}\, \eta^{-2/3} \ll 1 .
\ee

Likewise, the fluid is assumed to be well modeled by Newtonian hydrodynamics 
(as seen in the FNC frame).  Prior to tidal encounter, the star has vanishing 
or minimal internal fluid velocities in this frame.  The internal stellar 
sound speed $a_s$ and stresses will be small and comparable to the 
self-gravity,
\be
a_s^2 \simeq \frac{p}{\rho} \lesssim |\Phi_{*}| \simeq 
\varepsilon^2 = \frac{M_{*}}{R_{*}} \ll 1 ,
\ee
where $\rho$ is the rest mass density, $p$ is the isotropic pressure, and 
$\varepsilon$ is the (stellar) PN velocity scale.  Post 
encounter, fluid velocities will also be small (e.g., a few multiples 
of stellar escape speed $|v^i| \simeq \varepsilon$) and subrelativistic 
provided the region of interest is restricted in size.  In our models, we 
take the domain size to be 
\be
\mathcal{L} \lesssim 8 \times R_* .
\ee

For a star immersed in an external tidal field we expect the gravitational 
field to take the form
\be
\label{eqn:interaction}
g_{ab} = \eta_{ab} + h_{ab}^{\text{tidal}} + h_{ab}^{*} + h_{ab}^{IC} .
\ee
The self-gravity $h_{ab}^{*}$ depends only on the Newtonian potential 
$\Phi_*$ and is weak, $|h_{ab}^{*}| \ll 1$.  The tidal field 
$h_{ab}^{\text{tidal}}$, as seen in FNC, is determined by equations 
(\ref{eqn:metric00})-(\ref{eqn:metricmn}) and for 
$|\vec{x}|\lesssim \mathcal{R}$ is also weak.  The full metric must satisfy 
the Einstein field equations and their nonlinearity requires that a term 
$h_{ab}^{IC}$ be present that represents the field interaction and 
higher-order self-gravity corrections.  In the absence of the tidal field 
we would have
\ba
h_{00}^{*} &=& -2 \Phi_{*} + \mathcal{O}(\varepsilon^4) ,
\\
\nn
h_{0i}^{*} &=& \mathcal{O}(\varepsilon^3) ,
\\
\nn
h_{ij}^{*} &=& \mathcal{O}(\varepsilon^2) ,
\ea
where the Newtonian potential satisfies
\be
\nabla^2 \Phi_{*} = 4 \pi \rho ,
\ee
and where the missing corrections are (stellar) 1PN terms.  Neglect of these 
corrections sets a floor on the accuracy of our method.  In white 
dwarf/IMBH encounters, most white dwarfs will have 
$M_{*}/R_{*} \simeq 10^{-4}$, so that $\varepsilon \simeq 0.01$.  
Thus, our method has intrinsic relative errors at the level of $10^{-4}$.  
In what follows, in analyzing $h_{ab}^{\text{IC}}$ and 
$h_{ab}^{\text{tidal}}$, we neglect any term whose contribution to the 
fluid acceleration is at or below this error level.

We have defined two small velocity parameters, $\delta$ and $\varepsilon$.
These two scales are not necessarily comparable.  They are related by
\be
\delta = \varepsilon \, \mu^{-1/3} \, \eta^{-1/3} ,
\ee
and for a small mass ratio $\mu$ we find $\delta \gg \varepsilon$.  As an 
example, in our application with $\varepsilon \simeq 10^{-2}$, if we take 
$\mu = 10^{-4}$ and $\eta = 1$, we have a much higher orbital velocity scale: 
$\delta \simeq 0.22$.  This highlights one of the real advantages of Fermi 
normal coordinates.  We could never use Newtonian hydrodynamics in a frame 
fixed with respect to the black hole.  In combining self-gravity and the
tidal field, the simultaneous expansions in $\delta$ and $\varepsilon$ 
make (orbital) and (stellar) PN contributions, respectively.

The external tidal field has a radius of curvature $\mathscr{R}$, an 
inhomogeneity scale $\mathscr{L}$, and a time scale for changes in curvature 
$\mathscr{T}$ \cite{thor1985}.  Each of these scales is time dependent as 
viewed from the FNC frame center.  They reach their minima at pericenter
\be
\mathscr{L} \simeq R_p , \quad \mathscr{R} \simeq \mathscr{T} \simeq 
\left(\frac{R_p^3}{M}\right)^{1/2} ,
\ee
where $\mathscr{L}/\mathscr{T} = \delta$.  The tidal field is 
dominated by the quadrupole moment, which reaches a maximum of
$|C_{ij}| \simeq |C_{ij}^{(0)}| \simeq \mathscr{R}^{-2} \lesssim M /R_p^{3}$. 
The star's self-gravity is dominated by its mass monopole.  At pericenter, 
for encounters near threshold for disruption ($\eta \simeq 1$), there is a 
near balance between the quadrupole tidal term and the star's gravitational 
potential:
\be
|C_{ij}^{(0)}\, x^i x^j| \simeq \frac{M}{R_p^3} R_*^2 
= \varepsilon^2 \eta^{-2} \lesssim \varepsilon^2 \simeq |\Phi_*| .
\ee

The size of the gravitational field correction $h_{ab}^{IC}$ can now be 
estimated without a full calculation.  Substituting (\ref{eqn:interaction}) 
into the Einstein field equations would yield a nonlinear contribution 
no larger than 
\be
\left| h_{ab}^{IC} \right| \lesssim \left| h_{00}^{*} \right|\, 
\left| h_{00}^{\text{tidal}} \right|
\simeq \frac{M_{*}}{R_{*}} \frac{M}{R_p^3} \, R_{*}^2 
\simeq \varepsilon^4 \eta^{-2} \lesssim \varepsilon^4 .
\ee
Thus the interaction terms are formally at or below the size of the (stellar) 
1PN corrections, which we have already chosen to neglect, and can be 
dropped as well.

At our level of approximation the metric in (\ref{eqn:interaction}) is just 
the sum of Newtonian self-gravity and the FNC tidal field, but with two 
caveats.  The first involves the assumption of a stationary black hole 
background.  While suitable for test-body motion, the finite mass of the 
white dwarf will cause the black hole to wobble relative to a common center.
This $\mathcal{O}(\mu)$ correction is easily dealt with in Newtonian 
mechanics.  In relativity one could in principle treat this effect 
as a conservative perturbation in the black hole's gravitational field 
\cite{detw2004} and correct the motion of the FNC frame and the tidal terms.  
Alternatively, we could use a PN calculation of the two-body
orbit, and then calculate the tidal field.  We have done neither, which 
introduces an added small source of relative error of magnitude $\sim \mu$.  

The second caveat is that not all of the terms in the tidal field in 
(\ref{eqn:metric00})-(\ref{eqn:metricmn}) are significant given 
our error floor.  The magnitudes attained by some of the contributions to 
$g_{00}$ are as follows:
\ba
& &\left|C_{ij}^{(0)}\, x^i x^j \right| 
\lesssim \frac{M}{R_p^3}\, R_{*}^2 
= \varepsilon^2 \eta^{-2} , 
\\
& &\left|C_{ij}^{(1)}\, x^i x^j \right| 
\lesssim \frac{M}{R_p^3}\, R_{*}^2 \, \delta^2
= \varepsilon^4 \, \mu^{-2/3} \, \eta^{-2} , 
\\
& &\left|C_{ijk}^{(0)}\, x^i x^j x^k \right| 
\lesssim \frac{M}{R_p^4}\, R_{*}^3 
= \varepsilon^2 \mu^{1/3} \eta^{-8/3} ,
\\
& &\left|C_{ijk}^{(1)}\, x^i x^j x^k \right| 
\lesssim\frac{M}{R_p^4}\, R_{*}^3 \, \delta^2 
= \varepsilon^4 \mu^{-1/3} \eta^{-10/3} ,
\\
& &\left|C_{ijk}^{(2)}\, x^i x^j x^k \right| 
\lesssim\frac{M}{R_p^4}\, R_{*}^3 \, \delta^4 
= \varepsilon^4 \, (\mu^{-1}\, \varepsilon^2 ) \eta^{-4} ,
\\
& &\left|C_{ijkl}^{(0)} x^i x^j x^k x^l \right| 
\lesssim \frac{M}{R_p^5}\, R_{*}^4  
= \varepsilon^2 \, \mu^{2/3} \, \eta^{-10/3} .
\ea
These are the only terms that exceed the error floor of $\varepsilon^4$ 
(where in all cases we use $\eta = 1$ to ascertain significance).  The 
(orbital) 2PN part of the octupole tide deserves mention.  It is worth 
retaining only if $\mu \ll \varepsilon^2$, which is possible for black holes 
on the higher end of our mass range.  Missing from the list is the (orbital) 
1PN contribution to the $l =4$ tide, which is at the level of the 
(stellar) 1PN error and therefore negligible.  The same is true of the 
nonlinear (squared Riemann tensor) term 
$C_{(ij}\, C_{kl)}\, x^i x^j x^k x^l$.  The squared term involving 
$B_{ijk}$ is well below $\varepsilon^4$.

The first (quadrupole) term in $g_{mn}$ (depending upon $R_{imnj}$) is of 
the same magnitude as the Newtonian quadrupole tidal term in $g_{00}$, 
i.e., $\lesssim \varepsilon^2 \eta^{-2}$.  It is therefore at the level 
of the discarded (stellar) 1PN term, and it and all of the rest of the 
terms in the expansion of $g_{mn}$ are negligible.

The leading term in $g_{0m}$ can attain a magnitude of
\be
\left| B_{ijm}^{(0.5)}\, x^i x^j \right| \lesssim 
\frac{M}{R_p^3} \, R_{*}^2 \, \delta
= \varepsilon^3 \, \mu^{-1/3} \, \eta^{-7/3} .
\ee
Because $\mu \ll 1$ this term is larger than the (stellar) 1PN contribution 
and provides the possibility that the gravitomagnetic potential derived from 
it may be significant.  The next term has magnitude
\be
\left| B_{ijm}^{(1.5)}\, x^i x^j \right| \lesssim 
\frac{M}{R_p^3} \, R_{*}^2 \, \delta^3
= \varepsilon^3 \, (\mu^{-1} \, \varepsilon^2 ) \, \eta^{-3} .
\ee
Surprisingly, this (orbital) 1PN correction may also be significant at 
the high end of the black hole mass range where $\mu \ll \varepsilon^2$.
All of the other terms in the expansion of $g_{0m}$ are negligible.

We can lump all of the surviving parts of $h_{00}^{\text{tidal}}$ into a 
tidal potential $\Phi_{\text{tidal}}$, given by
\ba
\label{eqn:tidalpotential}
\nn
\Phi_{\text{tidal}} &=& \tfrac{1}{2}C_{ij}^{(0)}\, x^i x^j + 
\tfrac{1}{2}C_{ij}^{(1)}\, x^i x^j + 
\tfrac{1}{6}C_{ijk}^{(0)}\, x^i x^j x^k 
\\
\nn
& &\quad + \tfrac{1}{6}C_{ijk}^{(1)}\, x^i x^j x^k + 
\tfrac{1}{6}C_{ijk}^{(2)}\, x^i x^j x^k 
\\ 
& &\quad + \tfrac{1}{24}C_{ijkl}^{(0)}\, x^i x^j x^k x^l .
\ea
The surviving parts of $h_{0m}^{\text{tidal}}$ contribute to the 
gravitomagnetic potential
\be
\label{eqn:gmpotential}
A_m = \tfrac{2}{3} B_{ijm}^{(0.5)}\, x^i x^j 
+ \tfrac{2}{3} B_{ijm}^{(1.5)}\, x^i x^j .
\ee
These are the only tidal terms we need in assembling the final form of the
metric
\ba
\label{eqn:finalmetric}
g_{00} &=& - 1 - 2 \Phi_{*} - 2 \Phi_{\text{tidal}} 
+ \mathcal{O}\left(\varepsilon^4 \right) ,
\\
\nn
g_{0m} &=& A_{m} + \mathcal{O}\left(\varepsilon^3 \right) ,
\\
\nn
g_{mn} &=& \delta_{mn} + \mathcal{O}\left(\varepsilon^2 \right) .
\ea
This is the same conclusion as Ishii {\it et al}.~\cite{ishi2005}, except that 
we have identified those terms in the FNC metric that should be 
consistently retained.

\subsection{Fluid equations and retained tidal terms}
\label{subsec:fluid_eqn_retained_tidal_terms}

We assume a perfect fluid with stress-energy tensor 
\be
T^{ab} = (\rho + \rho \, \Pi + p)\, u^a u^b + p\, g^{ab} ,
\ee
where $\Pi$ is the specific energy.  The four velocity $u^a$ as well 
as $T^{ab}$ are assumed expressed in the FNC frame.  The fluid satisfies 
\be
\label{eqn:initialfluid}
{T^{ab}}_{;b} = 0 , \qquad \left(\rho u^a \right)_{;a} = 0 .
\ee

We simplify these equations using the weak field expansion 
(\ref{eqn:finalmetric}) and a slow motion approximation with the two 
velocity expansion parameters $\varepsilon$ and $\delta$.
The usual conserved mass density $\rho^{*} = \rho \sqrt{-g} u^0$ 
satisfies an exact conservation law, while $\rho$ satisfies
\be
\label{eqn:massconservation}
\frac{\partial\rho}{\partial \tau} + \frac{\partial}{\partial x^k} 
\left(\rho v^k \right) = 0 + \mathcal{O}(\rho \varepsilon^3/\mathcal{L}) .
\ee
Here $v^k = u^k/u^0$.  To obtain the fluid equation of motion (Euler 
equation), we note first that under our assumptions (i.e., FNC frame, 
$|\vec{x}| \lesssim \mathcal{L}$, $\eta \gtrsim 1$) we can still expect
\ba
T^{0i} &=& \rho v^i + \mathcal{O}(\rho \varepsilon^3) ,
\\
T^{ij} &=& \rho v^i v^j + \delta^{ij}p + \mathcal{O}(\rho \varepsilon^4) .
\ea
Neglect of the (stellar) 1PN corrections sets the floor on accuracy.  We 
then expand the connection, retaining only terms that will exceed the error
floor.  We find
\ba
\label{eqn:connection}
\nn
\Gamma^i_{00} &=& \delta^{ij} \frac{\partial}{\partial x^j} 
\left(\Phi_{*} + \Phi_{\text{tidal}}\right) 
+ \delta^{ij} \frac{\partial}{\partial \tau} A_j 
+ \mathcal{O}(\varepsilon^4/\mathcal{L}) ,
\\
\Gamma^i_{0k} &=& \frac{1}{2} \delta^{ij} \left(
\frac{\partial A_j}{\partial x^k} - \frac{\partial A_k}{\partial x^j} \right) 
+ \mathcal{O}(\varepsilon^3/\mathcal{L}) ,
\ea
and $\Gamma^i_{jk} = \mathcal{O}(\varepsilon^2/\mathcal{L})$, which is 
negligible.  The fluid equation then follows,
\be
\label{eqn:euler}
\frac{\partial v_i}{\partial \tau} + 
v^k \frac{\partial v_i}{\partial x^k} + \frac{1}{\rho} 
\frac{\partial p}{\partial x^i} + \frac{\partial\Phi_{*}}{\partial x^i} 
= a_i^{\text{tidal}} + \mathcal{O}(\varepsilon^4/\mathcal{L}) ,
\ee
with tidal acceleration
\be
\label{eqn:tidalaccel}
a_i^{\text{tidal}} = - \frac{\partial\Phi_{\text{tidal}}}{\partial x^i} 
- \frac{\partial A_i}{\partial \tau} 
+ v^k \left( \frac{\partial A_k}{\partial x^i} - 
\frac{\partial A_i}{\partial x^k} \right) .
\ee

The fluid equation given here is identical to that used by Ishii {\it
et al}.~\cite{ishi2005} (see also \cite{fish1973}) with the exception that 
we differ in the form of the tidal and gravitomagnetic potentials.  The 
truncated forms of these potentials expressed in equations 
(\ref{eqn:tidalpotential}) and (\ref{eqn:gmpotential}) contain only those 
terms that should be consistently retained given our level of approximation.

%%%%%%%%%%%%%%%%%%%%%%%%%%%%%%%%%%%%%%%%%%%%%%%%%%%%%%%%%%%%%%%%%%%%%%%%%%%%%
\begin{figure}
{ \begin{center} 
\includegraphics[scale=1.04]{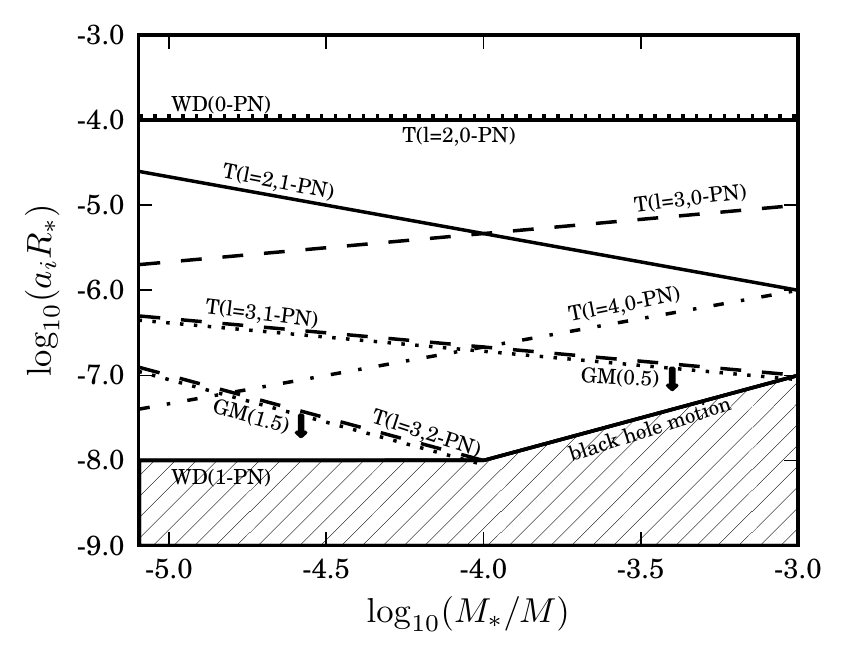} 
\end{center} 
\caption{\label{figure:PN_scales}
Scaling of acceleration terms from an expansion of the combined tidal and 
self-gravity fields.  The dominant two terms are the Newtonian self-gravity and 
Newtonian quadrupole tide.  Higher tidal moments ($l = 3$ and $l = 4$) 
are shown, as well as orbital PN corrections.  Also displayed is the upper
limit on the gravitomagnetic acceleration.  Neglect of stellar 1PN corrections 
sets a floor on accuracy, as does neglect of the motion of the black hole.
}  }
\end{figure}
%%%%%%%%%%%%%%%%%%%%%%%%%%%%%%%%%%%%%%%%%%%%%%%%%%%%%%%%%%%%%%%%%%%%%%%%%%%
Prior to tidal encounter the stellar gravitational acceleration can be
estimated by $|R_{*} \nabla\Phi_{*}| \simeq \varepsilon^2$.  For a white 
dwarf this dimensionless measure is of order $\simeq 10^{-4}$.  The (stellar) 
1PN errors will be at a level of $\mathcal{O}(\varepsilon^4)$, or 
$\simeq 10^{-8}$.  In Fig. \ref{figure:PN_scales} we show the 
order-of-magnitude size of the various tidal acceleration terms (at 
pericenter and assuming $\eta \simeq 1$) as functions of mass ratio $\mu$.  
The mass ratio runs from $\mu=10^{-3}$ 
down to just less than $10^{-5}$, where a white dwarf would cross the 
horizon before disrupting.  The two largest acceleration contributions, 
independent of $\mu$, are due to the stellar self-gravity and the Newtonian 
part of the quadrupole tide, $C_{ij}^{(0)}$.  These are denoted by WD(0PN) 
and T($l=2$,0PN), respectively.  The next two most important terms are 
the Newtonian octupole $C_{ijk}^{(0)}$ [T($l=3$,0PN)] and the 
(orbital) 1PN correction [T($l=2$,1PN)] to the quadrupole tide.  We 
plot also the Newtonian part of the $l = 4$ tide ($C_{ijkl}^{(0)}$) and 
first and second orbital PN corrections to the octupole tide.  Also plotted
are upper limits on the two (potentially) significant parts of the 
gravitomagnetic potential.  The upper limits are only achieved if we assume 
the star reaches breakup angular velocity.  All other terms in the tidal 
field are ignored since they will contribute accelerations at or below the 
magnitude of neglected (stellar) 1PN corrections.  The error floor is 
somewhat higher for $\mu > 10^{-4}$, as we have not accounted for 
nonstationarity of the black hole background.

\section{Initial white dwarf model and orbits approaching an IMBH}
\label{sec:models-orbits}

We now begin to apply the analytic approach (Sec.~\ref{sec:formalism}) and 
our numerical method (see Appendix~\ref{sec:method}) to investigate tidal 
interactions between a white dwarf and IMBHs in the mass range 
$M\sim 500$ to $1.7 \times 10^4 M_\odot$.  Only encounters that are at the 
threshold of disruption or weaker ($\eta \simeq 1 - 6$) are considered in 
this paper.  In this section we give the properties of the white dwarf model 
and the range of inbound orbits.  

\begin{table}[h]
\begin{center}
\caption{
\label{table:star_parameters_cgs}
Properties of polytropic white dwarf model.  Fundamental parameters are the 
mass $M_*$, radius $R_*$, polytropic index $n=3/2$, and adiabatic index 
$\gamma = 5/3$. }
\begin{tabular*}{8.6cm}{lrl@{\extracolsep{\fill}}l}
%    {@{\extracolsep{\fill}}lrll}
\hline
\hline
White dwarf parameters & & & Units \\
\hline
$M_*$ & $0.64$& & $M_\odot$\\
$R_*$ & $8.62$& $\times 10^8$ & cm \\
$L_*$ & $3.44$& $\times 10^{50}$ & g cm$^2$ s$^{-1}$ \\
$I_*$ & $9.67$& $\times 10^{49}$ & g cm$^2$\\
$\Phi_*$ & $1.10$& $\times 10^{-4}$ & \\
$\rho_c$ & $2.84$& $\times 10^6$ & g cm$^{-3}$ \\
$p_c$ & $1.51$& $\times 10^{23}$ & erg cm$^{-3}$ \\
$\tau_0$ & $1.05$& $\times 10^{1}$ & s \\
$E_{\text{tot}}$ & $-5.40$& $ \times 10^{49}$ & erg \\
\hline
\hline
\end{tabular*}
\end{center}
%\vspace{-0.6cm}
\end{table}
%%%%%%%%%%%%%%%%%%%%%%%%%%%%%%%%%%%%%%%%%%%%%%%%%%%%%%%%%%%%%%%%%%%

%%%%%%%%%%%%%%%%%%%%%%%%%%%%%%%%%%%%%%%%%%%%%%%%%%%%%%%%%%%%%%%%%%%
\begin{table}[h]
\begin{center}
\caption{
\label{table:star_parameters_M}
Stellar parameters in terms of black hole mass $M$ for the three mass
ratios $\mu=M_*/M$ studied.
 }
\begin{tabular*}{8.6cm}%
     {@{\extracolsep{\fill}}cllll}
\hline
\hline
$\mu$ & $1.28\times 10^{-3}$ & $4.21 \times 10^{-4}$ & $3.77 \times 10^{-5}$ 
& Units \\
\hline
$M_*$ & $1.28\times 10^{-3}$ & $4.21\times 10^{-4}$ &  $3.77\times 10^{-5}$ 
& M  \\
$R_*$ & $1.17\times 10^1$ & 3.84 & $3.44\times 10^{-1}$ & M \\ 
$L_*$ & $1.56\times 10^{-4}$ & $1.69 \times 10^{-5}$ & $1.36\times 10^{-7}$ 
& M$^2$ \\ 
$I_*$ & $1.78 \times 10^{-2}$ & $6.35 \times 10^{-4}$ & $4.56 \times 10^{-7}$ 
&  M$^3$\\
$\Phi_*$ & $1.10 \times 10^{-4}$& $1.10 \times 10^{-4}$& $1.10 \times 10^{-4}$ 
\\
$\rho_c$ & $1.15 \times 10^{-6}$ &$1.06 \times 10^{-5}$ &$1.33 \times 10^{-3}$ 
& M$^{-2}$\\
$p_c$ & $6.80 \times 10^{-11}$ & $6.28\times 10^{-10}$ & $7.85 \times 10^{-8}$ 
& M$^{-2}$\\
$\tau_0$ & $4.25 \times 10^3$ & $1.40\times 10^3$ & $1.25\times 10^2$ & M \\
$E_{\text{tot}}$ & \hspace{-2.5mm}$-6.00 \times 10^{-8}$ 
&\hspace{-2.5mm}$-1.98 \times 10^{-8}$ 
&\hspace{-2.5mm}$-1.77\times 10^{-9}$ & M \\
\hline
\hline
\end{tabular*}
\end{center}
%\vspace{-0.6cm}
\end{table}
%%%%%%%%%%%%%%%%%%%%%%%%%%%%%%%%%%%%%%%%%%%%%%%%%%%%%%%%%%%%%%%%%%%

%%%%%%%%%%%%%%%%%%%%%%%%%%%%%%%%%%%%%%%%%%%%%%%%%%%%%%%%%%%%%%%%%%%
\begin{table}[h]
\begin{center}
\caption{
\label{table:encounter_parameters}
Orbital parameters for different mass ratios $\mu$ and tidal parameters 
$\eta$.  The pericentric distance $R_p$ and starting distance $R_i$ are 
given.  Also shown is the cumulative geodetic frame precession experienced 
in the simulations. }
\begin{tabular*}{8.6cm}%
     {@{\extracolsep{\fill}}cccccc}
\hline
\hline
$\mu$ & $\eta$ & $\tilde{L} [M]$ & $R_p [M]$ & $R_i [M]$ & $\Delta\varphi$ \\
\hline
$1.28\times 10^{-3}$ & 1 & 14.8 & 107.5 & 1167 & 4.48e-02\\
-                   & 2 & 18.6 & 170.6 & 1120 &  2.79e-02\\
-                   & 3 & 21.2 & 223.6 & 1086 &  2.12e-02\\
-                   & 4 & 23.4 & 270.9 & 1060 &  1.73e-02\\
-                   & 5 & 25.2 & 314.3 & 1041 &  1.48e-02\\
-                   & 6 & 26.7 & 354.9 & 1027 &  1.30e-02\\
\hline
$4.21\times 10^{-4}$ & 1 & 10.3 & 51.2 & 555.4 & 9.64e-02\\
-                   & 2 & 12.9 & 81.3  & 532.8 & 5.95e-02\\
-                   & 3 & 14.7 & 106.6 & 516.7 & 4.49e-02\\
-                   & 4 & 16.2 & 129.1 & 504.7 & 3.67e-02\\
-                   & 5 & 17.4 & 149.8 & 495.6 & 3.13e-02\\
-                   & 6 & 18.5 & 169.2 & 488.8 & 2.75e-02\\
\hline
$3.77\times 10^{-5}$ & 1 & 5.0 & 10.2 & 109.0 &  6.07e-01\\
-                   & 2 & 6.1 & 16.3 & 104.9 &  3.40e-01\\
-                   & 3 & 6.9 & 21.3 & 101.8 &  2.48e-01\\
-                   & 4 & 7.5 & 25.8 & 99.6  &  1.99e-01\\
-                   & 5 & 8.0 & 30.0 & 97.9  &  1.68e-01\\
-                   & 6 & 8.5 & 33.8 & 96.6  &  1.46e-01\\
\hline
\hline
\end{tabular*}
\end{center}
%\vspace{-0.6cm}
\end{table}
%%%%%%%%%%%%%%%%%%%%%%%%%%%%%%%%%%%%%%%%%%%%%%%%%%%%%%%%%%%%%%%%%%%

%%%%%%%%%%%%%%%%%%%%%%%%%%%%%%%%%%%%%%%%%%%%%%%%%%%%%%%%%%%%%%%%%%%
\begin{figure*}[!htbp]
{ 
  \begin{tabular}{cc} 
      \includegraphics[scale=1.02]{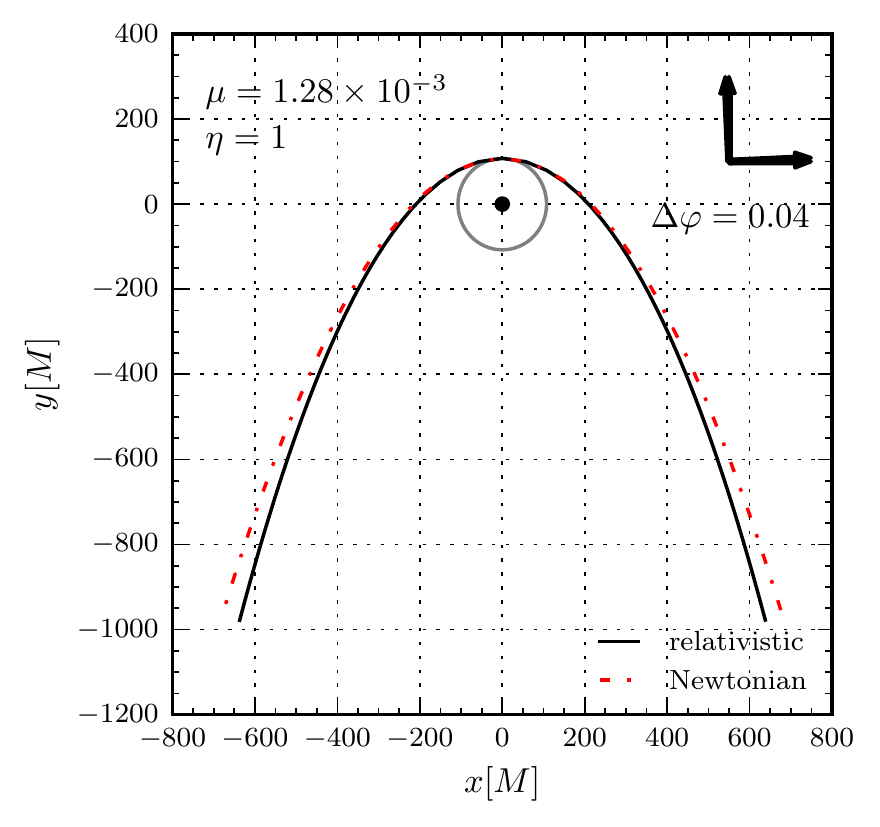} &
      \includegraphics[scale=1.02]{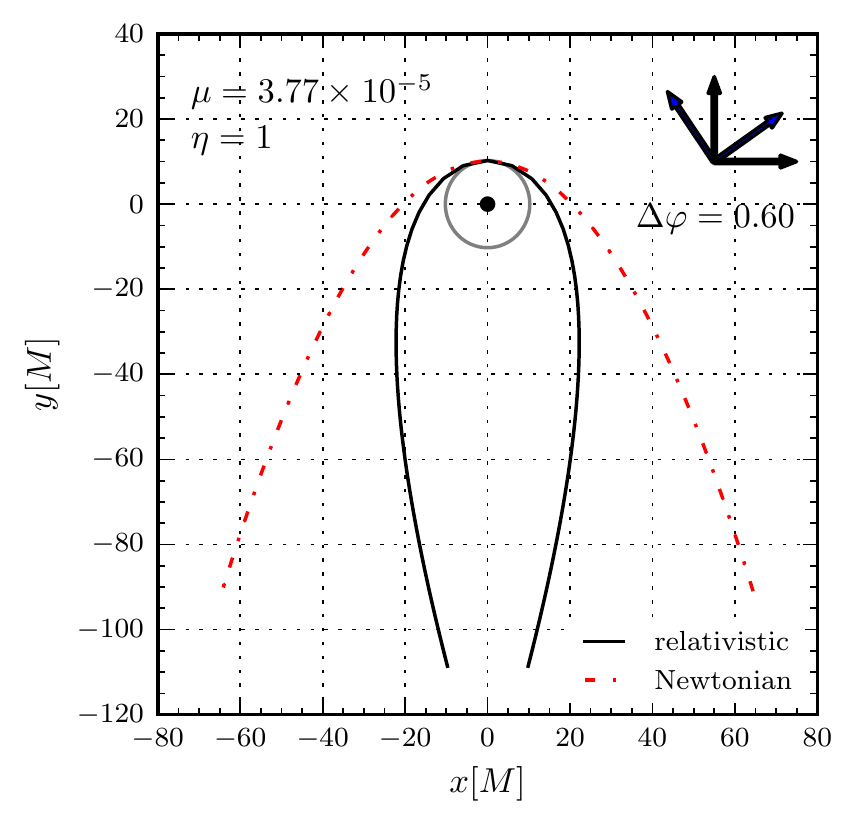} \\
  \end{tabular}                                                  
    
    \caption {\label{figure:compare_orbits_eta1} Trajectories followed by 
the FNC domain for $\eta=1$ encounters.  Positions are given in units of 
black hole mass $M$.  FNC domain motion for $\mu = 1.28 \times 10^{-3}$ 
is on the left and $\mu = 3.77 \times 10^{-5}$ is on the right.  
Relativistic precession of the frame through an angle $\Delta\varphi$ is 
indicated in the upper right corner (more pronounced for the 
$R_p \simeq 10 M$ case).  For reference the Newtonian parabolic orbit with 
the same pericentric distance is shown as the dashed curve. }}

\end{figure*}
%%%%%%%%%%%%%%%%%%%%%%%%%%%%%%%%%%%%%%%%%%%%%%%%%%%%%%%%%%%%%%%%%%%

\subsection{Stellar model}
\label{subsec:stellar_model}

The initial white dwarf is modeled as a nonrotating polytrope.  The 
Lane-Emden equation with polytropic index $n=3/2$ provides initial density 
and pressure profiles, which are then mapped onto a three-dimensional Cartesian grid.  
The adiabatic index is taken to be $\gamma=5/3$, making the star 
neutrally stable against convection.  We choose a mass 
$M_{*}=0.64M_\odot$ and radius $R_{*}=8.62\times 10^8$ cm, from 
which follows the remaining stellar properties.  These properties are 
assembled in Table \ref{table:star_parameters_cgs}.  They include the central 
density $\rho_c$, central pressure $p_c$, and stellar total energy 
$E_{\text{tot}} = -(3/7)\, G M_*^2/R_* = - E_{\text{int}} = \tfrac{1}{2} 
E_{g}$.  We also calculate the moment of inertia 
$I_* = \tfrac{1}{3} \int x_i x_i \rho \, d^3x $ and the estimated break-up 
angular momentum $L_* = \sqrt{GM_*^3 R_*}$.  The fundamental radial 
pulsation period is given by $\tau_0 = 2\pi \sqrt{R_*^3/(\alpha M_*)}$, where 
$\alpha = 2.712$ results from our choice of $n=3/2$ and 
$\gamma = 5/3$~\cite{cox1980}.  A cold, degenerate white dwarf equation of 
state is not used, though we note the central density 
$\rho_c\sim 3 \times 10^6$ g cm$^{-3}$ implies relativistic degeneracy 
would play a role.  The size of the dimensionless gravitational potential 
($\simeq 10^{-4}$) determines the accuracy of our formalism 
(Sec.~\ref{subsec:self_grav_tidal_field}).

We investigate white dwarf--black hole encounters using three different mass 
ratios $\mu \equiv M_*/M$: $1.28\times 10^{-3}$, $4.21 \times 10^{-4}$, and 
$3.77 \times 10^{-5}$, corresponding to IMBH masses of $M = 500 \, M_\odot$, 
$M = 1.52 \times 10^3\, M_\odot$, and $M = 1.70 \times 10^4 \, M_\odot$, 
respectively.  The code uses black hole mass $M$ as the fundamental 
unit of mass, length, and time.  The stellar parameters, when written in 
terms of $M$, therefore are functions of $\mu$.  Their values are gathered 
in Table \ref{table:star_parameters_M}.

\subsection{Orbits}
\label{subsec:orbits}

The duration of each simulation is set equal to $10\, \tau_0$, with the star 
reaching pericenter at $\tau = 0$.  At pericenter $r = R_p$ the radial phase 
is $\chi=0$.  The starting separation $R_i$ is chosen, which determines 
$\chi_i$ and the azimuthal angle $\phi_i$.  In the FNC frame, the black hole 
appears to swing about through an angle $\Psi(\tau)$.  The initial orientation 
of the frame can be freely chosen.  The geodesic equations and the Eq.  
(\ref{eqn:FNC_frame_rotation}) for $\Psi$ are integrated.  In the black hole 
frame the FNC frame vectors precess by an angle $\varphi$ that is the 
difference between $\phi$ and $\Psi$.  These orbital parameters and the 
cumulative frame precession $\Delta\varphi$ are summarized in 
Table \ref{table:encounter_parameters}.

Figure \ref{figure:compare_orbits_eta1} plots trajectories (as seen in
the black hole frame in Schwarzschild coordinates) of the FNC frame
center for a pair of $\eta = 1$ encounters.  Relativistic apsidal advance 
and frame precession are evident in passing both the $M = 500 M_\odot$ 
(left) and $17,000 M_\odot$ (right) black holes, though both effects are 
more pronounced in the latter case.  Plotted for comparison is the 
Newtonian parabolic orbit with the same pericentric distance.

\subsection{Hydrodynamic parameters, resolution, and runtimes}
\label{subsec:hydro_parameters}

The PPMLR hydro method (see Appendix~\ref{sec:method}), like most grid-based 
schemes, requires that some tenuous atmosphere surround the star.  The 
initial density $\rho_{\rm{atm}}$ and pressure $p_{\rm{atm}}$ are set low 
enough to not affect the dynamics of the star.  To ensure this, we choose 
the atmospheric density to be $\rho_{\rm{atm}} = \rho_c \times 10^{-15}$.  
To set the pressure, we first assume a value for the initial atmospheric 
sound speed, taking it to be equal to the virial velocity at $r = 2R_*$: 
$c_{\rm{atm}}^2 = M_*/(2 R_*)$.  The atmospheric pressure is then 
set equal to $p_{\rm{atm}} = c_{\rm{atm}}^2 \rho_{\rm{atm}}/\gamma$.  

It is also useful in the hydrodynamic scheme to set minimum values for the 
density and pressure ($\rho_{\rm{floor}}, p_{\rm{floor}}$) that cannot be 
breached.  Such a floor sometimes proves necessary, as a strong shock 
wave encountering another large discontinuity in density might otherwise 
yield a zone with negative density or pressure.  The floor density 
and pressure can be set quite low.  We take 
$\rho_{\rm{floor}} = \rho_c \times 10^{-25}$ and
$p_{\rm{floor}} = c_{\rm{atm}}^2 \rho_{\rm{floor}}/\gamma.$  

While an elongated domain might be used, all of our simulations involved 
a box with equal side lengths and an evenly-spaced Cartesian grid.  We 
tested the degree of convergence of all of our results by using meshes 
with $128^3$, $256^3$, or $512^3$ total numbers of zones.  The length of 
the side of the computational domain was set to $L = 4 R_*$ for stellar 
equilibrium tests and weak ($\eta = 4-6$) encounters.  For stronger or 
disruptive encounters the domain is taken to be larger, with $L=8R_*$. 
Resolution depends primarily on the number of zones across the radius of 
the initial star.  We refer to the different resolutions by 
$\Delta_A$, $\Delta_B$, and $\Delta_C$, with $\Delta_A = R_*/32$, 
$\Delta_B = R_*/64$, $\Delta_C = R_*/128$.  Also, zero-gradient outflow 
boundary conditions are used on the domain surface.  

The domain is split into slabs for computing on a cluster.  For simplicity 
we took the slabs to be one-zone thick sheets and allocated one cluster core
per sheet.  Thus the number of cores is locked to the number of zones in 
one direction.  Our highest resolution runs used 512 processors.  At this 
resolution simulations lasting ten dynamical times required between 88 and 
127 hours of wall-clock time.

\subsection{Equilibrium configurations and tests of hydro plus gravity}
\label{subsec:equil_config}
%%%%%%%%%%%%%%%%%%%%%%%%%%%%%%%%%%%%%%%%%%%%%%%%%%%%%%%%%%%%%%%%%%%%%
\begin{figure}[!htbp]
{ 
\begin{center} 
\includegraphics[scale=1.02]{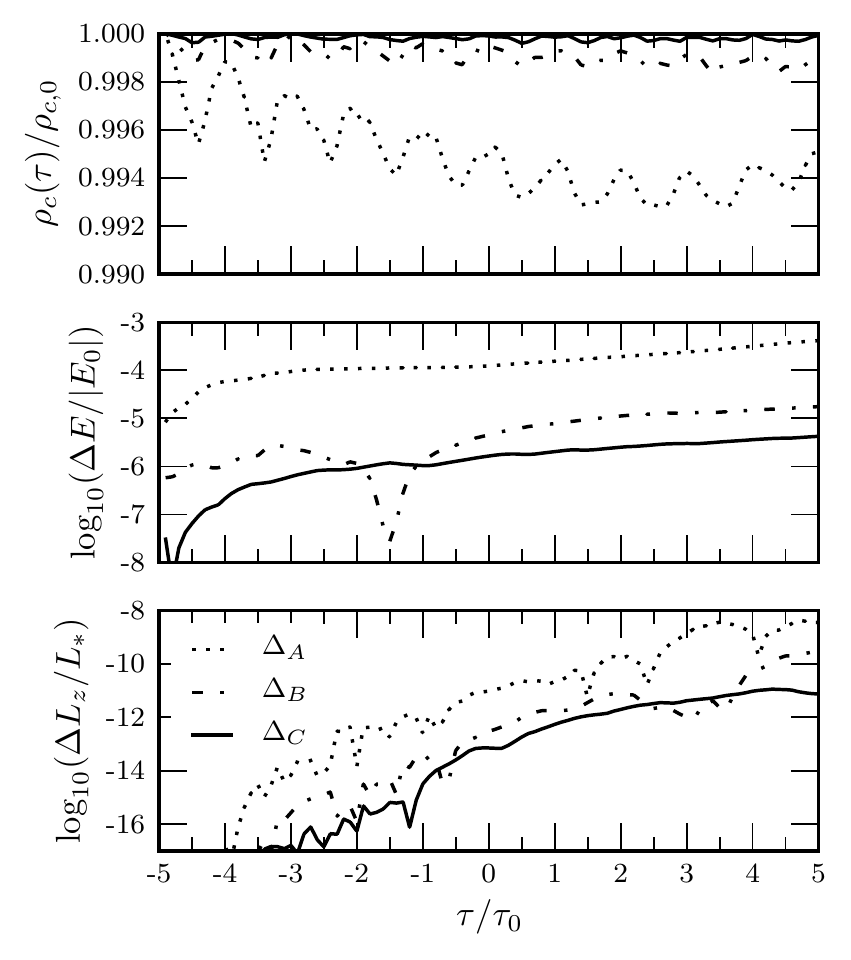} 
  \caption{
  \label{figure:equilibrium} 
Tests of hydrostatic equilibrium with a polytropic star having $n=3/2$ and 
$\g = 5/3$.  In the top panel we plot the central density $\rho_c(\tau)$ 
as a function of time, normalized by its initial value.  The time $\tau$ 
on the horizontal axis (shared by all three panels) is given in units of
the fundamental radial pulsation period $\tau_0$.  Low amplitude pulsations 
are evident at exactly the $F$-mode period, which last $>10$ periods.  
Models with three different resolutions, $\Delta_A$, $\Delta_B$, $\Delta_C$ 
(see text) are compared.  Convergence to equilibrium is evident.  Small 
secular drift in the central density occurs, a result of weak spurious 
entropy generation in the hydro code.  The second panel plots fractional 
change in total energy $E_{\rm tot}(\tau)$.  Higher resolution models 
conserve energy to $\lesssim 10^{-5}$.  The bottom panel plots the change 
in total spin angular momentum $L_{\rm tot}(\tau)$ scaled relative to a 
dimensional estimate of break-up angular momentum $L_*$.  
} \end{center} 
}
\end{figure}
%%%%%%%%%%%%%%%%%%%%%%%%%%%%%%%%%%%%%%%%%%%%%%%%%%%%%%%%%%%%%%%%%%%%%

An important test is how close to exact stellar equilibrium can
three-dimensional simulations be held.  Equilibrium models serve as a
control, especially for weak tidal encounters.  The Lane-Emden
equation is integrated with a fine one-dimensional mesh.  The
resulting hydrodynamic radial profiles are mapped onto the
three-dimensional Cartesian mesh of chosen resolution.  The Poisson
solver (see Appendix~\ref{sec:method}) is then called to find the
self-gravitational potential in the three-dimensional domain.

The resulting stellar model is found to be close to but not exactly in
equilibrium.  The top panel of Fig. \ref{figure:equilibrium} reveals a
small fractional oscillation and drift in the central density.  Three
main sources of error contribute to breaking equilibrium.  First,
there is discretization error in mapping the well-resolved
one-dimensional Lane-Emden radial profiles onto a three-dimensional
Cartesian grid.  Second, there are inaccuracies in the gravitational
field obtained with the Poisson solver.  These two effects combine to
place the initial star slightly out of hydrostatic equilibrium and in
response the star oscillates, primarily in the fundamental radial mode
(Fig. \ref{figure:equilibrium}).  Third, there is a weak spurious
generation of entropy within the star, a byproduct of the PPM
algorithm (and many other hydro schemes \cite{bale2002}).  In PPM, any
gradient in density and pressure, even when balanced by gravitational
acceleration in hydrostatic equilibrium, is viewed by the method as a
discontinuity, or small shock.  Small secular increases in entropy
occur, leading to slight expansion of the star and decrease in central
density.  As Fig. \ref{figure:equilibrium} shows, at low resolution
the effect is an average decrease in density of $\sim 0.5$ \% over
five to ten dynamical times.

Both of these effects are reduced with higher resolution.  In these tests 
we used domains with $128^3$, $256^3$, and $512^3$ zones.  In each case 
the domain length was four times the radius of the star, $L=4R_*$, and so the
three resolutions considered had 32, 64, and 128 zones per $R_*$.  Based on 
this test, we take the lowest resolution of interest to be $\Delta_A$, with 
our best results requiring resolutions of $\Delta_B$ and $\Delta_C$.  At 
the higher resolutions we hold total energy conservation to 
$\lesssim 10^{-5}$ and the equilibrium models have effectively no tendency 
to generate spurious angular momentum.  

During a tidal encounter, however, a star will be set into nonradial (and 
radial) oscillations.  The amplitude and persistence of these oscillations 
is an important result to be derived from the numerical simulations.  The 
tests in this section show that stellar pulsations are well maintained by 
our code even at modest resolution.  The radial mode is seen to damp in 
amplitude over time but with $Q \simeq 40$.  

\section{Hydrodynamic features and mass stripping in weak tidal encounters}
\label{sec:features-massloss}

%%%%%%%%%%%%%%%%%%%%%%%%%%%%%%%%%%%%%%%%%%%%%%%%%%%%%%%%%%%%%%%%%%%%%
\begin{figure*}[!htbp]
{
\begin{tabular}{ccc}
\includegraphics[scale=1.05]
{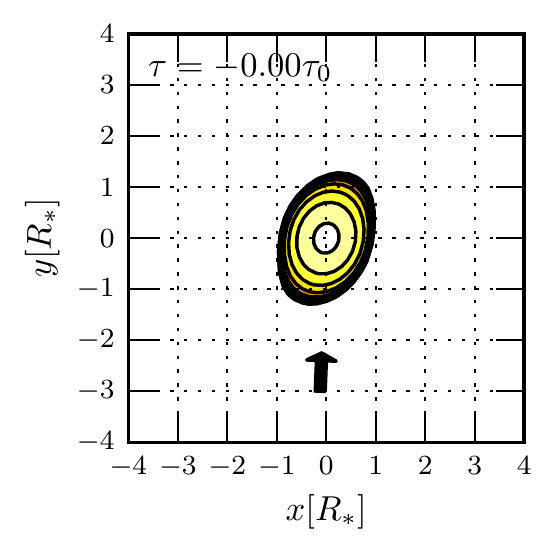} &
\includegraphics[scale=1.05]
{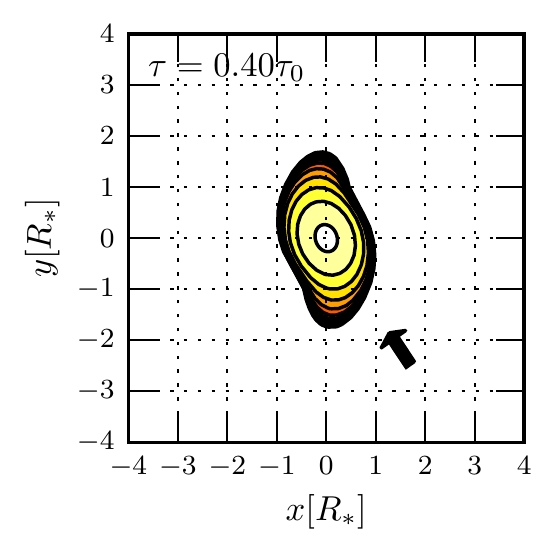} &
\includegraphics[scale=1.05]
{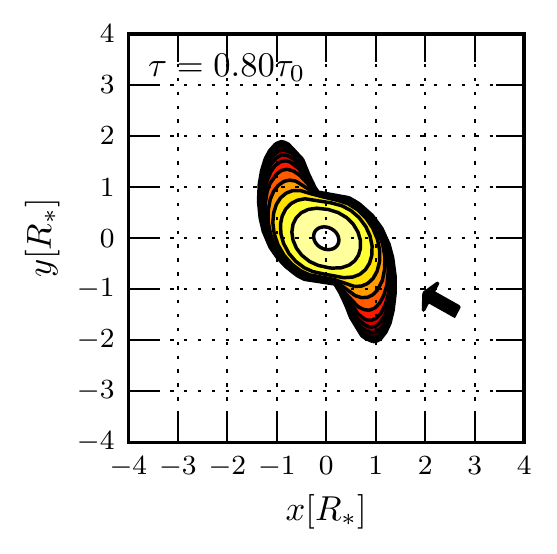} \\
\includegraphics[scale=1.05]
{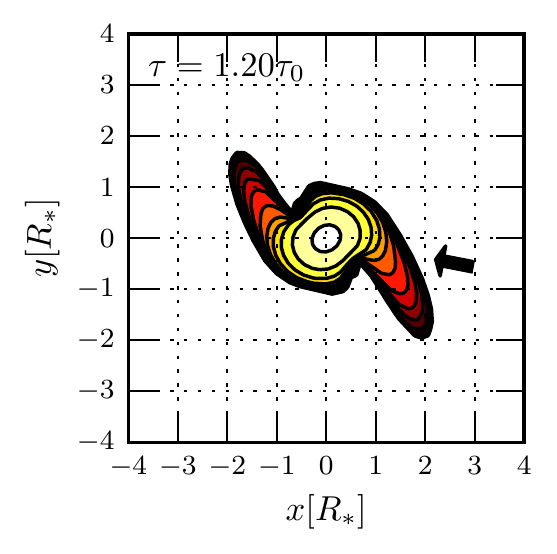} &
\includegraphics[scale=1.05]
{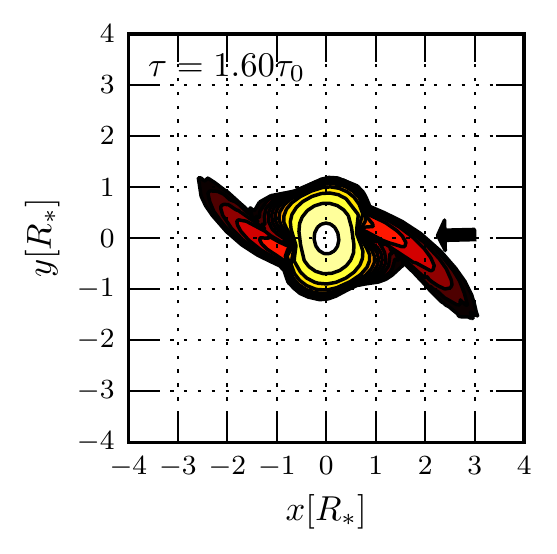} &
\includegraphics[scale=1.05]
{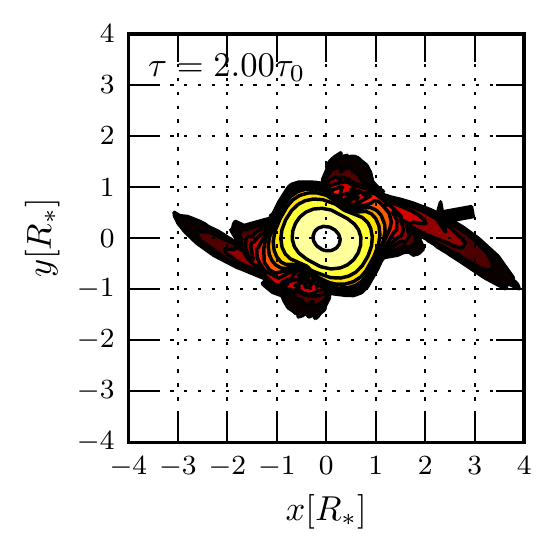} \\
%% \includegraphics[scale=1.05]
%% {eta3_contours/star_density_contour_mu1m5/pdf_files/1050.pdf} &
%% \includegraphics[scale=1.05]
%% {eta3_contours/star_density_contour_mu1m5/pdf_files/1054.pdf} &
%% \includegraphics[scale=1.05]
%% {eta3_contours/star_density_contour_mu1m5/pdf_files/1058.pdf} \\
%% \includegraphics[scale=1.05]
%% {eta3_contours/star_density_contour_mu1m5/pdf_files/1062.pdf} &
%% \includegraphics[scale=1.05]
%% {eta3_contours/star_density_contour_mu1m5/pdf_files/1066.pdf} &
%% \includegraphics[scale=1.05]
%% {eta3_contours/star_density_contour_mu1m5/pdf_files/1070.pdf} \\
      \end{tabular}
	\caption
	{\label{figure:eta3_density_contour}
        Density contour plots of an $\eta=3$ encounter with
        $\mu = 3.77\times 10^{-5}$.  Positions are in units of $R_*$.  
        The contour lines are of $\log_{10}\rho$, ranging from $-8$ to 
        $-2$ in steps of $0.5$ (as seen in the equatorial plane).  Besides
        the Newtonian quadrupole, the octupole tidal term and 
        (orbital) relativistic correction to the quadrupole term have been 
        included in the calculation. 
\\
\\
}}
%\end{figure*}
%%%%%%%%%%%%%%%%%%%%%%%%%%%%%%%%%%%%%%%%%%%%%%%%%%%%%%%%%%%%%%%%%%%%%%%

%%%%%%%%%%%%%%%%%%%%%%%%%%%%%%%%%%%%%%%%%%%%%%%%%%%%%%%%%%%%%%%%%%%%%%%
%\begin{figure*}[b]
{
      \begin{tabular}{ccc}
	\includegraphics[scale=1.05]{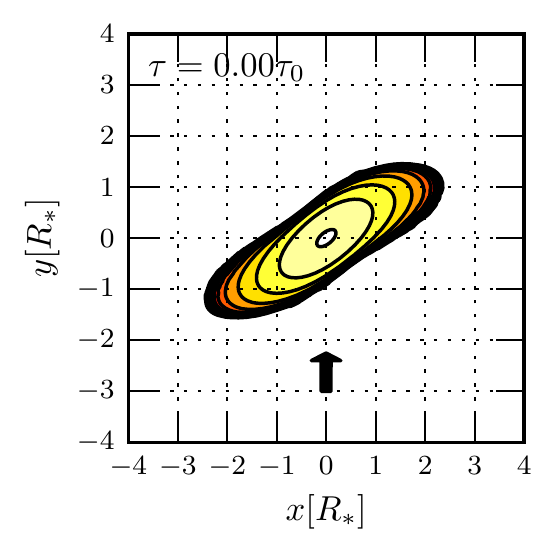} &
	\includegraphics[scale=1.05]{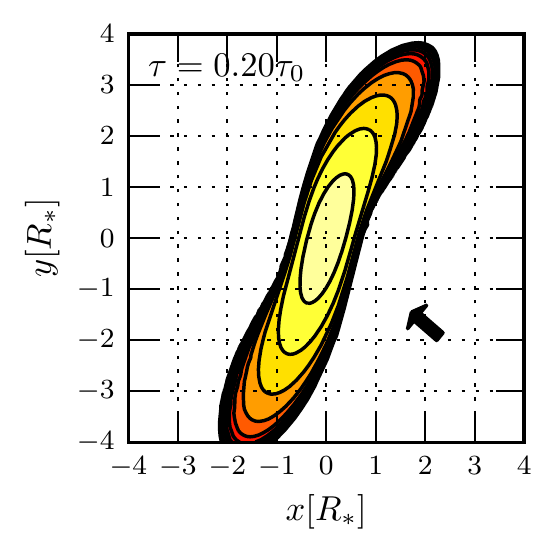} &
	\includegraphics[scale=1.05]{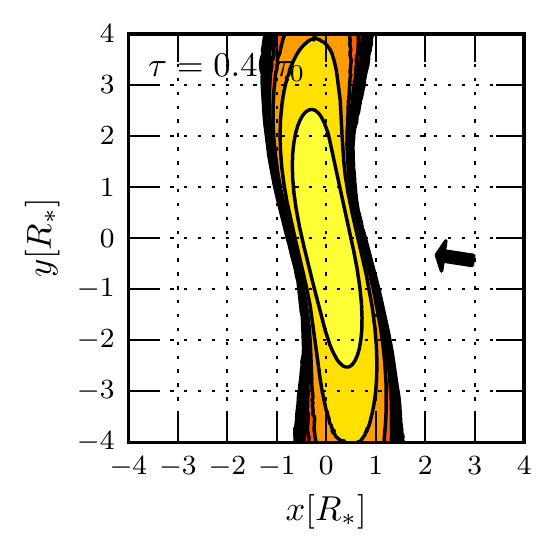} \\
      \end{tabular}
	\caption
	{\label{figure:eta1_density_contour}
        Density contour plots of an $\eta=1$ encounter with 
        $\mu = 3.77\times 10^{-5}$.  Positions are in units of $R_*$.  
        The simulation begins at $\tau=-5\,\tau_0$, reaches pericenter at 
        $\tau=0$, and ends at $\tau=5\,\tau_0$.  The contour lines are
        given for $\log_{10}\rho$ from $-8$ to $-2$ in steps of $0.5$.  
        Note slight deflection of the CM in this and preceding 
        figure.  }}
\end{figure*}
%%%%%%%%%%%%%%%%%%%%%%%%%%%%%%%%%%%%%%%%%%%%%%%%%%%%%%%%%%%%%%%%%%%%%%%

%%%%%%%%%%%%%%%%%%%%%%%%%%%%%%%%%%%%%%%%%%%%%%%%%%%%%%%%%%%%%%%%%%%%%%

\begin{figure*}[!htpb]
{ 
\begin{tabular}{cc} 
  \includegraphics[scale=1.075]{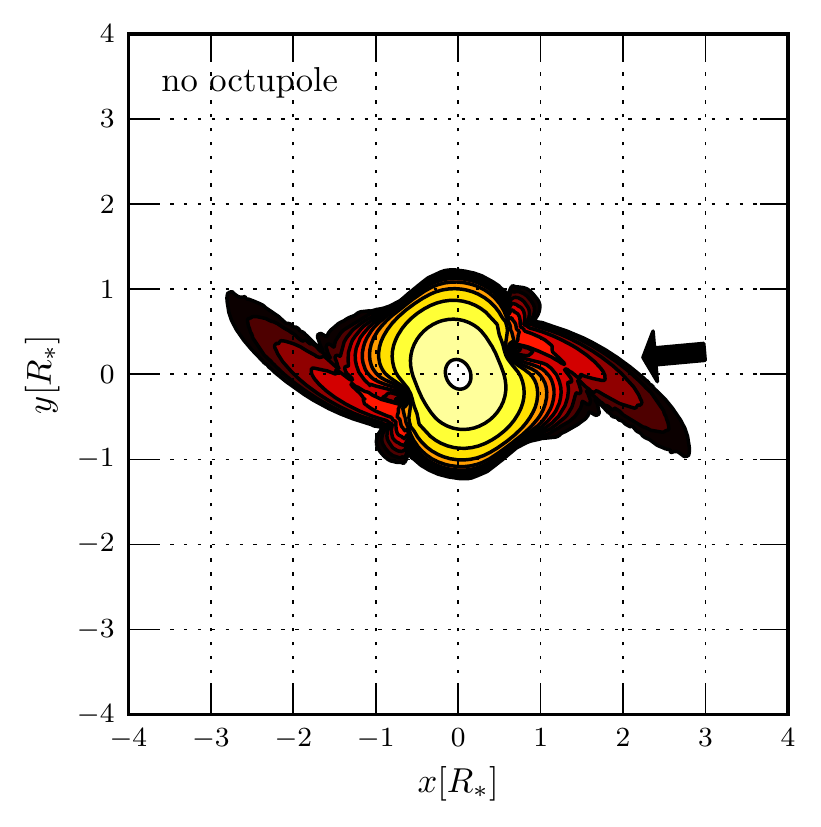}  & 
  \includegraphics[scale=1.075]{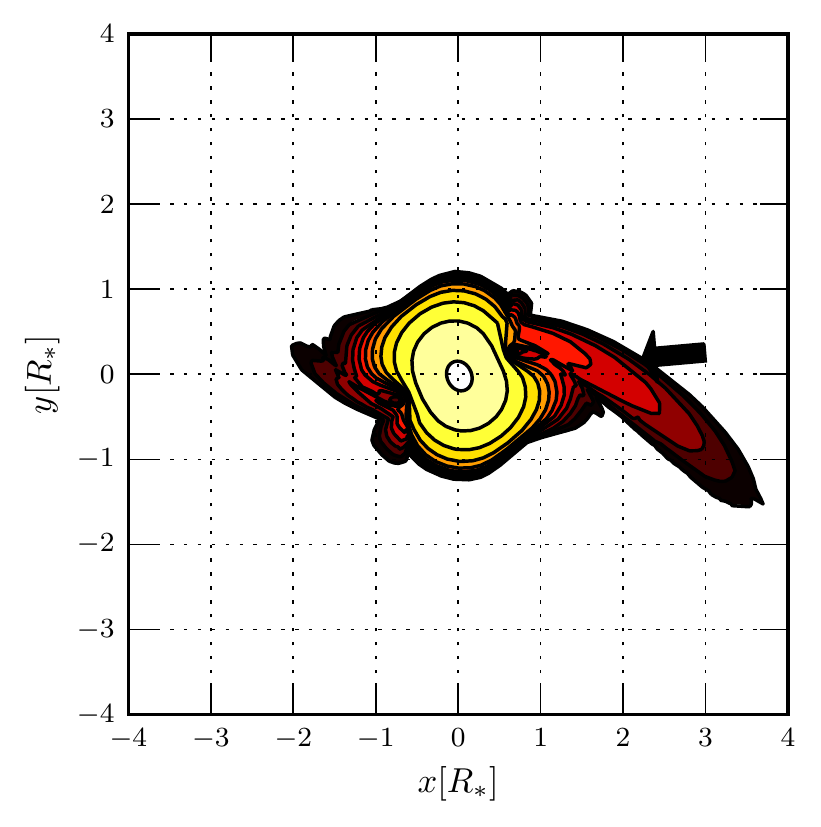} \\ 
\end{tabular} 
\caption
  { \label{figure:density_contour_eta3} 
  Contour plots of density of the star in the x-y plane at $\tau =
  1.7\, \tau_0$ for an $\eta=3$ encounter with $\mu=1.28\times
  10^{-3}$.  Positions are in units of $R_*$.  The contour lines are
  given for $\log_{10}\rho$ from $-11$ to $-5$ in steps of 0.5.
  Comparison is made between a simulation with only the quadrupole
  ($l=2$) tidal terms (left) and one with both the quadrupole and
  octupole ($l=3$) tidal terms included (right).  Resolution of these
  simulations is $\Delta_B$.  The effect of the octupole tidal terms
  in driving a deflection of the CM is evident in the asymmetry of the
  tidal lobes.  } }
\end{figure*}

%%%%%%%%%%%%%%%%%%%%%%%%%%%%%%%%%%%%%%%%%%%%%%%%%%%%%%%%%%%%%%%%%%%%%%%

%%%%%%%%%%%%%%%%%%%%%%%%%%%%%%%%%%%%%%%%%%%%%%%%%%%%%%%%%%%%%%%%%%%%%%%

\begin{figure*}[!htbp]
{
\begin{tabular}{cc}
\includegraphics[scale=1.05]{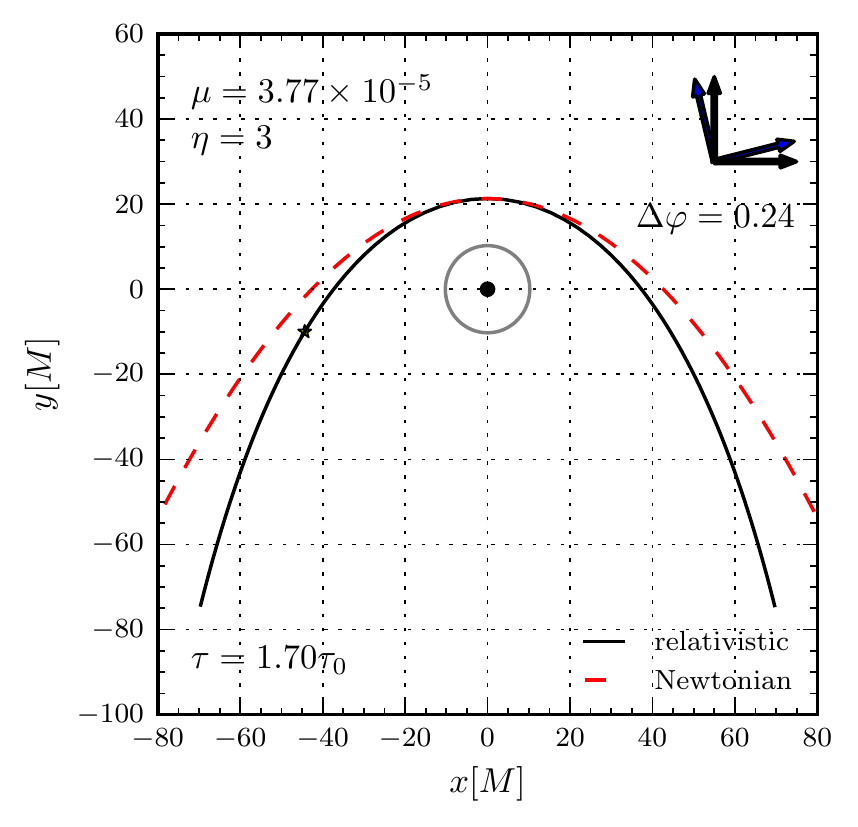} &
\includegraphics[scale=1.05]{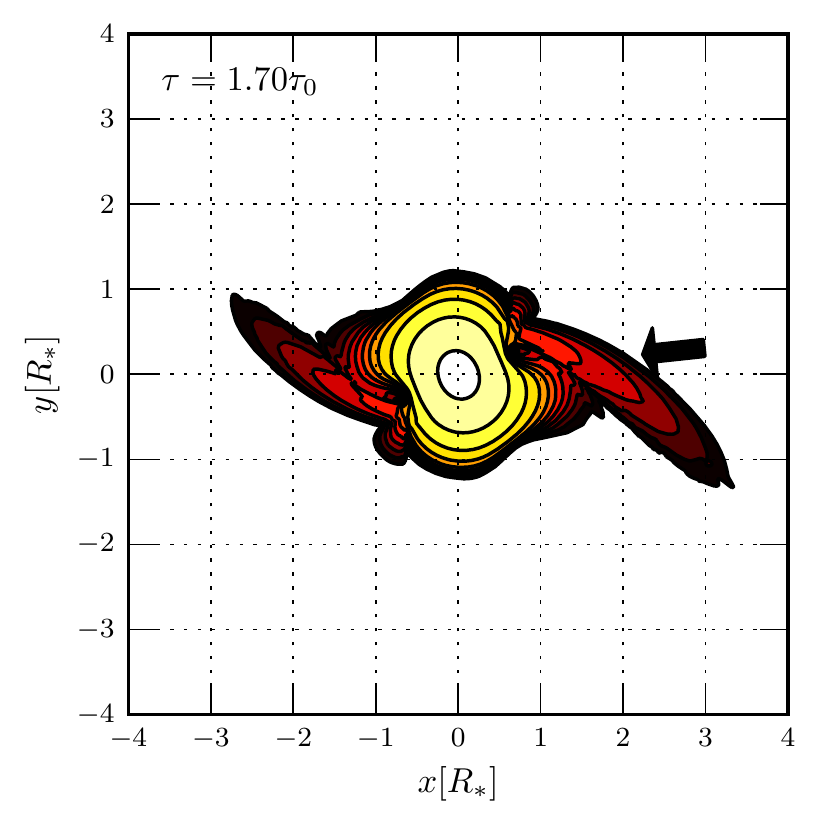} \\
\end{tabular}
\caption
{\label{figure:orbit_mu1m5_eta3}
   An $\eta=3$ encounter with more extreme mass ratio 
   $\mu = 3.77\times 10^{-5}$.  Results shown at time $\tau=1.7\, \tau_0$.  
   On the left, position of the frame center is marked on the orbit.  
   On the right are shown density contours in the x-y plane.  The contour
   lines are given for $\log_{10}\rho$ from $-8$ to $-2$ in steps of $0.5$.  
   The quadrupole and octupole tidal terms have been included.  For this
   more extreme mass ratio, the effects of the octupole tide are less 
   pronounced (compare to Fig.~\ref{figure:density_contour_eta3}). 
}}
\end{figure*}

%%%%%%%%%%%%%%%%%%%%%%%%%%%%%%%%%%%%%%%%%%%%%%%%%%%%%%%%%%%%%%%%%%%%%%

In this section we consider some of the qualitative hydrodynamic features 
that are seen in tidal encounters, focusing especially on our inclusion of 
a higher-order tidal moment.  In addition, we calculate and show the 
amount of mass loss that occurs as a function of parameter $\eta$ for weak 
tidal encounters between a white dwarf and an IMBH.  

\subsection{Octupole tidal term}
\label{sec:octupole}

In general our simulations could include all of the tidal acceleration 
terms identified as potentially significant in Sec.~\ref{sec:formalism}.  
However, it is useful to turn various terms on or off and compare simulations
to see the resulting effects.  One result of doing so is that we find that 
our mass ratios are too small to make inclusion of the $l = 4$ tide 
worthwhile.  Consequently we have not included $l = 4$ in any of the 
results in this paper.  The same is not true of the octupole tide 
($l = 3$), which generates interesting physical effects.

In Figs.~\ref{figure:eta3_density_contour} 
and~\ref{figure:eta1_density_contour} we consider encounters at our most 
extreme mass ratio ($\mu=3.77\times 10^{-5}$) and with encounter strengths 
$\eta = 3$ and $\eta = 1$, respectively.  In these simulations we have 
included the full tidal field.  With $\eta=3$ the star is tidally disturbed 
with a small fraction of mass stripped from the star, as can be seen at a 
sequence of times in Fig.~\ref{figure:eta3_density_contour}.  In contrast, 
the star is fully tidally disrupted when $\eta=1$, as seen in 
Fig.~\ref{figure:eta1_density_contour}.  Similar but not identical 
results from high-resolution mesh-based calculations can be found in 
Refs.~\cite{khok1993,khok1993strong,frol1994}.  The primary difference is our 
inclusion of the octupole tidal term, whose effect shows up in the asymmetry 
of the tidal lobes in Fig.~\ref{figure:eta3_density_contour} and the 
deflection of the CM that is also evident in both 
Fig.~\ref{figure:eta3_density_contour} and 
Fig.~\ref{figure:eta1_density_contour}. 

Figure~\ref{figure:density_contour_eta3} even more clearly shows the effects
of the octupole tide.  In this case we plot an $\eta=3$ encounter with less 
extreme mass ratio $\mu = 1.28\times 10^{-3}$.  The full tidal field is 
incorporated in the simulation shown in the right panel, while the left panel 
shows the same simulation except for switching off the octupole tide.  The 
arrow represents the direction from the black hole to the origin of the 
FNC frame.  For contrast Fig.~\ref{figure:orbit_mu1m5_eta3} shows on the 
right side an $\eta = 3$ encounter at the same dynamical time but from a 
simulation with our most extreme mass ratio.  The asymmetry is still 
present but much less pronounced, as reference to Fig.~\ref{figure:PN_scales} 
and the analysis of Section \ref{subsec:fluid_eqn_retained_tidal_terms} 
would suggest.

\subsection{Mass loss in weak tidal encounters}
\label{sec:massloss}

%%%%%%%%%%%%%%%%%%%%%%%%%%%%%%%%%%%%%%%%%%%%%%%%%%%%%%%%%%%%%%%%%%%%%%%
\begin{figure}[!htbp]
{ 
\includegraphics[scale=1.075]{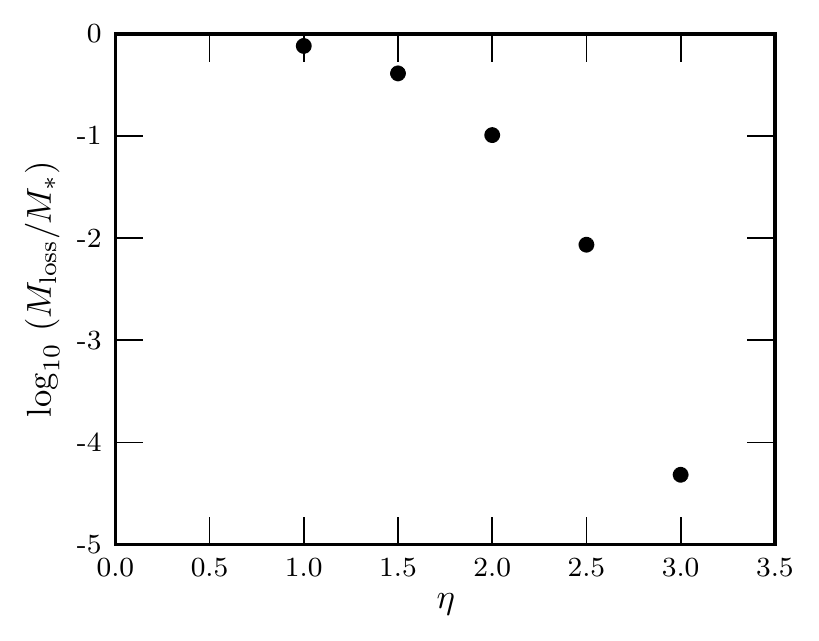}
\caption
 { \label{figure:outflow_mass} 
Fraction of mass stripped from white dwarf as a function of $\eta$ for 
weak tidal encounters.  In each simulation the mass ratio was fixed 
at $\mu = 1.28\times 10^{-3}$ and we plot the fraction of mass lost 
from the domain at $\tau = 5\, \tau_0$ (the end of the simulations).  The 
fraction of mass lost for $\eta=4$ was also computed and found to be 
$\simeq 10^{-10}$ (off scale). }}
\end{figure}
%%%%%%%%%%%%%%%%%%%%%%%%%%%%%%%%%%%%%%%%%%%%%%%%%%%%%%%%%%%%%%%%%%%%%%%

Weak tidal encounters may result as stars diffuse into the loss cone of a 
SMBH or IMBH.  Successive passages may heat the star and the induced 
oscillations may be resonant with the orbit~\cite{rath2005}.  Gravitational 
bremsstrahlung may be important as well for compact systems.  The combination 
of these effects spurs a reduction in $\eta$ with each passage.

Another important effect of weak encounters is partial mass stripping.  
In the absence of competing effects, any mass loss will be reflected in 
the star having a lower average central density and larger average radius 
during the next encounter.  This effect in turn shifts $\eta$ to a lower 
value and enhances the likelihood of disruption~\cite{krol2011}.

We have calculated the amount of mass loss in a set of weak tidal encounters.
In Fig. \ref{figure:outflow_mass}, the fractional amount of mass stripped 
from the star and lost from the computational domain is given for a range of 
encounter parameters from $\eta=1$ through $\eta = 3$.  (The $\eta = 4$ 
case was computed as well but the measured mass loss fraction of 
$\simeq 10^{-10}$ may be low enough to be affected by the ``atmosphere'' 
we are forced to add to the hydrodynamic simulations.)   The $\eta = 1$ 
case involved complete disruption.  In each case we held the mass ratio 
fixed at $\mu=1.28\times 10^{-3}$.  

For $\eta \gtrsim 2$ ($R_p \gtrsim 1.6\, R_t$) the amount of mass loss is 
small enough that multiple passages with partial stripping should occur.  
If the star were not heated (but see Sec.~\ref{sec:tidal-energy}), then 
the response of the mean radius would be determined by
$\delta R_*/R_* = -(1/3) \delta M_*/M_*$ and 
$\delta\eta/\eta = \delta M_*/M_*$.  The $1.0$ \% mass loss at 
$\eta = 2.5$, for example, would induce a radius increase of only $0.3$ \% 
and a drop in $\eta$ of 1 \%.  At $\eta = 2$ however, the mass loss is 
$10$ \% and the mean radius increase would be $3$ \%, leading to a 
10 \% drop in $\eta$.  As we will show in the next section, shock heating 
is important as well and the effects of mass loss just set a lower bound 
on effective reduction of $\eta$.

\section{Tidal energy deposition, relativistic effects, and capture}
\label{sec:tidal-energy}

In this section we discuss the transfer of energy into the star that occurs 
as a result of tidal interaction.  This topic has been addressed 
previously (see e.g., \cite{pres1977,cart1985,lee1986,khok1993}).  We 
consider the subject again for several reasons.  First, our FNC system is 
tailored to follow small relative motions with respect to the initial 
timelike geodesic.  Second, the FNC expansion we use
retains higher-order moments and orbital relativistic effects.  Finally, we 
are able to apply relatively high resolution ($512^3$) and assess convergence
of our numerical results. 

The equation of motion (\ref{eqn:euler}) of the fluid in the FNC frame is 
nearly Newtonian.  As such we can carry over much of the standard 
understanding of tidal heating \cite{cart1985} with only minimal modification.
If we contract Eqs. (\ref{eqn:euler}) and (\ref{eqn:tidalaccel}) with 
$v^i$, use the continuity equation and first law of thermodynamics, we 
obtain the equation of energy conservation of the star (as seen in the FNC 
frame)
\ba
\label{eqn:heating}
\nn
\frac{\partial}{\partial \tau}\left( \rho \Pi + \tfrac{1}{2} \rho v^2 \right) 
&+& \frac{\partial}{\partial x^k}
\left[ \rho v^k \left( \Pi + \frac{p}{\rho} + \tfrac{1}{2} v^2 \right)\right]
+ \rho v^i \frac{\partial \Phi_{*}}{\partial x^i} \\
& & \quad = -\rho v^i \frac{\partial \Phi_{\text{tidal}}}{\partial x^i}
- \rho v^i \frac{\partial A_i}{\partial \tau} .
\ea
Integrating over a volume encompassing all of the material, we can define 
the internal energy $U_{*} = \int \rho \Pi \, d^3x$, 
self-gravitational energy
$\Omega_{*} = \tfrac{1}{2} \int \rho \Phi_{*} \, d^3x $, 
and kinetic energy $T_{*} = \tfrac{1}{2} \int \rho v^2 \, d^3x $.  The 
equation of energy conservation of the star is then
\be
\frac{d E_{*}}{d \tau} = 
- \int \rho v^i \frac{\partial \Phi_{\text{tidal}}}{\partial x^i} \, d^3x
- \int \rho v^i \frac{\partial A_i}{\partial \tau} \, d^3x ,
\ee
where the total energy $E_{*} = U_{*} + \Omega_{*} + T_{*}$ would be 
conserved in the absence of tides.  We have assumed that the fluid 
remains adiabatic (no radiative cooling) and that no mass or energy fluxes 
through a sufficiently large bounding surface.

To gain a physical picture, imagine dropping the gravitomagnetic potential 
$A_i$ \eqref{eqn:gmpotential} and retaining only Newtonian terms in the
tidal potential $\Phi_{\rm tidal}$ \eqref{eqn:tidalpotential}.  Then the
tidal energy transfer~\cite{cart1985} would reduce to
\be
\frac{d E_{*}}{d \tau} = 
- \tfrac{1}{2}C_{ij}^{(0)}\, \dot{I}_{ij}  
- \tfrac{1}{6}C_{ijk}^{(0)}\, \dot{I}_{ijk}  
+ \cdots ,
\ee
where $I_{ij} = \int \rho x_i x_j d^3x$ and 
$I_{ijk} = \int \rho x_i x_j x_k d^3x$ are the second and third
mass moments, dot refers to the time derivative, and of course only the
trace-free parts of $I_{ij}$ and $I_{ijk}$ contribute.  While our code 
calculates all of the moments and relativistic corrections we enumerated in 
Sec.~\ref{sec:formalism}, the quadrupole tide still dominates the 
energy transfer.

\subsection{Total energy deposition and comparison with linear theory}
\label{subsec:compare_linear_theory}

\begin{figure}[!htbp]
{
\includegraphics[scale=1.09]{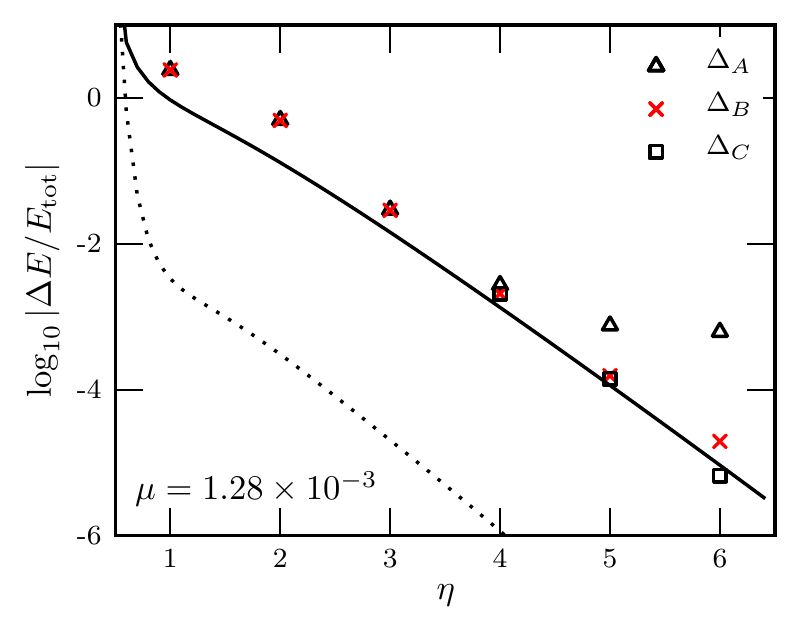} 
\caption { \label{figure:linear_theory} 
Measured energy deposition versus encounter parameter $\eta$ and comparison 
with linear theory.  Each simulation was made with 
$\mu = 1.28 \times 10^{-3}$.  The points indicate measured fractional 
increases in the total energy of 
the star at the end of each simulation.  Results are given at three different 
resolutions for each $\eta$.  The measured energies are numerically well 
converged for $\eta = 1$ through $\eta = 4$, and probably $\eta = 5$ as 
well.  Our range of resolution is not adequate to measure the deposited 
energy for $\eta = 6$, which is at the level of $10^{-5}$ of the initial 
stellar energy.  The solid curve gives the prediction of linear theory for
the contributions of the quadrupole ($l = 2$) and octupole tides and 
the nonradial oscillations they induce.  The dotted curve is the prediction 
for the octupole ($l = 3$) tide alone, which is two orders of magnitude 
smaller than the quadrupole at this mass ratio.
} }
\end{figure}

A series of simulations were run with a fixed mass ratio of 
$\mu=1.28\times 10^{-3}$ but with varying encounter parameters 
($\eta = 1$ through $\eta = 6$) and at three different resolutions.
In each case we measured the final total energy of the configuration 
as seen in the FNC frame and compared it to the initial energy of the 
inbound star.  The tidal field was seen to do work on the star and the 
gain in energy is depicted in Fig.~\ref{figure:linear_theory}.

%%%%%%%%%%%%%%%%%%%%%%%%%%%%%%%%%%%%%%%%%%%%%%%%%%%%%%%%%%%%%%%%%%%%%%
%%%%%%%%%%%%%%%%%%%%%%%%%%%%%%%%%%%%%%%%%%%%%%%%%%%%%%%%%%%%%%%%%%%%%%%

%%%%%%%%%%%%%%%%%%%%%%%%%%%%%%%%%%%%%%%%%%%%%%%%%%%%%%%%%%%%%%%%%%%%%
\begin{table}[h]
\begin{center}
\caption{
\label{table:energy_excess}
Energy deposited into the star or debris as a function of tidal parameter 
$\eta$.  The second column gives the fractional increase in the energy of 
the star relative to the magnitude of the star's initial total energy.  
The third column gives the expected fractional increase in energy based on 
linear theory of excitation of nonradial modes.  The fourth column gives 
the difference, which is a fractional excess energy deposition.  The 
fifth column estimates the amount of the excess energy that might be 
attributed to shock heating and the sixth column estimates the amount of the 
excess that could be explained by excitation of the fundamental radial mode. 
Each encounter was simulated with a mass ratio of 
$\mu=1.28\times 10^{-3}$.  Tabulated numbers in column two were drawn from
our highest resolution runs. }
%% %2       & 2.16e-8  & 0.539 & 0.228  \\
%% 2       & 2.16e-8  & 0.908 & 0.228  \\
%% 3       & 8.75e-10 & 1.523 & 0.140  \\
%% 4       & 4.62e-10 & 0.261 & 0.003 \\
\begin{tabular*}{8.6cm}%
     {@{\extracolsep{\fill}}ccccccc}
\hline
\hline
\noalign{\smallskip}
$\eta$ & $\frac{\Delta E_{\rm dep}}{|E_{\text{tot}}|}$ & 
$\frac{\Delta E_{\rm lin}}{|E_{\text{tot}}|}$ & 
$\frac{\Delta E_{\rm ex}}{|E_{\text{tot}}|}$ & 
$\frac{\Delta E_{\rm shock}}{|E_{\text{tot}}|}$  &
$\frac{\Delta E_{\rm rad}}{|E_{\text{tot}}|}$ \\
\noalign{\smallskip}
\hline
1       & 2.41     & 9.39e-01 & 1.48 & & \\
2       & 4.89e-01 & 1.30e-01 & 3.59e-01 & 3.3e-01 & 8.9e-02  \\
3       & 2.89e-02 & 1.44e-02 & 1.45e-02 & 2.4e-02 & 1.6e-03  \\
4       & 2.11e-03 & 1.34e-03 & 7.68e-03 & 2.0e-03 & 1.4e-05 \\
5       & 1.50e-04 & 1.15e-04 & 3.49e-05 & & \\
6       & 1.16e-05 & 9.23e-06 & 2.40e-06  & & \\
\hline
\hline
\end{tabular*}
\end{center}
%\vspace{-0.6cm}
\end{table}
%%%%%%%%%%%%%%%%%%%%%%%%%%%%%%%%%%%%%%%%%%%%%%%%%%%%%%%%%%%%%%%%%%%%%

The energy gain is also tabulated in Table \ref{table:energy_excess}, 
but expressed as a ratio to the magnitude of the total energy 
$E_{\text{tot}} = -(3/7)\, G M_*^2/R_*$ of the initial star.  We see that for 
$\eta = 1$ the star has come apart.  Progressively less energy is 
deposited for weaker encounters.  By comparing simulations made at three 
different resolutions it is apparent that the results for $\eta = 1$ 
through $\eta = 5$ have converged.  The result for $\eta = 6$ is less 
well known.  In any event, we appear able to determine accurately fractional
energy depositions as small as $10^{-4}$.

Energy observed in the simulations to be deposited onto the star can be 
compared to the predictions of linear theory.  Press and Teukolsky 
\cite{pres1977} and Lee and Ostriker \cite{lee1986} calculated the
amount of energy that a time-dependent linear perturbation in the 
gravitational potential would induce via the excitation of nonradial modes.  
The interaction is decomposed into spherical harmonics, and each tidal 
multipole will excite the corresponding lowest frequency nonradial $l$ 
mode.  The total energy deposited is a sum over each $l$ mode 
contribution and is given by
\be \label{energy_deposition_linear_theory}
\Delta E_{\text{lin}} = \left
(\frac{GM_*^2}{R_*} \right ) \left (\frac{M}{M_*} \right )^2
\sum_{l=2,3,...} \left (\frac{R_*}{R_p} \right )^{2l +2} T_l(\eta),
\ee 
where the key to the theory is calculating the dimensionless functions 
$T_l(\eta)$~\cite{pres1977}, which depend on $\eta$ and also the 
polytropic index $n$.  We use the results for $T_l$ obtained 
previously~\cite{lee1986,port1993} for an $n=3/2$ polytrope.  

The mass ratio (taken to be $\mu=1.28\times 10^{-3}$ in this section)
affects the relative magnitude of the $l=2$ and $l=3$ excitations, as 
reference to \eqref{energy_deposition_linear_theory} makes clear.  In 
Fig.~\ref{figure:linear_theory} we plot the curves for both the $l=2$ 
and $l=3$ contributions to $\Delta E_{\text{lin}}$.  The linear 
contribution of the octupole tide is two orders of magnitude below that 
of the quadrupole. 

We see a clear convergence with linear theory in the limit of weak 
encounters (up to $\eta = 5$).  As the strength of the encounter grows 
the energy actually deposited is seen to exceed the predictions of 
linear theory.  This excess in energy deposition appears to be real 
(based on numerical convergence) and confirms results discovered 
previously~\cite{khok1993}.  As Table \ref{table:energy_excess} indicates, 
the excess can be as much as 50\% to 75\% of the total.  The result has 
important implications for predictions of the cross section for tidal 
capture of stars.

%%%%%%%%%%%%%%%%%%%%%%%%%%%%%%%%%%%%%%%%%%%%%%%%%%%%%%%%%%%%%%%%%%%%%%
\begin{figure}[!htbp]
{\begin{center} 
\includegraphics[scale=1.04]{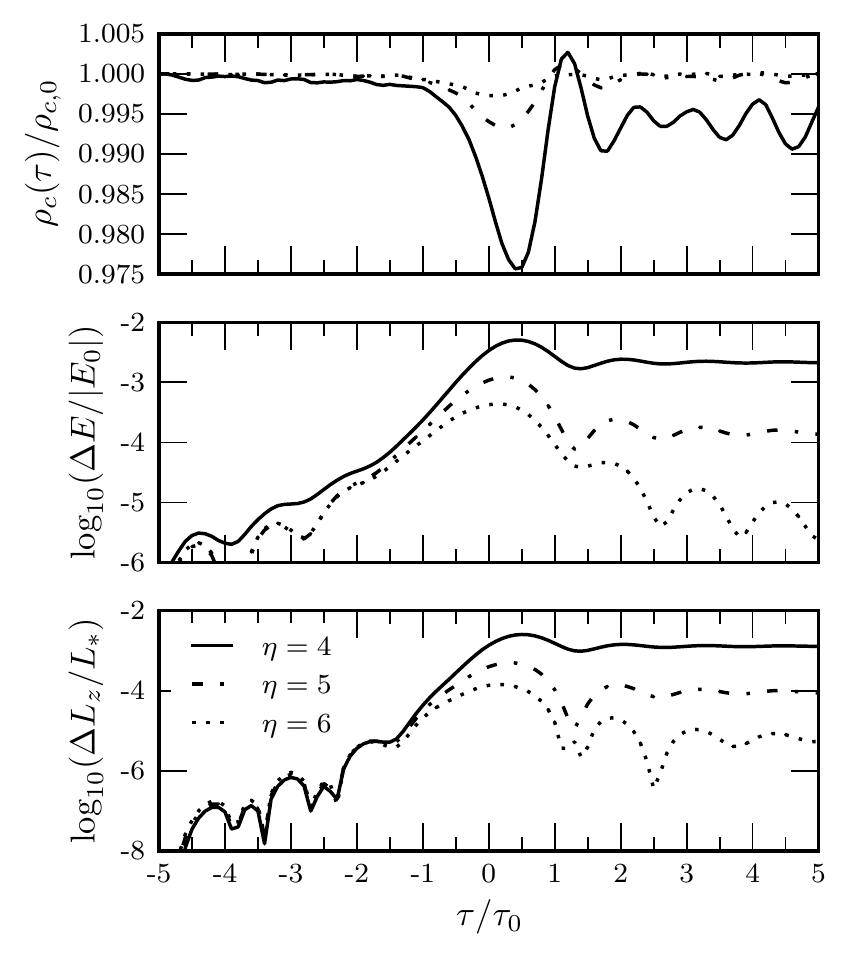} 
  \caption{
  \label{figure:weak_encounters} 
Weak tidal encounters.  Top panel shows fractional changes in central 
density for $\eta = 4$, $\eta =5$, and $\eta =6$ encounters.  
Mid panel shows (normalized) energy deposited for each encounter.
Bottom panel displays (normalized) spin angular momentum deposited for 
each encounter.  The mass ratio is $1.28 \times 10^{-3}$.  Each curve is 
drawn from the highest resolution ($\Delta_C$) simulations.
} \end{center} 
}
\end{figure}

%%%%%%%%%%%%%%%%%%%%%%%%%%%%%%%%%%%%%%%%%%%%%%%%%%%%%%%%%%%%%%%%%%%%%%%
%%%%%%%%%%%%%%%%%%%%%%%%%%%%%%%%%%%%%%%%%%%%%%%%%%%%%%%%%%%%%%%%%%%%%%%

\begin{figure}[!htbp]
{\begin{center} 
\includegraphics[scale=1.04]{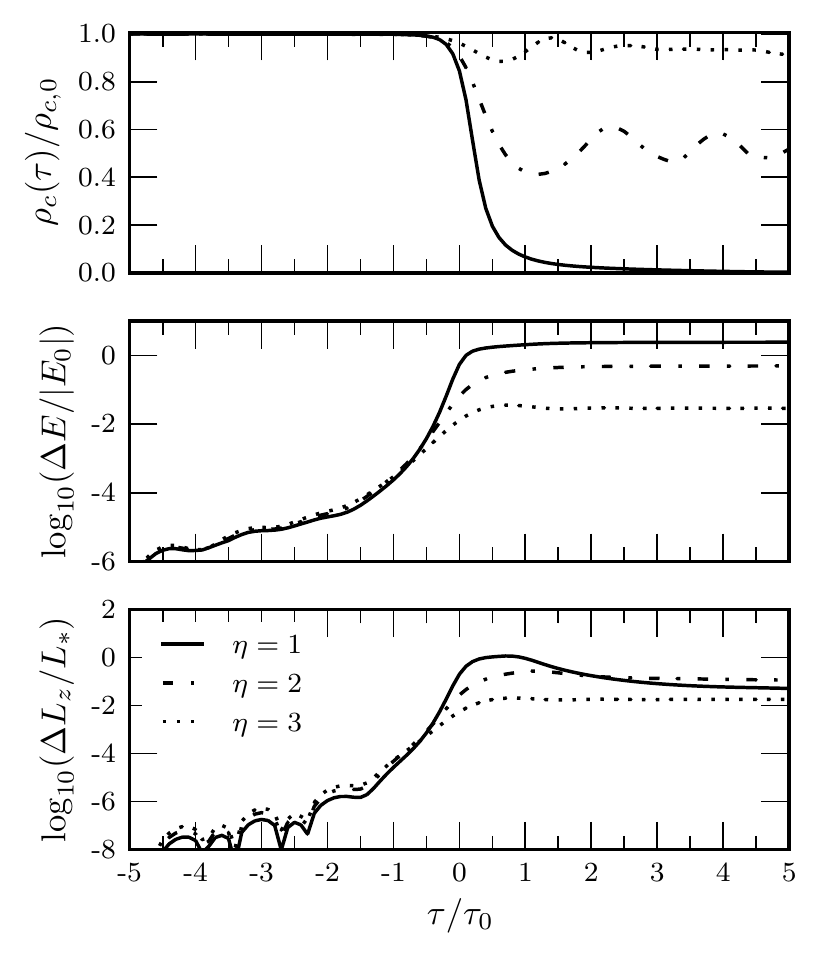} 
  \caption{
  \label{figure:disruptions} 
Partial and complete disruptions.  Same as previous figure except displaying 
results for $\eta =1$, $\eta =2$, and $\eta =3$ encounters.  The star 
disrupts completely when $\eta =1$ as seen by behavior of central density.  
Energy deposition continues to rise with the strength of the encounter, 
while angular momentum deposition saturates.  The mass ratio is 
$1.28 \times 10^{-3}$.  Results are drawn from highest resolution simulations.
} \end{center} 
}
\end{figure}

%%%%%%%%%%%%%%%%%%%%%%%%%%%%%%%%%%%%%%%%%%%%%%%%%%%%%%%%%%%%%%%%%%%%%%

\subsection{Shock heating and the energy excess}
\label{subsec:energy_excess_shock_heating}

The time dependence of the central density of the white dwarf provides 
clues on where the excess energy resides.  In Fig. 
\ref{figure:weak_encounters} and \ref{figure:disruptions} the top 
panels show $\rho_c (t)$ normalized by the initial central density of the 
star.  Figure \ref{figure:weak_encounters} shows the behavior for 
weak tidal encounters ($\eta=4$, $\eta=5$, and $\eta=6$) and Fig. 
\ref{figure:disruptions} displays stronger encounters ($\eta=1$, $\eta=2$, 
and $\eta=3$).  All models were computed with $\mu=1.28\times 10^{-3}$. 
Complete disruption is evident for $\eta = 1$.  For weaker encounters 
the central density typically decreases to a new lower average value and 
oscillates.  The average normalized central density, post-encounter, is 
(1) $\rho_c'/\rho_c \simeq 0.5$ for $\eta = 2$, 
(2) $\rho_c'/\rho_c \simeq 0.93$ for $\eta = 3$, and 
(3) $\rho_c'/\rho_c \simeq 0.995$ for $\eta = 4$.  

For a polytropic model, a reduction in the central density can arise either 
by reducing the mass of the star or by increasing $K = p/\rho^{1+1/n}$.  
As we have already seen, weak encounters involve some loss of mass.  
Furthermore, as our sequence of contour plots indicates, weakly disturbed 
stars are affected by formation of shocks in the outer layers.  The 
heating is not uniform but we can get an approximate sense of the effects 
by assuming it is.  For the nonce we assume that $K \rightarrow K' > K$ 
following the encounter.  For a $n = 3/2$ polytrope, the following scaling 
laws hold
\be
K \sim M_*^{2/3} \, \rho_c^{-1/3} , \quad
R_* \sim M_*^{1/3} \, \rho_c^{-1/3} , \quad
|E_{\text{tot}}| \sim M_*^{5/3} \, \rho_c^{1/3} .
\ee 
We treat the mass loss and change in central density as observables that 
indicate a new approximate polytropic state.  Using the scalings, we can 
estimate that the change in total energy of the star would follow
\be
\Delta E \simeq |E_{\text{tot}}| 
\left[1 -\left(\frac{M_*'}{M_*}\right)^{5/3} 
\left(\frac{\rho_c'}{\rho_c}\right)^{1/3} \right] .
\ee
Except for the case $\eta = 2$ where 10\% of the mass is lost, the resulting 
changes in energy are mostly the result of shock heating.  Using the values 
obtained from Figs. ~\ref{figure:outflow_mass}, \ref{figure:weak_encounters}, 
and \ref{figure:disruptions}, we estimate the effects of (assumed) uniform 
heating and list the results in the fifth column of 
Table \ref{table:energy_excess}.  The correspondence with the fractional 
excess energy gain in column four is suggestive that shock heating provides 
the sink for most of the excess.

\subsection{Radial mode excitation}
\label{subsec:radial_mode}

After weak tidal encounters with the black hole, the white dwarf is 
observed to oscillate not only in the quadrupole ($l =2$, $m = \pm 2$) $f$ 
modes but also in the fundamental radial $F$ mode.  This can best be seen 
by the oscillations in central density in 
Figs.~\ref{figure:weak_encounters} and \ref{figure:disruptions} for 
the $\eta = 3$ and $\eta = 4$ encounters.  In those two cases the 
observed oscillations match the period of the linear radial $F$ mode 
almost exactly.

The excitation of the radial mode was noted by Khoklov {\it et al}.
(1993)~\cite{khok1993,khok1993strong}.  Excitation of this mode by a 
tidal field is not possible at linear order.  Khoklov {\it et al}. attributed 
it to a nonlinearity in the post-encounter hydrodynamics.  They further 
suggested that it might be a locus of some of the excess energy gain.

To examine this idea, we simulated a set of dynamical stellar models that 
were deliberately set into radial oscillation.  In these tests no tidal 
field was included.  By varying the amplitude of the radial oscillation 
we sought to correlate the increase in energy in the star with the 
amplitude of oscillation in the central density.  The observed oscillations
in the central density that occur in our tidal encounters could then 
be used to estimate the amount of energy in the radial mode.

%%%%%%%%%%%%%%%%%%%%%%%%%%%%%%%%%%%%%%%%%%%%%%%%%%%%%%%%%%%%%%%%%%%%%%

\begin{figure}[!htbp]
{ \includegraphics[scale=1.07]{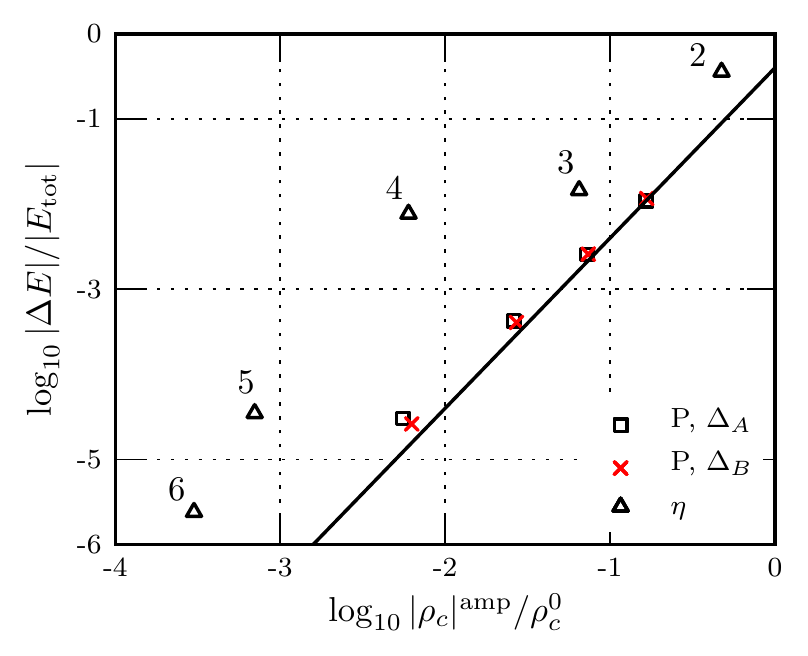}
\\ 
\caption
{ \label{figure:radial_pulsation_test} 
  Excitation of the fundamental radial mode.  We compare radial 
  pulsation models with those of tidal encounters for $\eta=2-6$ and 
  for mass ratio $\mu=1.28\times 10^{-3}$.  Squares and crosses indicate 
  results for radial oscillation simulations, measuring the added energy 
  of the star versus amplitude of the oscillation in the central density.  
  The line indicates a quadratic dependence.  Individual triangles are 
  points constructed 
  from tidal encounter simulations, plotting the excess deposited energy 
  versus observed amplitude of the central density oscillation.  For weak 
  encounters the observed energy excess is an order of magnitude or two 
  higher than can be explained by excitation of the fundamental radial 
  mode.  Energy in the radial mode can be 11\% to 25\% of the excess in 
  encounters with $\eta = 3$ and $\eta = 2$.} }
\end{figure}

%%%%%%%%%%%%%%%%%%%%%%%%%%%%%%%%%%%%%%%%%%%%%%%%%%%%%%%%%%%%%%%%%%%%%

We generate radially pulsating models by introducing a homologous scaling 
of the Lane-Emden density profile as an initial condition.  Consider a 
homologous mapping of the star that takes the original radius $R_*$ to 
$R_*'=R_*/\lambda$ via scaling parameter $\lambda$.  Assume that the 
equilibrium stellar profile is $\rho(r)$.  Map the original density 
profile to a new one using
\be
\label{radial_pulsation_density_profile}
\bar\rho(r) = \lambda^3 \rho(\lambda r) .
\ee
With this scaling the mass is unaffected by the transformation.  Then 
we assume that $K$ does not scale and calculate the altered pressure 
profile from the new density taking the polytropic index fixed also.  
The scaled density is used to calculate a new gravitational 
potential, which no longer provides hydrostatic equilibrium.  Technically, 
the initial radial displacement is linear, which would not match the shape 
of the fundamental radial mode amplitude.  Accordingly, we might expect a 
set of radial overtones to be excited.  Practically, though, most of the 
excitation is observed to be in the $F$ mode. 

We compute models with parameter range $\lambda=[0.9, 0.95, 0.98,
0.995, 1.005, 1.02]$.  We compare the change in total energy between
the radially pulsating models and the equilibrium model, 
$\Delta E^\lambda_{\rm tot} - \Delta E^{\lambda=0}_{\rm tot}$, with the 
observed amplitude of oscillation in the central density.  The resulting 
points form an approximately quadratic power-law relationship, as can be 
seen in Fig.~\ref{figure:radial_pulsation_test}.  

To compare these radially pulsating models to the tidal encounters, we
read off the amplitude of central density oscillations from
Figs.~\ref{figure:weak_encounters} and \ref{figure:disruptions} for
different $\eta$.  These oscillation amplitudes and the observed tidal
excess energy for the same $\eta$ are used to plot points in
Fig.~\ref{figure:radial_pulsation_test} also.  They are marked in the
figure to indicate their associated value of $\eta$.  In all tidal
encounter cases the excess energy is greater than can be explained by
energy in the excitation of the radial mode, though for $\eta = 2$ and
$\eta = 3$ the radial mode may be a non-negligible contributor at the
level of 25\% to 11\%, respectively.  For weaker encounters the radial
mode is as much as two orders of magnitude smaller.  Some numerical
values are compared in Table \ref{table:energy_excess}.

\subsection{Relativistic effects on energy deposition}

\begin{figure}[!htbp]
{
\includegraphics[scale=1.1]{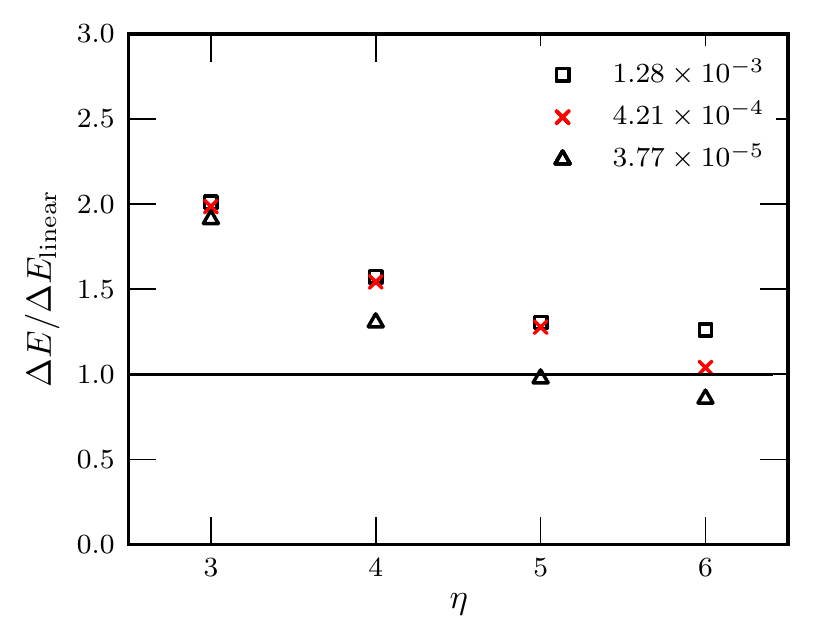} 
\caption { \label{figure:linear_theory_mu} 
Energy deposition normalized to expected values from linear theory.  
Three mass ratios were examined.  All of the simulations were made at 
highest resolution, $\Delta_C$.  Growth in energy transfer with increasing 
tidal strength is evident.  Also evident is the effect of relativistic 
motion and relativistic tidal corrections in the most extreme mass ratio 
case (triangles).  A slight suppression in energy transfer appears at 
$\eta = 3$, $\eta = 4$, and $\eta = 5$ for $\mu = 3.77 \times 10^{-5}$ 
that differs from the mutually consistent values seen at less extreme 
mass ratios.  
} }
\end{figure}

Another way to see the excess in energy deposited on the star is to 
plot its value normalized to the value expected from linear theory.  
Figure~\ref{figure:linear_theory_mu} shows the data for weak encounters 
in this fashion, but does so for all three mass ratios.  The plot shows 
clearly how the excess energy grows with increasing tidal encounter 
strength and yet approaches the linear result nicely for the weakest 
encounters.

%%%%%%%%%%%%%%%%%%%%%%%%%%%%%%%%%%%%%%%%%%%%%%%%%%%%%%%%%%%%%%%%%%%%%%
%%%%%%%%%%%%%%%%%%%%%%%%%%%%%%%%%%%%%%%%%%%%%%%%%%%%%%%%%%%%%%%%%%%%%

More importantly, this plot shows the presence of relativistic effects 
on the energy transfer.  All of these simulations were done at our 
highest resolution, $512^3$.  The only variables were $\eta$ and $\mu$.  
As we discussed previously, the $\eta = 6$ results are not numerically
converged but the results for $\eta = 3$ through $\eta = 5$ are.  
For the most extreme mass ratio (triangles in the plot), the star passes 
much closer to the black hole and the effects of relativistic motion and 
relativistic corrections in the tidal field will be more pronounced.  We 
observe a slight suppression in the energy transfer for $\eta = 3$ 
through $\eta = 5$.  Simulations run at several resolutions indicate 
the effect is real.  Notice the level of consistency in the energy 
transfer that occurs for the other two, less extreme mass ratios.
  
An explanation for the suppression likely can be found in a modification 
of the linear theory.  As Press and Teukolsky~\cite{pres1977} show, the 
overlap integral for the tidal interaction at each mode $nlm$ involves 
a product between two terms, $Q_{nl}$ (which depends upon the normal mode 
amplitudes) and $K_{nlm}$ (which depends upon the time dependence of the 
amplitude and phase of the $lm$ part of the tidal field).  In the FNC 
frame (with approximately Newtonian self-gravity and a weak tidal field), 
the integrals for $Q_{nl}$ are indifferent to whether the orbital motion 
is relativistic or not.  The same is not true of $K_{nlm}$ [Press and 
Teukolsky's Eq. (39)].  For example with $l = 2$, the tidal amplitude 
will not only vary as $r(t)^{-3}$ but will be corrected by an (orbital) 1PN 
term that behaves as $r(t)^{-5}$~\eqref{tidal_tensor_Cij}.  Furthermore, the 
relativistic shape of the orbit will be important and the motion of the 
black hole in the FNC frame requires that the azimuthal angle $\phi(t)$ 
in their equation be replaced by $\Psi(t)$.  In effect, $T_l(\eta)$ has 
to be replaced by a function $T_l(\eta,\Phi_p)$ of $\eta$ and a measure of 
the depth of the relativistic potential, $\Phi_p=M/R_p$.

\subsection{Capture orbits: effects of tides and gravitational bremsstrahlung}
\label{sec:bremss}

The energy transferred into the star by the tides comes at the expense of 
orbital energy.  If we assume the inbound white dwarf has zero total orbital 
energy ($\tilde{E} = 1$ in the relativistic sense), the orbital energy 
after the encounter becomes $E_{\text{orb}} = - \Delta E$ and can be 
used to estimate the semi-major axis $a = - M_* M/2 E_{\text{orb}}$ of the 
initial capture orbit.  For compact systems, and especially for higher 
mass black holes, the effect of gravitational wave bremsstrahlung should be 
included.  The orbital energy is then 
$E_{\text{orb}} = - \Delta E_{\text{tidal}} - \Delta E_{GW}$.

To calculate the gravitational bremsstrahlung, we approximate the orbital 
motion as Newtonian and first use the classic result \cite{pete1964} for 
the rate of energy loss averaged over one period of a bound orbit
\be
\left < \frac{dE}{dt} \right > 
= \frac{32}{5} \frac{G^4 (M + M_*) M^2 M_*^2}{c^5 a^5} f(e),
\ee
where the eccentricity $e$ determines
\be
f(e) = \left ( 1-e^2 \right )^{-7/2} 
\left ( 1 + \frac{73}{24} e^2 + \frac{37}{96} e^4 \right ).
\ee
Multiplying this luminosity by the orbital period 
$T = 2\pi a^{3/2}/[G (M + M_*)]^{1/2}$ yields the amount of energy radiated 
in one period
\be
\Delta E_{GW} = \frac{64\pi }{5} 
\frac{G^{7/2} (M + M_*)^{1/2} M^2 M_*^2}{c^5 } f(e)  a^{-7/2} .
\ee
To get the burst associated with a parabolic orbit, we replace $a$ in the 
above formula with pericentric distance $R_p = a (1-e)$ and then take the 
limit as $e \rightarrow 1$ 
\be
\Delta E_{GW} =  \frac{85\pi \sqrt{2} }{24}\frac{G^{7/2}}{c^5} 
\frac{(M + M_*)^{1/2} M^2 M_*^2 }{R_p^{7/2}} .
\ee

%%%%%%%%%%%%%%%%%%%%%%%%%%%%%%%%%%%%%%%%%%%%%%%%%%%%%%%%%%%%%%%%%%%
\begin{table}[h]
\begin{center}
\caption{
\label{table:GW bremsstrahlung}
Capture orbits from the combined effects of tidal energy transfer and
gravitational bremsstrahlung.  A range of $\eta$ for weak encounters in 
which the white dwarf survives is considered.  Three mass ratios $\mu$ are 
examined.  Tidal energy transfer and gravitational wave loss are separately 
listed.  Energies are compared to the scale of kinetic energy 
$T_p = M M_*/R_p$ at pericenter.  The importance of gravitational 
bremsstrahlung rises with increasing black hole mass.  The resulting 
capture orbits are given in terms of the ratio $R_{\text{max}}/R_p$ between 
apocentric and pericentric distances. 
}
\begin{tabular*}{8.6cm}%
     {@{\extracolsep{\fill}}cccccc}
\hline
\hline
\noalign{\smallskip}
$\mu$ & $\eta$ & $\frac{\Delta E_{\rm GW}}{T_p} $ 
& $\frac{\Delta E_{\text{tidal}}}{T_p}$ 
& $\frac{R_{\rm max}}{R_p}$\\
\noalign{\smallskip}
\hline
$1.28\times 10^{-3}$ & 2 & 5.30e-08 & 3.92e-03 & 2.54e+02 \\
-                   & 3 & 2.70e-08 & 3.04e-04 & 3.29e+03 \\
-                   & 4 & 1.67e-08 & 2.69e-05 & 3.72e+04 \\
-                   & 5 & 1.15e-08 & 2.21e-06 & 4.50e+05 \\
\hline
$4.21\times 10^{-4}$ & 2 & 1.11e-07 & 1.90e-03 & 5.25e+02\\
-                   & 3 & 5.65e-08 & 1.43e-04 & 6.99e+03\\
-                   & 4 & 3.50e-08 & 1.26e-05 & 7.94e+04\\
-                   & 5 & 2.41e-08 & 1.03e-06 & 9.48e+05\\
\hline
$3.77\times 10^{-5}$ & 2 & 5.53e-07 & 4.59e-04 & 2.18e+03 \\
-                   & 3 & 2.83e-07 & 2.75e-05 & 3.60e+04 \\
-                   & 4 & 1.75e-07 & 2.12e-06 & 4.35e+05 \\
-                   & 5 & 1.20e-07 & 1.58e-07 & 3.60e+06 \\
\hline
\hline
\end{tabular*}
\end{center}
%\vspace{-0.6cm}
\end{table}
%%%%%%%%%%%%%%%%%%%%%%%%%%%%%%%%%%%%%%%%%%%%%%%%%%%%%%%%%%%%%%%%%%%

We list results for tidal capture of white dwarfs in Table 
\ref{table:GW bremsstrahlung}.  There we tally the amount of tidal 
energy transfer observed in the simulations for different mass ratios 
$\mu$ and encounter strengths $\eta$ (for those cases in which the star 
survives).  Also given is the amount of gravitational wave energy loss 
during the encounter.  (There is also gravitational wave emission from 
the internal hydrodynamic motions of the star, but these can be shown to 
be orders of magnitude smaller.)  Energies are given relative to the 
scale of kinetic energy of the star at pericenter, which is easily 
obtained from Table~\ref{table:encounter_parameters}.  The effect of 
the tides dominates but gravitational bremsstrahlung increases in 
importance for more massive black holes.  Given the total energy loss 
from the orbit, the capture orbit can be described by the ratio 
$R_{\text{max}}/R_p = (1+e)/(1-e)$ between the apocentric and pericentric 
distances.

\section{Angular momentum }
\label{sec:tidal-angular}

A tidal encounter transfers not only energy but also angular momentum.  
Since the fluid equations are nearly Newtonian in the FNC frame, the 
standard analysis of self-gravitating fluids~\cite{cart1985} provides 
an approximate physical picture.  Let the spin tensor be 
\be
J_{ik} = \tfrac{1}{2} \int \rho (x_i v_k - v_i x_k) d^3x ,
\ee
and the first moment of the tidal force tensor be
\be
F_{ik} = -\int \rho x_i \partial_k \Phi_{\text{tidal}} d^3x 
\simeq - I_{ij} C_{jk}^{(0)} .
\ee
The antisymmetric part of the tensor virial theorem expresses conservation
of angular momentum and we find that the tidal field exerts a torque given
by
\be
\dot{J}_{ik} = - \tfrac{1}{2} 
\left( I_{ij} C_{jk}^{(0)} - I_{kj} C_{ji}^{(0)} \right) 
+ \cdots .
\ee
A torque results whenever the bulge drawn up dynamically in $I_{ij}$ lags 
(or leads) the principal axis of the tidal field.  Numerically we compute 
the action of the full tidal field, including higher order moments and 
(orbital) relativistic corrections.  But the most important contributor to 
the torque remains the Newtonian part of the quadrupole tide.  

In our models, the black hole appears to move (in the FNC frame) in the 
$x$-$y$ plane, which induces changes in the $z$ component of the white 
dwarf's spin angular momentum, $L_z$.  The bottom panels of 
Figs.~\ref{figure:weak_encounters} and \ref{figure:disruptions} show the 
growth in $L_z$ (normalized to an estimate of breakup angular momentum) 
during tidal encounters of varying strength.  The total amount of angular 
momentum deposited in the star varies over four orders of magnitude in 
models that range from $\eta = 6$ to $\eta = 1$.  In several cases the 
spin overshoots before settling back~\cite{nolt1982}, an effect of the 
black hole racing out more than 90 degrees ahead of the principal axis of 
the star.

\subsection{Center of mass deflection}
\label{sec:tidal-cm}

The deposition of angular momentum has been seen in many past numerical 
studies.  What is new in this paper is calculating the effects of the 
octupole tide, $C_{ijk}$.  We can again get a physical picture by considering
Newtonian behavior.  Let the first moment of the mass distribution be 
$D_k = \int \rho \, x_k \, d^3 x$.  Then take the momentum equation 
\eqref{eqn:euler}, restrict it to Newtonian terms, and construct an equation
of motion for $D_k$,
\be
\ddot{D}_k = - C_{ki}^{(0)} D_i - \tfrac{1}{2} C_{kij}^{(0)} I_{ij} 
- \tfrac{1}{6} C_{kijl}^{(0)} I_{ijl} + \cdots.
\ee
If the octupole and other higher-order moments vanish, and if the 
star is initially centered on the frame $D_k = 0$, then the CM
has no tendency to move.  If however the octupole tide is present, it 
couples to the second mass moment and drives an acceleration of the CM.  
Once the CM shifts the quadrupole tide plays a role 
also.  

%%%%%%%%%%%%%%%%%%%%%%%%%%%%%%%%%%%%%%%%%%%%%%%%%%%%%%%%%%%%%%%%%%%%%%
\begin{figure}[!ht]
{ \begin{center}
      \includegraphics[scale=1.0]{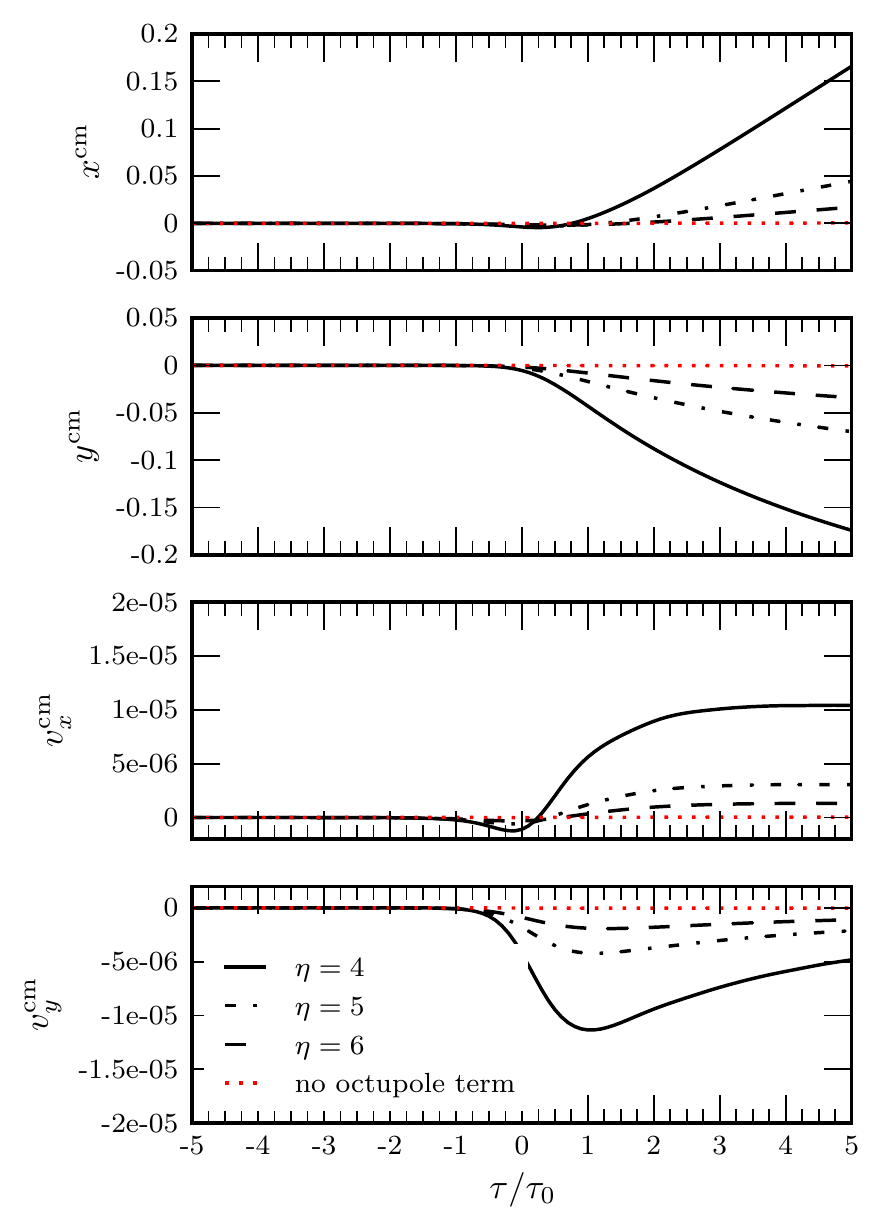}
  \caption
      {  \label{figure:center_of_mass} 
Deflection of the center of mass.  Positions (top two panels) and 
velocities (bottom two panels) of the CM in simulations with 
$\eta=4,5,6$ and mass ratio $\mu = 1.28\times 10^{-3}$.  A simulation with 
no octupole tidal term is shown (dotted red line) for comparison.  The 
resolution in each case was $512^3$.
} \end{center} }
\end{figure}
%%%%%%%%%%%%%%%%%%%%%%%%%%%%%%%%%%%%%%%%%%%%%%%%%%%%%%%%%%%%%%%%%%%%%%

Since our models include the octupole tide, we see its effect on CM 
motion.  Some of these effects are apparent in the contour 
plots shown earlier in Figs. \ref{figure:eta3_density_contour},
\ref{figure:eta1_density_contour}, and \ref{figure:density_contour_eta3}, 
primarily in the asymmetry of the tidal lobes.  More quantitatively, 
we compared simulations that included only the quadrupole ($l=2$) tide
with ones that included both the quadrupole ($l=2$) and octupole ($l=3$) 
terms.  In Fig. \ref{figure:center_of_mass}, the octupole tide can be 
seen to cause a deflection of the CM of the star away from 
the FNC origin (here we show actual shifts in position, not components 
of $D_k$).  No deflection is seen in the quadrupole-only model.  The size 
of the deflection can also be seen to depend upon the strength of the 
encounter.  These small positional changes and velocities are well 
determined at our highest resolution.

\subsection{Orbital angular momentum change}

Surprisingly (from a numerical standpoint) the CM position and 
velocity are determined well enough that we can compute from them the change 
in orbital angular momentum.  To our knowledge this has not been tried or 
seen in previous simulations of tidal encounters.  

To provide a simple physical picture again, consider a Newtonian isolated 
fluid body of mass $M_*$ moving about a heavy (stationary) mass.  Let 
coordinates for the fixed frame be $X_k$ and $t$ and let the 
motion of a second frame following the object be described by 
$X_k^{(0)}(t)$ and $V_k^{(0)}(t)$.  The (total) angular 
momentum tensor seen in the fixed frame is
\be
J_{ij}^{\prime} = \tfrac{1}{2} \int \rho (X_i V_j - V_i X_j) d^3X ,
\ee
and for motion about a stationary mass $M \gg M_*$ (conservative central 
force), $dJ_{ij}^{\prime}/dt = 0$.  Consider coordinates more suited to 
following internal positions and velocities
\be
x_k = X_k - X^{(0)}_k (t), \qquad  v_k = V_k - V_k^{(0)}(t).
\ee
We can then use the moving frame to decompose the angular momentum into 
orbital and spin (internal) parts
\be
J_{ij}^{\prime} = L_{ij}^{(0)} + \Delta L_{ij} + J_{ij} ,
\ee
where
\be
\label{equation:orbitalpart}
L_{ij}^{(0)} = 
\tfrac{1}{2} M_* \left(X_i^{(0)} V_j^{(0)} - V_i^{(0)} X_j^{(0)} \right) ,
\ee
and
\be
\label{eqn:deltalnewt}
\Delta L_{ij} =
  \tfrac{1}{2} \left( X_i^{(0)} \dot{D}_j - \dot{D}_i X_j^{(0)} \right) 
+ \tfrac{1}{2} \left( D_i V_j^{(0)} - V_i^{(0)} D_j \right) .
\ee
If the moving frame follows the CM, then $D_k = \dot{D}_k = 0$ 
and $\Delta L_{ij} = 0$.  In that case, changes in the spin angular 
momentum will be compensated by changes in the orbital angular momentum, 
$dJ_{ij}/dt = -dL_{ij}^{(0)}/dt $.  If instead the moving frame is set to 
follow the orbit of a test mass (similar to FNCs), then $L_{ij}^{(0)}$ is
conserved and the changes in (internal) $J_{ij}$ are compensated by 
$\Delta L_{ij}$,
\be
\frac{dJ_{ij}}{dt} = -\frac{d\Delta L_{ij}}{dt} .
\ee

\begin{figure}[!htbp]
{ 
\includegraphics[scale=1.04]
{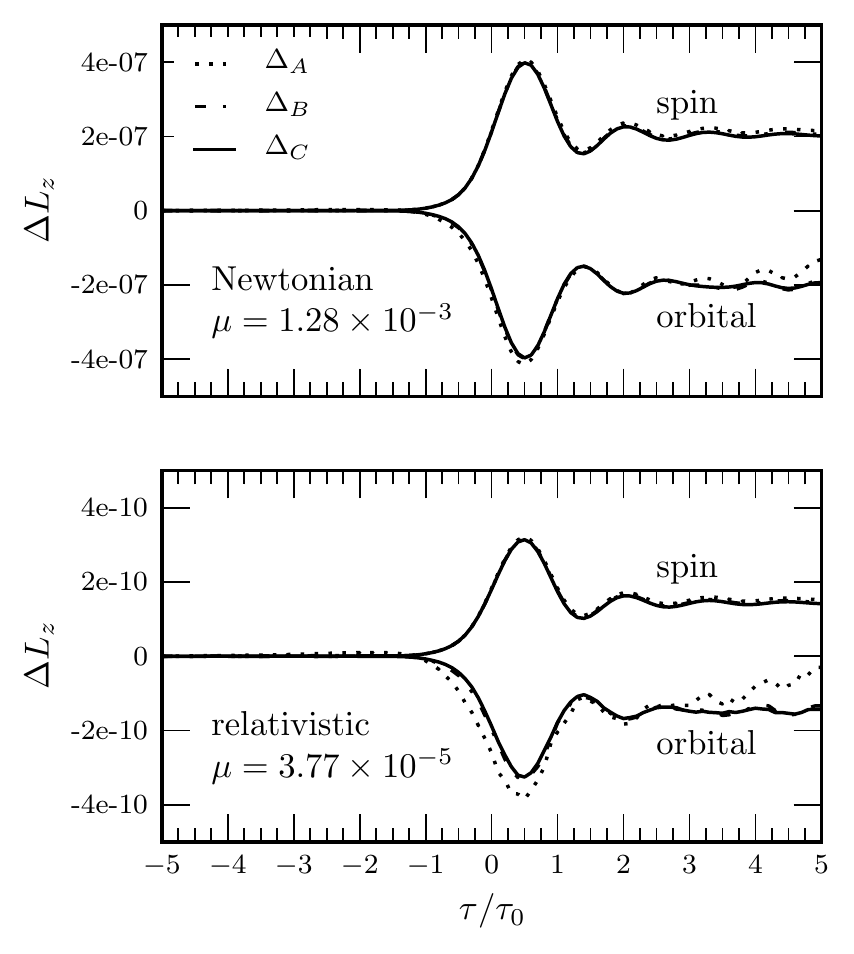}
\caption
      { \label{figure:total_angular_momentum} 
Total angular momentum conservation in Newtonian and relativistic
simulations.  The top panel shows a strictly Newtonian model while the 
bottom panel shows a full FNC model with our most relativistic orbit.  
We use an encounter strength $\eta=4$ and a mass ratio for the
Newtonian case of $\mu = 1.28 \times 10^{-3}$ and 
$\mu=3.77\times 10^{-5}$ for the relativistic case.  Three resolutions 
($\Delta_A, \Delta_B, \Delta_C$) are shown.  
 }}
\end{figure}

To test our ability to track these complementary effects in our numerical 
models, we set up a strictly Newtonian version of our code (Newtonian tidal
field and orbit) and simulated a tidal encounter with mass ratio 
$\mu = 1.28 \times 10^{-3}$.  The upper curve in the top panel of 
Fig. \ref{figure:total_angular_momentum} shows the gain in the $z$ 
component of angular momentum $L_z^{\text{spin}}$ of the star in an 
$\eta=4$ encounter.  The lower curve shows the independently determined 
history of $\Delta L_z^{\text{orbital}}$ and its remarkable 
(numerical) balance with the increase in spin.  The balance is only possible 
because we have included both the quadrupole and octupole tides.  Three 
different resolutions are shown to provide a sense of the convergence.  

%%%%%%%%%%%%%%%%%%%%%%%%%%%%%%%%%%%%%%%%%%%%%%%%%%%%%%%%%%%%%%%%%%%%%%
%%%%%%%%%%%%%%%%%%%%%%%%%%%%%%%%%%%%%%%%%%%%%%%%%%%%%%%%%%%%%%%%%%%%%%

\subsection{Relativistic angular momentum transfer}

There are several obstacles to duplicating the above result in our full 
simulations (i.e., in general relativity).  These include (1) defining 
angular momentum rigorously (i.e., in an asymptotically flat spacetime), 
(2) the difficulty in defining a split between spin and orbital angular 
momenta, and (3) the difficulties in transferring angular momentum between 
coordinate systems ~\cite{thor1985,misn1973}.

The encounter between a white dwarf and IMBH can be regarded as isolated 
in an asymptotically flat spacetime.  So total angular momentum can be 
defined (asymptotically) and is conserved.  Furthermore, consistent with 
the approximations we have used, the IMBH is sufficiently massive to be 
regarded as stationary (Schwarzschild).  The spacetime thus has Killing 
vectors and in particular has a rotational Killing vector 
$\xi_{(\phi)}^{\mu} = (\partial/\partial \phi)^{\mu}$ about the axis normal 
to the orbital plane.  We may then form the conserved four-current 
$\mathscr{J}^\mu = T^{\mu\nu} \xi_{(\phi)\nu}$ and from it derive a 
conserved total angular momentum of the star (and debris)
\be
\label{total_angular_momentum_conservation_Schwarzschild}
J_z = \int \mathscr{J}^0 \sqrt{-g}\, d^3 X 
= \int d^3 X \sqrt{-g}\, \rho\, h\, r^2\sin^2\theta\, U^{\phi}\, U^{0} ,
\ee
where $h=1 + \Pi + p/\rho$ is the specific enthalpy and the integral 
is over a volume that encompasses all of the material.  

To approximately split orbital and spin angular momenta, we view the fluid 
as confined to a small volume (FNC domain).  The center of the frame moves 
on $X^{\mu} = X_{(0)}^{\mu}(\tau)$, which has constants of motion 
$\tilde{E}$ and $\tilde{L}$.  To the extent 
that the gravitational mass of the star does not change and can be 
approximated by $M_*$ (order $\varepsilon^2$ errors), the initial orbital 
angular momentum will be 
$L_z^{(0)} = M_* r_{(0)}(t)^2 U_{(0)}^{\phi}(t) = M_* \tilde{L}$, which
coincides with $J_z$.  Thus $L_z^{(0)}$ is the analogue 
of \eqref{equation:orbitalpart}.  

As the star passes through pericenter internal motions develop and spin 
is deposited in the fluid (which we compute in the FNC frame).  The CM is 
also deflected, which affects the orbital angular momentum.  If the 
velocity and position of the CM are transferred from the FNC frame 
to Schwarzschild coordinates, they can be used to form corrections 
$\delta X^{\mu}$ and $\delta U^{\mu}$ relative to the geodesic 
$X_{(0)}^{\mu}(\tau)$.  We can then compute the change in the orbital 
angular momentum using
\be
\label{eqn:deltal}
\delta L_z^{\text{orbital}} = M_* \left(r_{(0)}^2 \delta U^{\phi} 
+ 2 r_{(0)} U_{(0)}^{\phi} \delta r \right) .
\ee

An event $(\tau,x^i)$ in FNCs will have a Schwarzschild coordinate 
location $X^{\mu}$.  By assumption, $X^{\mu}$ is close to $X_{(0)}^{\mu}$ 
and we can form an expansion
\be
\delta X^{\mu} = X^{\mu} - X^{\mu}_{(0)}(\tau) 
= x^i \left (\frac{\partial X^{\mu}}{\partial x^i} \right )_{(0)} + \cdots .
\ee
Then, recognizing the tetrad frame components,
\be
\label{eqn:deltax}
\delta X^{\mu} = x^i \lambda_i^{\mu} + \cdots .
\ee
In like fashion, the velocity $u^a$ at $(\tau,x^i)$ can be transformed to
components $U^{\mu}$ at the event $X^{\mu}$ by
\be
\label{eqn:utransform}
U^{\mu}(X^{\nu}) = 
\frac{\partial X^{\mu}}{\partial \tau}(\tau, x^k) \, u^0 +
\frac{\partial X^{\mu}}{\partial x^i}(\tau, x^k) \, u^i .
\ee
The transformation matrix is then expanded about FNC frame center, 
yielding
\ba
U^{\mu}(X^{\nu}) = 
\frac{\partial X^{\mu}}{\partial \tau}(\tau, 0) \, u^0 
&+&\frac{\partial X^{\mu}}{\partial \tau\, \partial x^i}(\tau, 0)\, u^0\, x^i 
\\
\nn
&+&
\frac{\partial X^{\mu}}{\partial x^i}(\tau, 0) \, u^i + \cdots .
\ea
We reduce this expression by recognizing first that 
$U_{(0)}^{\mu} = \partial X^{\mu}/\partial \tau(\tau, 0)$.  Secondly, in
the FNC frame $u^0 \simeq 1 + \mathcal{O}(\varepsilon^2)$ and 
$u^i = v^i + \mathcal{O}(\varepsilon^3)$.  Finally, 
$\lambda_i^{\ \mu} = \partial X^{\mu}/\partial x^i$, which allows us to
write 
\be
\label{eqn:deltau}
\delta U^{\mu} = U^{\mu} - U^{\mu}_{(0)} 
= x^i \frac{d \lambda_i^{\mu}}{d\tau}  
+ v^i \lambda_i^{\mu} + \cdots .
\ee
Then, we set $x^i = D_i/M_*$ and use \eqref{eqn:deltax} to calculate 
$\delta r$.  Next we set $v^i = \dot{D}_i/M_*$ and use \eqref{eqn:deltau} 
to find $\delta U^{\phi}$.  These are then both employed in 
\eqref{eqn:deltal} to obtain the shift in the relativistic orbital 
angular momentum.

The bottom panel in Figure \ref{figure:total_angular_momentum} shows results 
from an $\eta = 4$ encounter with our most relativistic orbit 
($3.77\times 10^{-5}$).  The top curve is the angular momentum deposited 
into the star.  The bottom curve is from our calculation of \eqref{eqn:deltal} 
for the change in the orbital angular momentum.  Three resolutions are 
shown and the result is well converged numerically.  It is worth noting 
that the remarkable balance between the two in this relativistic case 
depends heavily on the transformations shown above.  A straightforward 
application of the Newtonian expression \eqref{eqn:deltalnewt} fails 
to provide an accurate measure of the compensating change.

\subsection{Energy versus angular momentum and comparison with the 
affine model}

\begin{figure}[!htbp]
{
\includegraphics[scale=1.09]{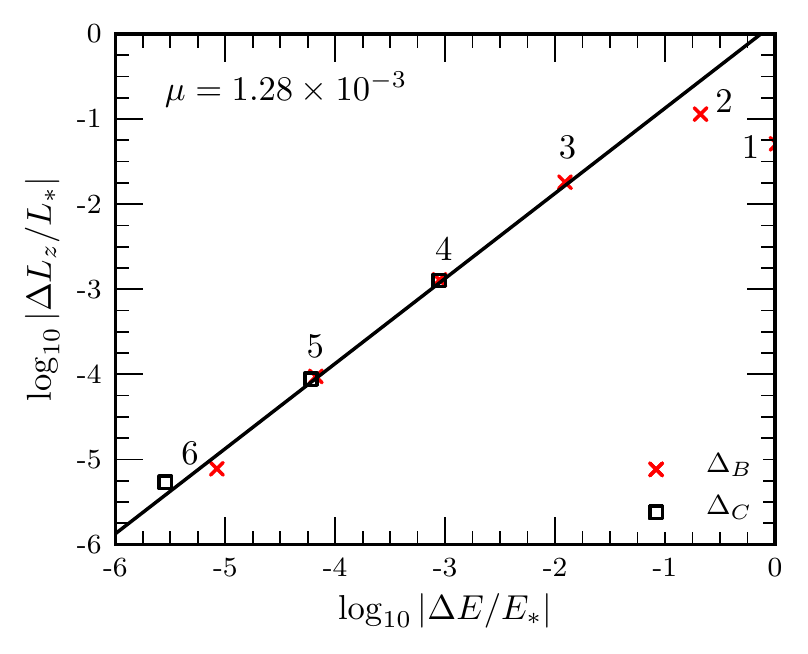} 
\caption { \label{figure:linear_theory_kochanek} 
Tidally transferred angular momentum $\Delta L_z$ versus energy $\Delta E$. 
Angular momentum and energy normalized by $L_* = \sqrt{G M_*^3 R_*}$ and 
$E_* = G M_*^2/R_*$, respectively.  Solid line is from analysis of 
ellipsoidal models \cite{koch1992}.  Simulations used mass ratio 
$\mu=1.28\times 10^{-3}$.  
} }
\end{figure}

Having obtained both the energy and the angular momentum that are 
transferred in a tidal encounter, we compare the two in Fig.
\ref{figure:linear_theory_kochanek}.  We confirm the previously known 
linear relationship \cite{koch1992,nolt1982}, which holds over a broad 
range of encounter strengths.  Kochanek has analyzed \cite{koch1992} the 
relationship within the context of ellipsoidal models and tested it through 
use of an affine code \cite{cart1985,lumi1986}.  His analysis found the 
proportionality to be
\be 
\Delta E_{\rm tot} 
= \frac{|E_g|}{\sqrt{15}} \frac{\Delta L_z}{\sqrt{I_* |E_g|}} , 
\ee 
where $E_g = 2 E_{\text{tot}}$ is the gravitational potential energy of the 
star.  In Fig. \ref{figure:linear_theory_kochanek} we plot this 
ellipsoidal model relationship (solid line) for comparison.  It fails to 
fit full simulation results at the upper end (especially for full disruption 
at $\eta = 1$) where linear analysis ceases to be a good approximation.
Each of our simulations used mass ratio $\mu=1.28\times 10^{-3}$ and results
are shown for higher resolutions $\Delta_C$ and $\Delta_B$.

%%%%%%%%%%%%%%%%%%%%%%%%%%%%%%%%%%%%%%%%%%%%%%%%%%%%%%%%%%%%%%%%%%%%%%

\section{Conclusions}
\label{sec:conclusions}

In this paper we investigated the tidal interactions between a white dwarf
and an intermediate mass black hole.  We used a Fermi normal coordinate 
system that provides a local moving frame roughly centered on the star.  
The FNC approach yields an expansion of the black hole tidal field that 
contains quadrupole and higher multipole moments and orbital relativistic 
effects.  It also allows a simpler, nearly-Newtonian treatment of the 
star's hydrodynamic motions and self-gravity, at least in a sufficiently 
confined FNC domain.  We detailed which terms in the tidal field expansion
are consistent with this approximation to the hydrodynamics and 
self-gravity.  A new numerical code was constructed based around this 
formalism.  It utilizes the well-developed PPMLR hydrodynamics method and 
a three-dimensional spectral method approach to solve for the self-gravitational potential.

We simulated a set of tidal models, both weak encounters and those at the 
threshold of disruption.  At the outset, simulations were computed of our 
stellar equilibrium models, without a superposed tidal field, to demonstrate
how ``quiet'' the initial models are and to establish their value as a 
control.  We then examined the overall hydrodynamic features in tidal 
encounters and computed the mass loss from a white dwarf as a result of 
different weak encounters.  For the range of black hole masses we considered 
in this paper, we found the $l = 4$ part of the tidal field to have 
negligible impact.  The same would not be true had we modeled white dwarf 
encounters with $10 M_{\odot}$ to $50 M_{\odot}$ black holes.  The tidal 
field expansion also includes the gravitomagnetic term.  Its effects are 
subtle and we intend to address those in a subsequent paper. 

Besides computing the mass loss, one of the principal focuses of this paper 
was on transfer of energy from the orbit into the white dwarf and total 
energy losses from the orbit.  We computed accurately the deposition of 
tidal energy onto the star.  We then investigated where that energy resides.
After comparing to the results of linear theory, we found that a combination 
of excitation of nonradial and radial modes and surface layer heating 
accounts for the energy transfer to the star.  Stars that survive a tidal 
encounter (1) are oscillating violently in the fundamental (rotating) 
quadrupole mode, (2) suffer some mass loss, (3) are shock heated in their 
outer layers, (4) see an average reduction in their central density and an 
increase in their radius, and (5) develop nonlinearly an oscillation in their 
fundamental radial mode.  We also identified a slight relativistic 
suppression of tidal energy transfer in the encounters with the most massive 
black hole we considered.  All of these effects seen in our numerical 
models were shown to be accurately determined by considering several 
different finite difference resolutions.  Several of these effects would 
make disruption more likely upon a second passage.  With energy transfer 
to the star computed, we separately calculated the amount of energy loss 
from the orbit due to gravitational bremsstrahlung.  We combined these 
losses to estimate the range of tidal capture orbits that result following 
weak encounters.

Lastly, we turned attention to transfer of angular momentum from the 
orbit into spin of the white dwarf.  We computed the tidal torquing of 
the star and debris, and confirmed with our full numerical models the 
previously predicted linear relationship between transferred energy and 
deposited angular momentum.  Then we demonstrated the result of including the 
octupole part of the tide in driving a deflection of the center of mass 
of the star (and debris), which to our knowledge is the first instance of 
this effect being computed in finite difference numerical models.  
Furthermore, we were able to take the observed CM deflection and compute 
directly from it the change in orbital angular momentum.  The increase in 
spin angular momentum in the star is seen to balance nicely with decrease 
in orbital angular momentum.  While expected physically, this result is 
only possible to see once enough terms are included in the tidal field.  
Furthermore, it is necessary to include an approximate relativistic 
calculation to transform effects seen in the FNC frame into changes in 
orbital angular momentum seen in the black hole frame.

\acknowledgments

R.M.C.~acknowledges support from a U.S. Department of Education GAANN
fellowship, under Grant No. P200A090135, a NC Space Grant Graduate 
Research Assistantship, and a Dissertation Completion Fellowship from the 
UNC--Chapel Hill Graduate School.  C.R.E. acknowledges support from the
Bahnson Fund at the University of North Carolina--Chapel Hill.

\appendix

\section{Numerical Method}
\label{sec:method}

Our numerical method for calculating tidal interactions between a
massive black hole and a star consists of three parts: a module that
computes the motion of the FNC frame and the tidal acceleration terms,
a hydrodynamics solver, and a self-gravity module.  Overall, the method
combines a three-dimensional finite difference approach for the
hydrodynamics and a spectral co-location technique for the self-gravity.

\subsection{Motion of the FNC frame and tidal accelerations}

The FNC frame center follows a timelike geodesic of the Schwarzschild
background spacetime.  We integrate the Darwin form of the geodesic
Eqs., (\ref{eqn:darwin1})--(\ref{eqn:darwin3}), using a Runge-Kutta 
routine for some set of orbital parameters (i.e., $e$ and $p$) and some 
initial azimuthal position $\phi$ or radial distance $R_i$.  The position of 
the frame center in Schwarzschild coordinates is obtained as functions of 
the radial phase $\chi$.  We can invert $\tau = \tau(\chi)$ and take the 
proper time $\tau$ instead as the curve parameter.  Proper time along 
the geodesic becomes the time coordinate within the entire FNC frame.

Most of the effort in computing the tidal accelerations is accomplished
via the derivations in Sec. \ref{sec:formalism}.  We must integrate 
Eq. (\ref{eqn:FNC_frame_rotation}) for the frame precession angle 
$\Psi$ given some initially chosen orientation.  The components of the 
Riemann tensor in the black hole frame are computed at the instantaneous 
position of the frame center and projected into the FNC frame using the 
time-dependent components of the FNC frame vectors.  From this the various 
tidal tensors are computed and finally the tidal potential 
(\ref{eqn:tidalpotential}) and gravitomagnetic potential 
(\ref{eqn:gmpotential}) are computed.

\subsection{PPMLR hydrodynamics algorithm}

We use a version of the explicit, time-dependent hydrodynamics code VH-1
\cite{blondin} to solve the equations for inviscid flow of an ideal
compressible gas with fixed adiabatic index $\gamma$ and with gravitational
acceleration terms.  The code is based on the piecewise parabolic method
(PPM) of Colella and Woodward \cite{cole1984} and uses the Lagrangian-remap
formulation of the method.  It is an extension of Godunov's method that
offers high-order accuracy in smooth regions of the flow (third order in
space and second order in time) while sharply capturing discontinuities.
Our version of VH-1 was recast in C and ported to parallel machines running
under MPI.  The coordinate topology in the hydrodynamics code is taken to be
Cartesian.  We use a zero-gradient outflow boundary condition.  Mass,
energy, and momentum are allowed to flux out of the domain provided the
local, instantaneous normal component of velocity is outward directed.
Otherwise, the fluxes are set to zero.

This particular hydrodynamics technique is very standard and verification of
the code follows a well-known procedure.  We ran the code against a battery
of standard test problems, including but not limited to (1) the Sod shock
tube \cite{sod1978}, (2) twin, colliding blast waves, (3) Mach 3 wind tunnel
with step, and (4) double Mach reflection of a strong shock (see Woodward
and Colella \cite{wood1984}).  The results \cite{chen2012} were
indistinguishable from those published previously.

The hydrodynamics code, as well as the other major elements, use domain
decomposition to facilitate parallel computing.  To make the algorithm
simple, we have in fact chosen to divide the three-dimensional domain into
slabs that are one zone deep, and farm each thin slab to an individual
processor.  Each processor executes the directionally split part of the
algorithm along the two directions of available data, and then partially
updated data is gathered to transpose the domain decomposition along another
direction.  A run with $N^3$ total spatial grid points will make use of 
$N$ processors.  We have run a few computations on a $1024^3$ grid with 
$1024$ processors.  Most of our highest resolution runs have $N=512$.

Several other tests of the hydrodynamics, when combined with the self-gravity
routine, were made and these are discussed in 
Sec.~\ref{subsec:hydro_parameters}.

\subsection{Pseudo-spectral self-gravity solver}

Our computation of the self-gravitational field follows closely a method 
developed by Broderick and Rathore \citep{brod2006}.  At our level of 
approximation, the gravitational field as a whole is determined by a 
gravitomagnetic potential $A_k$, a scalar tidal potential $\Phi_{\rm tidal}$, 
and the self-gravitational potential $\Phi$.  The field $\Phi$ satisfies
Poisson's equation
\be 
\label{eqn:poisson}
\nabla^2\Phi(\vec{x}) = 4\pi \rho(\vec{x}) , 
\ee 
where $\rho$ is the Newtonian mass density.  In principle, 
$\Phi + \Phi_{\rm tidal}$ satisfies (\ref{eqn:poisson}) subject to imposition 
of appropriate boundary conditions.  However, we separately compute the 
tidal potential and solve the Poisson equation only for the 
self-gravitational field $\Phi$ subject to the condition $\Phi \rightarrow 0$ 
as $r\rightarrow \infty$.

We solve (\ref{eqn:poisson}) with a discrete sine transform (DST) in three 
dimensions.  This spectral approach serves to rapidly invert the matrix 
resulting from finite differencing the elliptic equation.  Unfortunately, 
the way our boundary conditions are handled (see below) does not allow 
exponential convergence, one of the other primary benefits of spectral 
methods.  Instead our solutions of (\ref{eqn:poisson}) are second-order 
(algebraically) convergent, consistent with the other parts of the code.  

While a Fourier transform is natural on a Cartesian mesh, matching to an 
asymptotically vanishing boundary condition on $\Phi$ requires some effort.
Use of the DST implies that the field vanishes everywhere on the boundary 
of our rectangular domain.  To circumvent this, the method described by 
Broderick and Rathore \cite{brod2006} (see also \cite{pres1996}) crafts a 
one-zone-thick distribution of mass $\rho^B$ (image mass) in the outermost 
zone along each boundary face of the domain.  With the right distribution of 
image mass the solution for $\Phi$ will approach the boundary with a 
fall-off that is consistent with (extrapolated) vanishing at infinity and 
with correct multipole content.  The problem to be solved with the DST is then
\be
\nabla^2 \Phi = 4\pi \left (\rho + \rho^B \right) = 4 \pi \rho_{\rm total}, 
\ee 
once $\rho^B(\vec{x})$ is specified.  We provide details in what follows, 
especially how our cell-centering makes for slight differences with Broderick
and Rathore.

The grid consists of cell-centered data, so that the outermost points in 
any direction are a half zone away from the physical faces (boundaries) of 
the domain.  For example, let $I$ be the number of zones in the 
$x$ direction.  Let $\Delta x$ be the zone increment and $L_x = I \Delta x$ 
be the width of the domain.  Then $x_i = (i+1/2)\Delta x$ denote the locations
of the field values in the $x$ direction, with equivalent discrete locations 
$y_j$ and $z_k$ in the other directions.  Note that while the FNC 
system will have its origin at the center of the domain, in order to apply 
the DST we make a shift temporarily so that the origin in the DST calculation 
is placed at a corner of the domain.  The field and the source are assumed 
to be odd symmetric across any face.  Then, given the half zone 
cell-centering, the forward transform is a DST of type II \cite{fftw}, 
\be
\hat{f}_{l} = \sum_{i=0}^{I-1} f(x_{i})
\sin\left[\frac{\pi}{I} (i+1/2)(l+1)\right] ,
\ee
and the inverse transform is a DST of type III,
\be
f(x_{i}) = \frac{2}{I} \sum_{l=0}^{I-2} \hat{f}_{l} 
\sin\left[\frac{\pi}{I} (i+1/2)(l+1)\right] 
+ \frac{(-1)^i}{I} \hat{f}_{I-1} .
\ee

We can easily generalize from one dimension to three, but do so in a way 
that enables parallel computing.  Let $I,J,K$ be the number of zones in the 
$x$, $y$, and $z$ directions.  Let $i,j,k$ distinguish the spatial locations 
like before but now with $y_j = (j+1/2)\Delta y$ and $z_k = (k+1/2) \Delta z$.  
The corresponding discrete points in the transform space are indexed by 
$l,m,n$.  The discrete three-dimensional sine transform of, say, $\rho$ can 
be accomplished in three steps, each of which is a DST-II and can be computed 
in parallel given a slab decomposition of the domain: 
\ba\label{sine_transform_partial_sum}
u_{ljk}
\nn & = & \sum_{i=0}^{I-1} \rho_{ijk} 
\sin \left [\frac{\pi}{I} (i+1/2)(l+1)\right ],\\
v_{lmk}
\nn & = & \sum_{j=0}^{J-1}u_{ljk} 
\sin \left [\frac{\pi}{J} (j+1/2)(m+1)\right ],\\
\hat{\rho}_{lmn}
& = & \sum_{k=0}^{K-1}v_{lmk}\sin \left [\frac{\pi}{K}(k+1/2)(n+1)\right ].
\ea
Then, if from $\hat{\rho}_{lmn}$ we have determined $\hat{\Phi}_{lmn}$, we 
can reverse the process in parallel with a DST-III transform to find 
$\Phi_{ijk}$,
\ba
\label{inverse_sine_transform_partial_sum}
w_{lmk}
\nn & = & \frac{2}{K} \sum_{n=0}^{K-2} \hat{\Phi}_{lmn}
\sin \left [\frac{\pi}{K} (k+1/2)(n+1)\right ] \\
\nn & & + \frac{(-1)^k}{K}\hat{\Phi}_{lm,K-1}\\
y_{ljk}
\nn & = & \frac{2}{J} \sum_{m=0}^{J-2}w_{lmk}
\sin \left [\frac{\pi}{J} (j+1/2)(m+1)\right ] \\
\nn & & + \frac{(-1)^j}{J} w_{l,J-1,k}\\
\Phi_{ijk}
\nn & = & \frac{2}{I} \sum_{l=0}^{I-2}y_{ljk}
\sin \left [\frac{\pi}{I} (i+1/2)(l+1)\right ] \\
 & & + \frac{(-1)^i}{I} y_{I-1,j,k}.
\ea
We derive the algebraic connection between $\hat{\rho}_{lmn}$ and 
$\hat{\Phi}_{lmn}$ using a centered, second-order finite difference 
expression
\ba
\label{eqn:fdpoisson}
\left( \nabla^2 \Phi \right)_{ijk}
\nn & = & (\Phi_{i+1,jk} - 2\Phi_{ijk} + \Phi_{i-1,jk})/(\Delta x)^2\\
\nn & & + (\Phi_{i,j+1,k} - 2\Phi_{ijk} + \Phi_{i,j-1,k})/(\Delta y)^2\\
\nn & & + (\Phi_{ij,k+1} - 2\Phi_{ijk} + \Phi_{ij,k-1})/(\Delta z)^2\\
    & = & 4\pi \rho_{ijk} .
\ea
Upon substituting the Fourier transform we obtain 
\be
\hat{\Phi}_{lmn} = -4\pi \frac{\hat{\rho}_{lmn}}{\kappa_{lmn}^2},
\ee
where
\ba
\kappa_{lmn}^2
\nn & = & \frac{2}{(\Delta x)^2}
\left[1-\cos\left ( \frac{\pi (l+1)}{I}\right)\right] \\
\nn & & + \frac{2}{(\Delta y)^2}
\left[1-\cos\left ( \frac{\pi (m+1)}{J} \right ) \right ] \\
\nn & & +  \frac{2}{(\Delta z)^2}
\left[1-\cos\left ( \frac{\pi (n+1)}{K} \right ) \right ] ,
\ea
which is unchanged from Broderick and Rathore.  Note that determining 
$\kappa_{lmn}$ using (\ref{eqn:fdpoisson}) immediately makes the method 
algebraically convergent, but as we will see this is consistent with our 
handling of the boundary conditions.

We next consider how to trick the DST into providing a solution with an 
appropriate, asymptotically-falling $\mathcal{O}(1/r)$ field at the boundary 
of the domain.  This starts with deciding what the field at the boundary 
should be.  We rely on most of the mass being confined to the inner region 
of the domain and use a multipole expansion up to some order $l_{\rm max}$
\be
\label{multipole_expansion}
\Phi^B(\vec{x}) = -\sum^{l_{max}}_{l=0} \sum^l_{m=-l} 
\frac{4\pi }{2l+1} r^{-(l+1)} Q_{lm} Y_{lm}(\theta,\phi),
\ee
to give an approximation for the asymptotic field once we have obtained 
a set of source moments
$$
Q_{lm} = \int d^3 x' {r'}^l Y^*_{lm}(\theta',\phi') \rho(\vec{x}') .
$$
We find $l_{\rm max} = 5$ is typically sufficient.

Given the assumptions implicit in use of the DST (odd symmetry across each
face of the domain), with little mass density near the boundary the field 
will approach the boundary linearly and vanish.  In our discrete 
representation the field at the ultimate physical zone (say $\Phi_{0,jk}$) 
will be odd-symmetric with respect to the field in the neighboring ghost 
zone (in this case $\Phi_{-1,jk}$).  The trick is to introduce a boundary 
mass distribution $\rho^B$ in the outermost physical zones that generates 
just the right kink in the discrete field so that $\Phi \simeq \Phi^B$ in the 
ultimate physical zones while still being consistent with the odd symmetry 
at each domain face.

We can illustrate this in one dimension.  Let $i=0$ be the first physical 
cell at $x=+\Delta x/2$.  The DST requires that the field vanish at $x=0$, 
which is not at a field sample but corresponds to fractional location 
$i=-1/2$.  Instead, if we carry a neighboring ghost zone at $x=-\Delta x/2$ 
($i=-1$), the field there satisfies $\Phi_{-1}= -\Phi_{0}$.  The average 
between these two, the implied value on the domain boundary, is of course 
zero.  Now imagine instead that the first few field samples 
($\Phi_2$, $\Phi_1$, and $\Phi_0$) trend smoothly toward some (nonzero) 
$\Phi^B = \Phi^B_{-1/2}$ assumed to exist at $x=0$.  If we smoothly 
extrapolate to the first ghost zone we would have an implied value there of
\be
\label{eqn:phistarbc}
\Phi^*_{-1} = 2 \Phi^B_{-1/2} - \Phi_0 . 
\ee
If we evaluated (\ref{eqn:fdpoisson}) in the ultimate ($i=0$) location we 
would write
\ba
\label{eqn:phistarelliptic}
\Phi^*_{-1,jk} &-& 2\Phi_{0,jk} + \Phi_{1,jk} \\ &=& 
\Delta x^2 
\left( 4\pi \rho - \partial_y^2\Phi - \partial_z^2\Phi \right)_{0jk} .
\nn
\ea
In a solution to the elliptic system the last two equations could be combined
to ``close the mesh'' and encode the desired boundary condition.  In using the 
DST though, (\ref{eqn:phistarbc}) violates the required antisymmetry.  To 
get around this, substitute (\ref{eqn:phistarbc}) into 
(\ref{eqn:phistarelliptic}) and insert the DST-required condition 
$\Phi_{-1}= -\Phi_{0}$ to obtain
\ba
\Phi_{-1,jk} &-& 2\Phi_{0,jk} + \Phi_{1,jk} \\ &=& 
(\Delta x)^2 
\left( 4\pi \rho - \partial_y^2\Phi - \partial_z^2\Phi \right)_{0,jk} 
- 2 \Phi^B_{-1/2,jk} .
\nn
\ea
We can interpret this last piece on the right-hand side as a source
$\rho^B_{0,jk}$, where
\be
\rho_{0,jk}^B = -\frac{2\Phi_{-1/2,jk}}{4\pi (\Delta x)^2} ,
\ee
that adds to the real mass density $\rho$.  This image mass density is 
one zone thick.  The implied loss of differentiability in $\Phi$ is the 
primary reason why the spectral method will not converge exponentially, 
and is in this case algebraic and second order.

The discussion above was confined to one face of the domain.  We generalize 
by placing image mass density in the outermost zones on all six faces 
of the computational domain, 
\ba
\nn
4\pi \rho^B_{ijk} = &-&\frac{2}{(\Delta x)^2}
\left(\delta_{i,0}\Phi^B_{-1/2,jk}+\delta_{i,I-1}\Phi^B_{I-1/2,jk}\right) \\
\nn
&-&\frac{2}{(\Delta y)^2}
\left(\delta_{j,0}\Phi^B_{i,-1/2,k}+\delta_{j,J-1}\Phi^B_{i,J-1/2,k}\right) \\
\nn
&-&\frac{2}{(\Delta z)^2}
\left(\delta_{k,0}\Phi^B_{ij,-1/2}+\delta_{k,K-1}\Phi^B_{ij,K-1/2}\right) . \\
\ea
Given our cell centering, this image distribution differs from Broderick 
and Rathore by a factor of 2.  Once $\Phi$ has been determined within the 
domain, extrapolated values can be placed in the surrounding ghost zones 
[e.g., (\ref{eqn:phistarbc})] as needed.

To test the Poisson solver, a set of compact density distributions are 
constructed that have associated known analytic solutions for $\Phi$.  The 
mass is confined within a sphere of unit radius near the center of the 
domain.  The density is tapered to zero sufficiently smoothly so as to not 
affect the order of convergence of the method \cite{boyd2001}.  Each 
test distribution has a different angular multipole and since the equation 
is linear we use a superposition 
\be
\rho = \sum_{l=0} c_l r^l \left ( 1- r^2 \right )^3 P_l (\cos\gamma), 
\qquad {\rm for\ } r<1 ,
\ee
where $\rho=0$ for $r>1$.  Different orientations ($\gamma$) can be tested, 
as well as different relative multipole strengths $c_l$.  We tested 
configurations where $l=0$ dominated and ones where it did not contribute 
significantly.  In our highest resolution test with domain length $L=4$ 
in three dimensions with $512^3$ zones, the local error is less than 
0.1\% over most of the domain.

%%%%%%%%%%%%%%%%%%%%%%%%%%%%%%%%%%%%%%%%%%%%%%%%%%%%%%%%%%%%%%%%%%%%%
\begin{figure}[t]
{ 
\begin{center}
      \includegraphics[scale=1.03]{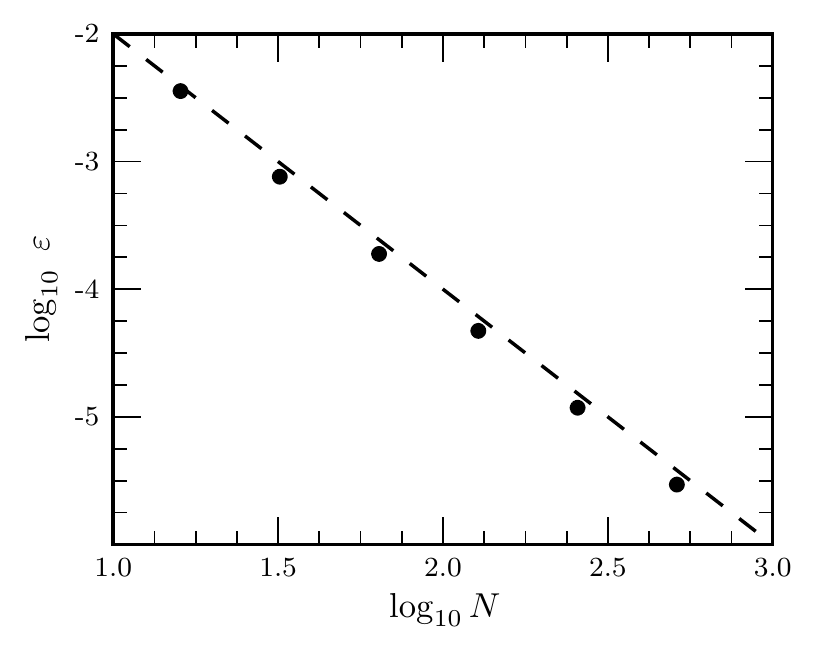} 
\end{center} \caption
      {\label{figure:convergence} 
Convergence of $L_2$ error between numerical $\Phi^{\rm num}$ and 
analytic $\Phi^{\rm analytic}$ solutions of a self-gravity test 
problem.  The $L_2$ error is plotted against number of zones, $N$, in one 
direction.  The total number of zones in a model is $N^3$.  The method 
exhibits second-order convergence.
} }
\end{figure}  
%%%%%%%%%%%%%%%%%%%%%%%%%%%%%%%%%%%%%%%%%%%%%%%%%%%%%%%%%%%%%%%%%%%%%

The solution should converge at second-order rate as the number of grid 
points or basis functions increases.  The behavior is illustrated in 
Fig. \ref{figure:convergence}.  Consider a domain with equal length sides 
$L$.  Take $I=J=K=N$.  The total number of grid locations is $N^3$.  The 
$L_2$ error is 
\be
\varepsilon = \left[\frac{1}{N^3} \sum_{ijk} 
\left(\Phi^{\rm num}_{ijk}-\Phi^{\rm analytic}_{ijk}\right)^2 \right]^{1/2} .
\ee
Figure \ref{figure:convergence} shows strict second-order convergence in 
the $L_2$ error over a range of five doublings of $N$.

\section{Octupole tidal tensor in the FNC frame}
\label{sec:cijk}

The nonzero components of the octupole tidal tensor in the FNC frame 
for parabolic $\tilde{E} = 1$ orbits are given by
\begin{widetext}
\ba\label{tidal_tensor_Cijk}
C_{111}
\nn  & = & \frac{3M}{4r^4} \Big[ 3\left( 1+\frac{7 \tilde{L}^2}{3r^2}\right)
\cos\Psi + 5  \left (1+\frac{\tilde{L}^2}{r^2} \right ) \cos 3\Psi
-  6 \frac{\tilde{L}}{r} U^r \left (1 + \frac{5\tilde{L}^2}{3r^2} \right )
\sin\Psi \\
\nn & - & 10  \frac{\tilde{L}}{r} U^r \left(1+\frac{\tilde{L}^2}{r^2}\right)
\cos 2\Psi \, \sin\Psi \Big ] V_2^{-1}\\
C_{131}
\nn & = & C_{311} = C_{113} =  \frac{M}{4r^4} \Big [  \frac{\tilde{L}}{r} U^r \left ( 1+\frac{5\tilde{L}^2}{r^2}\right )\cos\Psi
                                  + 15  \frac{\tilde{L}}{r} U^r \left ( 1 + \frac{\tilde{L}^2}{r^2} \right ) \cos 3\Psi  \\
\nn & + & 18 \left ( 1+\frac{11\tilde{L}^2}{9r^2} \right )\sin\Psi + 30 \left (1+\frac{\tilde{L}^2}{r^2} \right )\cos 2\Psi \, \sin\Psi \Big ] V_2^{-1}\\
C_{122}
\nn & = & C_{212} = C_{221} = \frac{M}{r^4} \Big [-3\left ( 1 + \frac{7\tilde{L}^2}{3r^2} \right ) \cos\Psi + \frac{\tilde{L}}{r} U^r \left (1+\frac{5\tilde{L}^2}{r^2} \right ) \sin\Psi \Big ] V_2^{-1}\\
C_{133}
\nn & = & C_{313} = C_{331} = \frac{M}{4r^4} 
\Big [ 3 \left (1 + \frac{7\tilde{L}^2}{3r^2} \right )\cos\Psi 
-15 \left (1+\frac{\tilde{L}^2}{r^2} \right )\cos 3\Psi \\
\nn & + & 14 \frac{\tilde{L}}{r} U^r \left (1+\frac{5\tilde{L}^2}{7r^2} 
\right ) \sin\Psi + 30 \frac{\tilde{L}}{r} U^r \left( 
1+\frac{\tilde{L}^2}{r^2}\right ) \cos 2\Psi \, \sin\Psi \Big ] V_2^{-1} \\
C_{322}
\nn & = &  C_{232} = C_{223} = -\frac{M}{r^4} \Big [\frac{\tilde{L}}{r} U^r \left (1+\frac{5\tilde{L}^2}{r^2} \right ) \cos\Psi + 3 \left (1+\frac{7\tilde{L}^2}{3r^2} \right )\sin\Psi \Big ] V_2^{-1}\\
C_{333}
\nn & = & \frac{3M}{4r^4} \Big [\frac{\tilde{L}}{r}U^r \left ( 1+ \frac{5\tilde{L}^2}{r^2} \right )\cos\Psi - 5\frac{\tilde{L}}{r}U^r \left (1+\frac{\tilde{L}^2}{r^2} \right) \cos 3\Psi  \\
& - & 12 \left (1+\frac{2\tilde{L}^2}{3r^2} \right )\sin\Psi + 20 \left (1+\frac{\tilde{L}^2}{r^2} \right )\sin^3\Psi \Big ] V_2^{-1}.
\ea
\end{widetext}
\bibliographystyle{apsrev4-1}
\bibliography{FNC}

\end{document}